# Meteoritic Proteins with Glycine, Iron and Lithium

Malcolm W. McGeoch[1], Sergei Dikler[2]
and Julie E. M. McGeoch[3*].

[1] PLEX Corporation, 275 Martine St., Suite 100, Fall River, MA 02723, USA.
[2] Bruker Scientific LLC, 40 Manning Rd, Billerica MA 01821.
[3] Department of Molecular and Cellular Biology, Harvard University, 52 Oxford St., Cambridge MA 02138, USA.
*Corresponding author. E-mail: mcgeoch@fas.harvard.edu

## ABSTRACT

We report that polymer amide [1] with a protein backbone of mainly glycine units and iron is present in the CV3 meteorites Acfer 086, Allende and KABA. The evidence for this is from particles of these meteorites after Folch extraction being analyzed by MALDI mass spectrometry and from the 3D physical structures that form in the various Folch solvent phases. The two physical forms we observe are branching rods and entrapping spheres on the 100micron scale. Two potential molecular structures of polymer amide are presented in this report. One we term hemolithin of mass 2320Da, contains glycine, hydroxy-glycine, Fe, O and Li. The other, we term hemoglycin of mass 1494Da is of glycine, hydroxy-glycine, Fe and O. Hemoglycin is connected covalently in triplets by silicon to form a triskelion. Analysis of the complete spectrum of isotopes associated with each molecular fragment shows very high $^{2}H$ enhancement above terrestrial averaging 25,700 parts per thousand (sigma = 3,500, n=15), confirming extra-terrestrial origin and hence the existence of these molecules within the asteroid parent body of the CV3 meteorite class. The hemoglycin triskelia join via silicon bonds into an extended lattice, as seen in mass spectrometry of lattice fragments at m/z 4641 and above [2]. The identification of hemolithin and hemoglycin required careful methodology at room temperature with the minimum of steps up to mass spectrometry analysis ensuring that only the laser step fragmented the molecules. The 3D structures were imaged through permanently closed Folch extraction V-vials at a magnification of 10.

## INTRODUCTION

Although individual amino acids have been found in abundance in carbonaceous meteorites there have been only two reports of polymers of amino acids, the first being of di-glycine [3], and the more recent report [1] being of large polymers of mainly glycine in the CV3 class carbonaceous chondrite Allende. In follow-on work to [1] a 4641Da molecular entity was discovered in Allende and Acfer 086 [2] with extra-terrestrial isotope enhancement that confirmed that these unexpected molecules were not artifacts due to terrestrial contamination. In the present study a state-of-the-art mass spectrometer [4] generated much-improved signal-to-noise ratios that allowed us to discover via the $^{54}Fe$ isotope side peak that iron was also present and bonded to the glycine and hydroxy-glycine, in specific arrangements. Also, via its $^{6}Li$ side peak, lithium was found to be a frequent component. Much of the present paper is devoted to the analysis via isotope satellites and the reconstruction of a 1494Da core molecular entity (hemoglycin) that gives rise to several higher m/z peaks in a mass spectrum. In the course of many visual observations of solution extractions (methods below) there appeared consistently a branched form on the 100micron scale with repeating 120degree angles, a form that could not be produced by any agent other than a protein. This appeared to be the macroscopic manifestation of much smaller-scale molecular shape, such as a triskelion. Spherical “vesicles” also were observed, suggesting a sheet-forming ability of the type offered by a triskelion molecule. Following the conclusion from mass spectrometry that a core hemoglycin molecule with mass 1494Da was present, it was possible to piece together how it formed triskelia. The sheet polymer of hemoglycin triskelia should have remarkable strength, comprising a 2D network of covalently bonded entities. To date this has presented a difficult target for mass spectrometric analysis, but the present results suggest directions for future X-ray and electron diffraction studies.
Following the presentation of methods, the smaller molecular fragments with m/z < 1000, generated in intense MALDI laser conditions, are reviewed. Less severe conditions are then applied in order to go to masses as high as 2,000 so as to understand the polymer structure. The discussion aims to place this molecule in context, whether its origin was in the proto-solar disc, or in molecular clouds well before the solar system condensed.

## METHODS

### Sample preparation

In summary: Micron particles of Acfer-086 and KABA were Folch extracted [1,5] into 4 phases that we label P1 –> P4: 1) top polar phase (P1); 2) above interphase (P2); 3) interphase (P3); 4) bottom chloroform phase (P4). Specifically, Acfer-086, a CV3 meteorite sample from Harvard Mineralogical and Geological Museum (Source: Agemour, Algeria, found 1989-90, TKW 173g) was delivered from the Museum in a sealed container to the Hoffman, Earth and Planetary Science clean room.
The KABA sample was sent to JEMMcG for analysis by Prof. Dr. Béla Baráth and Teofil Kovács PhD, Deputy Director, Museum of the Debreceni Reformatus Kollegium Kalvin ter 16, H4026, Debrecen.

**Meteorite Etch**: In a clean room extractor hood, at room temperature with high airflow, the samples were hand held with powder-free nitrile rubber gloves while being etched [1] to a total depth of 6mm with diamond burrs. The diamonds had been vacuum-brazed at high temperature onto the stainless-steel burr shafts to avoid the presence of glue of animal origin and organics in general. Etching on a fracture face (not an original exterior weathered face) was via slow steady rotation of a burr under light applied force via a miniature stepper motor that did not have motor brushes and did not contribute metal or lubricant contamination to the clean room. Two shapes of burr were used, the larger diameter type, in two stages, to create a pit of diameter 6mm and depth 6mm, and the smaller conical burr to etch a finely powdered sample, of approximately 1μm particles, from the bottom of the pit without contacting the sides. After each stage the pit contents were decanted by inversion and tapping the reverse side and a new burr was used that had been cleaned by ultrasonics in deionized distilled (DI) water followed by rinsing in DI water and air-drying in a clean room hood. The powder from the third etch was decanted with inversion and tapping into a glass vial and stored at -16C. Sample weights were in the range 2-8mg.

### Folch Extraction of polymer amide from meteorites for 3D analysis and mass spectrometry

Dry micron-scale meteorite particles are transferred to a 1 or 2ml borosilicate glass V-vial via a pipet and chloroform/methanol/water (3.3/2/1) are added and gently mixed via swirling the vial to form the 2 Folch phases. A dark torus layer forms at the interphase of the heavy chloroform bottom phase and the top polar phase. Within hours the interphase torus layer consists of 3D triskelion structures. Spherical 100micron diameter 3D structures also form from some Folch preparations at the interphase layer. The organic content of a given Folch extraction varies related to "hot spots" of organic material in the meteorite as some etched particles contain more than others. "Hot spots" of organic matter in meteorites have been noted by other investigators and are potentially due to the uneven accretion of the original asteroid from which the meteorites derive. Folch extractions for 3D structural analysis are imaged daily and left permanently capped at room temperature, destined for synchrotron X-ray analysis.

Folch extractions for mass spectrometry are handled as follows: After 1 day of extraction at room temperature the Folch extract phases were pipetted off as 50-100µl aliquots from the top downwards through the phases, each extract being labeled P1 through P4 in the above order of phases. In a second identical extraction experiment, only the above-interphase P2 was collected in view of its larger complement of the molecule in question. The residual phases P3 and P4 from this second experiment were re-established by addition of chloroform, methanol and water into a third experiment that was kept at room temperature for 7 more days (totaling 8). A flocculent precipitate appeared at the interphase and was pipetted off to become sample 2A. These aliquots were then analyzed by MALDI mass spectrometry.

**Sample Preparation for MALDI MS.**

The extracts were prepared with three MALDI matrices: α-cyano-4-hydroxycinnamic acid (CHCA), sinapinic acid (SA) and 2,5-dihydroxybenzoic acid (DHB). The CHCA solution was prepared at 10 mg/mL in 50% acetonitrile, 0.1% TFA. The saturated SA solution was prepared in ethanol. The second SA solution for double layer preparation at 20 mg/mL was prepared in 50% acetonitrile, 0.1% TFA. The DHB solution at 45 mg/mL was prepared in 30% acetonitrile, 0.1% TFA. The extract aliquots were mixed 1:1 with the CHCA solution or the DHB solution and 1 µL of the mixture was spotted in duplicate on a new, never used MTP 384 Ground Steel MALDI plate. The double layer SA preparation was accomplished by spotting 0.5 µL of the saturated SA solution first followed by 1 µL of 1:1 mixture of extract and the second SA solution at 20 mg/mL on the same MALDI plate.

**MALDI-TOF and MALDI-TOF/TOF measurements**

All spectra were acquired on a Bruker RapifleX MALDI-TOF/TOF system [4] equipped with a scanning Smartbeam 3D laser operating at 10 kHz. The instrument was operated in reflector positive ion mode. The 10-bit digitizer was set to a sampling rate of 5.00 GS/s. The ion source voltage 1 (acceleration voltage) was set to 20.00 kV, the ion source voltage 2 (extraction voltage) was at 17.44 kV and lens voltage was at 11.60 kV. The pulsed ion extraction time was set to 160 ns. Reflector voltage 1 was at 20.82 kV, reflector voltage 2 at 1.085 kV and reflector voltage 3 at 8.60 kV. Four thousand laser shots were acquired per spectrum.

The MS/MS spectra were acquired in TOF/TOF mode and 20000 laser shots were added per spectrum. In this mode parent ions were selected in an ion selector and then parent ions together with the corresponding fragment ions formed by monomolecular laser-induced dissociation were sent into a second source region located in the retractable MS/MS ion optics module. Ions were pulsed out of the second source into a field free region, followed by passage through a three-stage gridless reflector to reach the detector. Ion source voltage 1 was set to 20.00 kV, ion source voltage 2 was 19.45 kV and the lens voltage was at 18.00 kV. Drift tube voltage 1 was at 14.00 kV, drift tube voltage 2 at 18.70 kV and the MS/MS pulse voltage at 2.60 kV. Reflector voltage 1 was at 23.80 kV, reflector voltage 2 at 1.79 kV and reflector voltage 3 at 9.65 kV.

## RESULTS

### Visible 3D structures in Folch extraction V-vials

3D structures form over days to weeks in all phases of the Folch extraction and most are salt crystals when subject to X-ray diffraction. Other 3D forms that are present dominantly in phases P2 and P3 (see methods) suggest the presence of polymer amide or fiber-like protein with a C-C-N backbone, being alternately stiff and flexible, producing structures that repeat regularly. The protein fibers are mostly at the interphase between chloroform and water/methanol. Some grow on the wall of the V-vial (Figure 1), while others resemble triskelia (Figures 2 and 3) and/or a lattice entrapping air in a sphere on the 100micron diameter scale (Figure 4). The spheres entrapping air are stable for weeks. These 3D structures, analyzed briefly by X-ray diffraction at Harvard Chemistry Department, diffracted poorly indicating the need for synchrotron analysis which is planned to occur at APS Argonne. In some cases there were mineral-like identifications with weaker superimposed patterns indicating large unit cells as high as 54.68 Angstroms. We report this initial 3D structure now as it is important as an indicator of the presence of polymer in Acfer-086 and KABA that complements and supports the mass spectrometry findings. In the present report all mass spectrometry is from Acfer 086. 3D visible structures are from both KABA and Acfer 086.

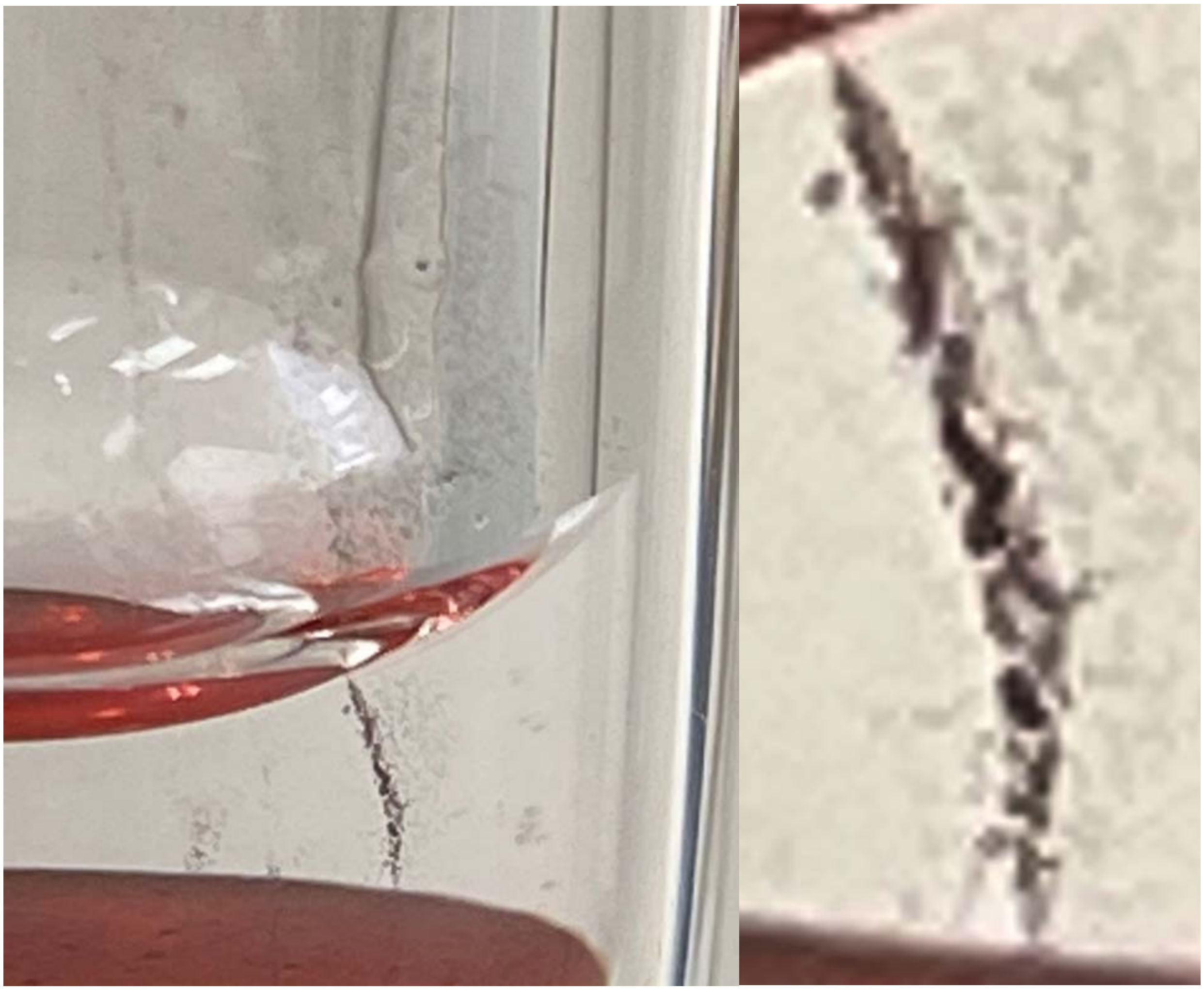

**FIGURE 1. 3D structure from KABA particles grows from the interphase layer into the polar phase and then into air above the polar layer (left ). Width of the structure in the enlargement on the right = 75µm. The red coloration is from Sudan III dye added to better visualize the 3D structures.**

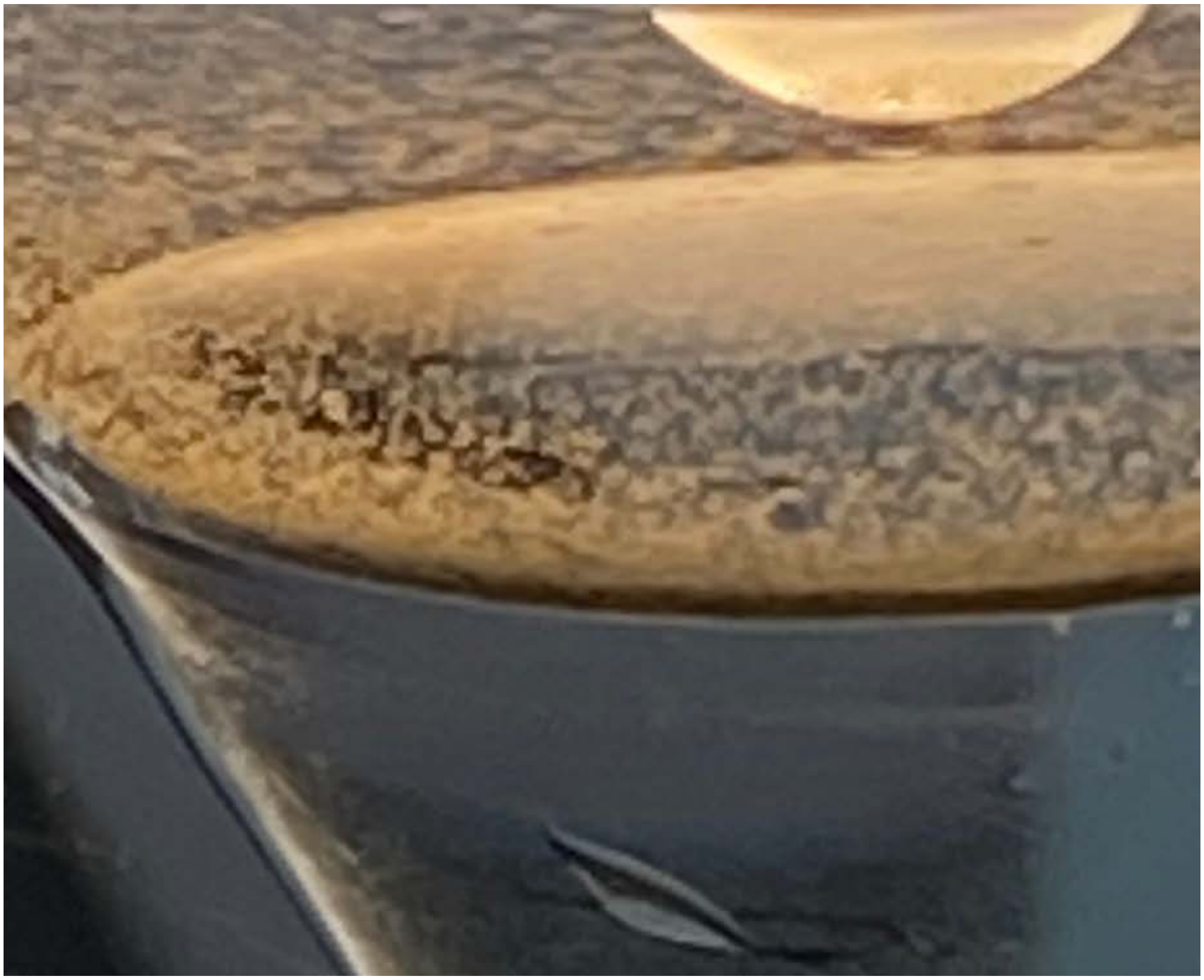

**Figure 2. Triskelion macrostructures at the interphase layer from Acfer-086. In the lower chloroform phase a salt crystal is visible. The interphase diameter is 9mm.**

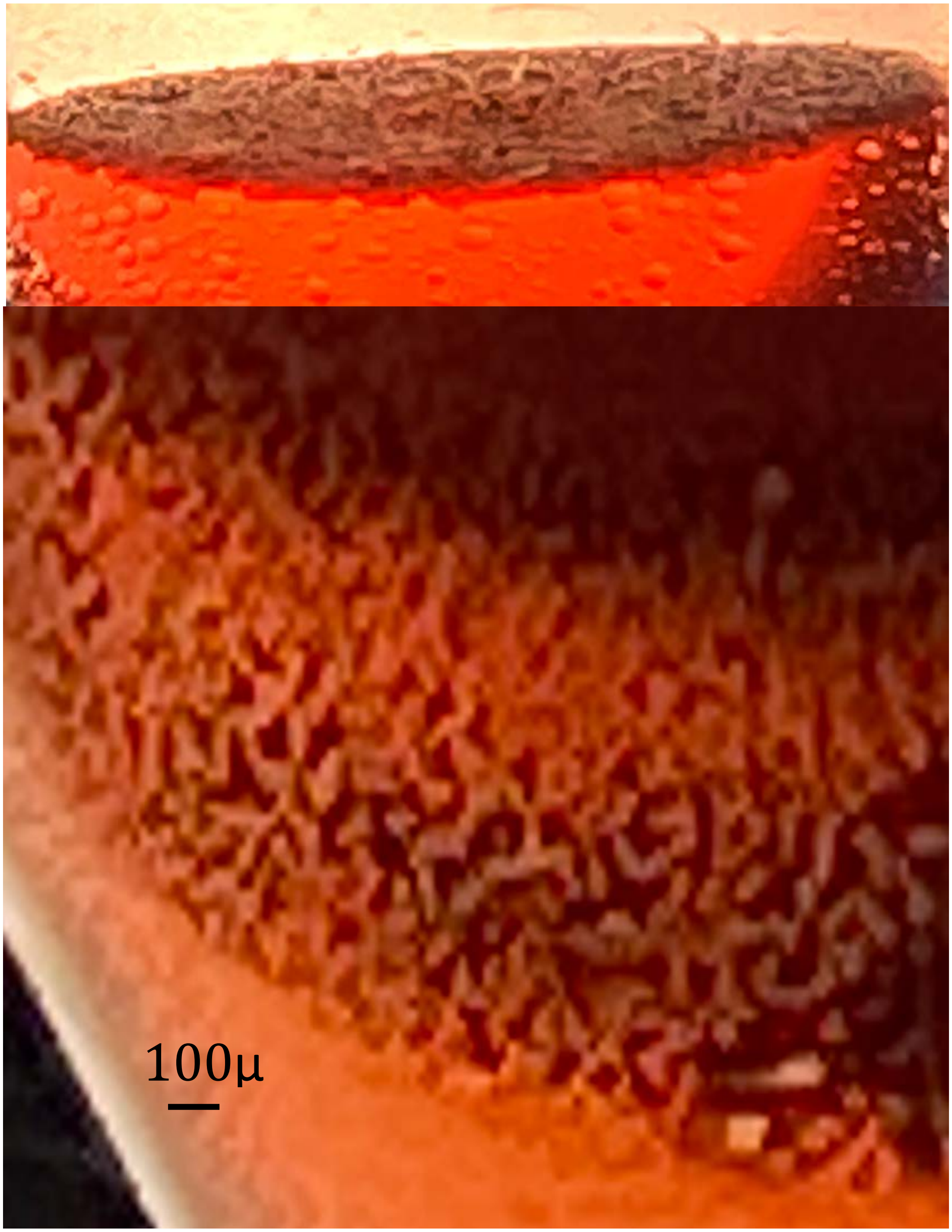

**Figure 3 Triskelion macrostructures from KABA at the interphase layer. The top image shows the entire interphase layer of diameter 9mm with many structures. The lower image is a horizontal view into the layer at higher magnification. The red coloration is from Sudan III dye added to better visualize the 3D structures.**

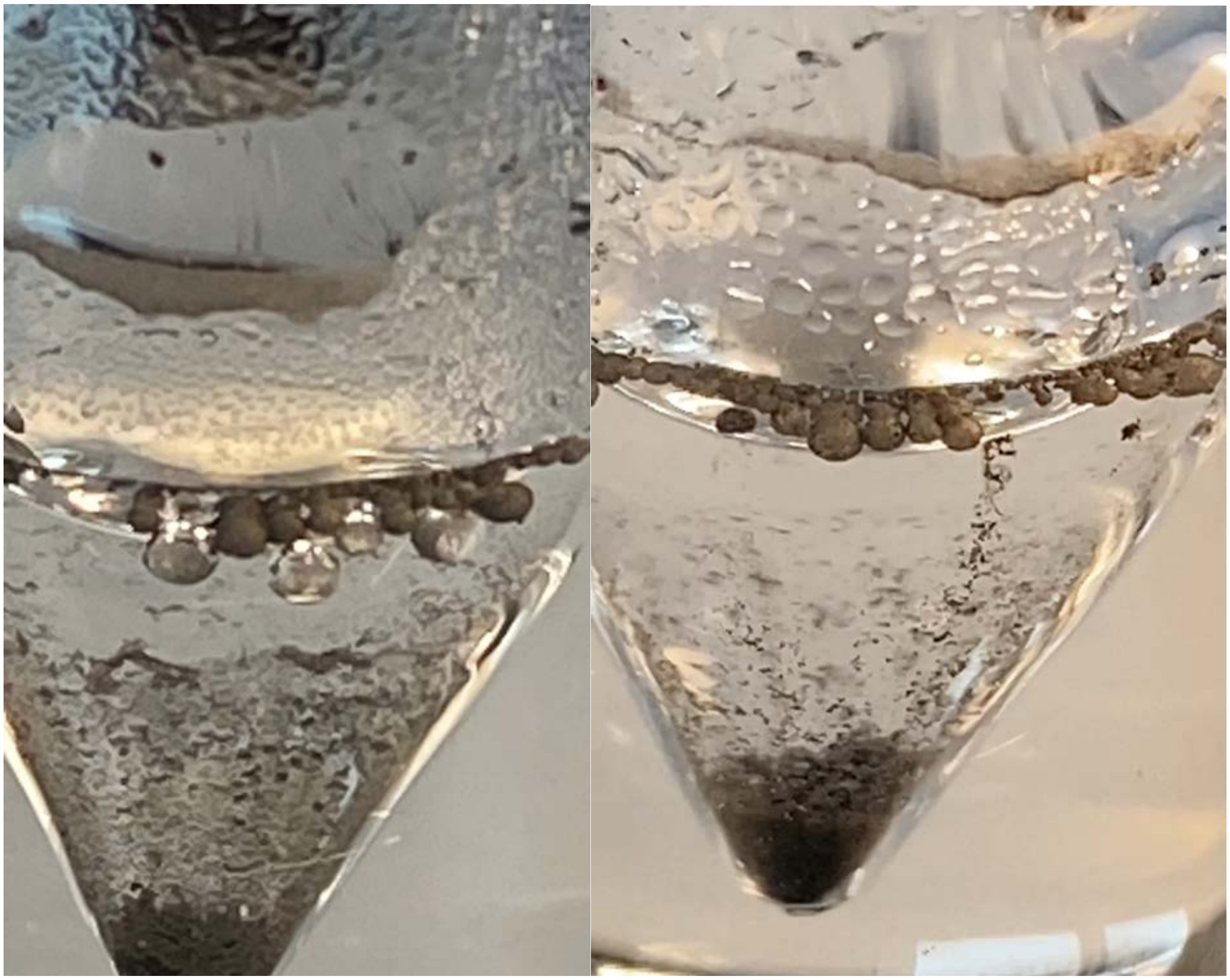

**Figure 4. Spherical hollow vesicles from KABA forming at the interphase between air and water. Vesicles immediately after formation on gently mixing KABA micron particles with water (left) are still intact and stable 19days later (right). Angled 3D forms are also on Vial wall. The largest vesicles have 500μm diameter.**

**Small molecular fragments: finding the dominant constituents.**

The identity of these unknown extraterrestrial proteins could not be established using existing software to identify the various peaks. This difficulty is compounded further when the molecule contains in addition to chains of glycine, hydroxylation on some of the glycine units and bonds to Fe, Si and Li. The hundreds of peaks from the fragmentation of polymer were identified by the following approach:

With the MALDI parameters optimized for fragmentation in the 1-1000 m/z range, a study was performed on phase P2, which already had been established as the phase carrying most of the molecules under study. The 1-1000 m/z range is cluttered with matrix clusters that are difficult to distinguish from real signals, although most matrix clusters can be predicted [6,7]. Our samples had a relatively high content of alkali atoms (described in [2]), which made clustering very severe, but rather than risk contamination by processing to remove alkalis we adopted a different stratagem. Instead of subtracting from a spectrum all the calculated clusters as in our prior analysis [1], we ran the same sample separately with each of the matrices CHCA, SA and DHB, and only selected the peaks that appeared in all three spectra for analysis, guaranteeing the elimination from this list of essentially all clusters. All such peaks for sample P2 in the 1-1000 range, with peak counts more than 15 times greater than noise, are listed in Table 1, together with their proposed composition in terms of only glycine and hydroxy-glycine amino acid residues, which have mono-isotopic residue masses 57.02146 and 73.01638 respectively. The matrix that yielded the largest peak count out of the three is listed. A large proportion of the peaks (16 out of 36) could be constructed from small polymers containing only these two residues provided that we included the possibility of three types of termination:

a) the bare combination of residues;

b) the combination of residues plus one proton;

c) the combination of residues plus an "aqueous" termination 'OH' + 'H'.

Other terminations that were close to a match are denoted by italics in the table, but these molecular species were not counted. By enumeration of the possibilities we determined that the probability of achieving a compositional match with any of these (type a, b or c) terminations by chance was 0.256, and found using the binomial distribution that the chance of our having obtained 16 or more matches with exactly the terminations (type a, b or c) out of 36 peaks was only 0.010. This confirmed the presence of these two amino acids, but did not exclude other possibilities. The average fragment length in this set was 4.5 residues. Lower temperature conditions within the MALDI plasma would allow larger molecular fragments to survive, as found in the following section.

**Table 1. Low molecular weight fragments from P2 with assignments**

| Observed m/z | Peak counts | Matrix | Assignment (Gly, $Gly_{OH}$) + t | $^2H$ fitted | $^{15}N$ preset | Calc. m/z |
|---|---|---|---|---|---|---|
| 57.074 | 132,670 | CHCA | (1,0) | | | 57.021 |
| 73.002 | 6,669 | DHB | (0,1) | | | 73.016 |
| 74.102 | 678,662 | CHCA | (0,1) + H | | | 74.024 |
| 114.096 | 101,559 | CHCA | (2,0) | | | 114.043 |
| 128.113 | 734,815 | CHCA | - | | | - |
| 130.166 | 172,225 | CHCA | (1,1) | | | 130.038 |
| 161.046 | 746,629 | DHB | - | | | - |
| 169.082 | 1,238,632 | CHCA | - | | | - |
| 172.045 | 1,511,547 | CHCA | (3,0) + H | | | 172.072 |
| 185.078 | 239,372 | CHCA | - | | | - |
| 197.071 | 299,125 | CHCA | - | | | - |
| 237.092 | 313,018 | SA | (0,3) + 'OH' + H | | | 237.060 |
| 244.271 | 260,294 | CHCA | (3,1) | | | 244.081 |
| 256.270 | 63,527 | CHCA | - | | | - |
| 272.300 | 72,300 | CHCA | - | | | - |
| 279.102 | 555,493 | CHCA | (2,2) + 'OH' + *2H* | | | 279.094 |
| 286.280 | 159,152 | CHCA | (5,0) + H | | | 286.115 |
| 293.134 | 248,245 | CHCA | (0,4) + H | | | 293.073 |
| 295.099 | 89,961 | SA | - | | | - |
| 304.312 | 454,322 | SA | (5,0) + 'OH' + *2H* | | | 304.126 |
| 316.329 | 178,473 | DHB | (3,2) - *H* | 30,000 | 1,015 | 316.089 |
| 332.342 | 239,731 | SA | (2,3) - *H* | 30,000 | 1,015 | 332.084 |
| 337.129 | 219,817 | CHCA | - | | | - |
| 375.142 | 43,128 | SA | (4,2) + H | | | 375.126 |
| 401.086 | 504,156 | CHCA | (7,0) + *2H* | 30,000 | 1,015 | 401.166 |
| 433.172 | 147,707 | SA | (6,1) + ‘OH’ + H | | | 433.156 |
| 449.168 | 316,351 | SA | (5,2) + ‘OH’ + H | | | 449.151 |
| 477.147 | 27,497 | SA | (8,0) + ‘OH’ + *4H* | | | 477.206 |
| 487.315 | 145,270 | DHB | (6,2) - *H* | 30,000 | 1,015 | 487.154 |
| 493.321 | 274,754 | CHCA | (7,1) + ‘OH’ + *4H* | | | 493.201 |
| 503.091 | 53,862 | SA | - | | | - |
| 509.296 | 213,767 | CHCA | (6,2) + ‘OH’ + *4H* | | | 509.196 |
| 515.302 | 105,500 | CHCA | (9,0) + *2H* | | | 515.209 |
| 531.275 | 77,837 | CHCA | (9,0) + ‘OH’ + H | | | 531.204 |
| 547.247 | 38,571 | CHCA | (8,1) + ‘OH’ + H | | | 547.199 |
| 563.217 | 25,057 | CHCA | (7,2) + ‘OH’ + H | | | 563.194 |

**Spectra in the high m/z range 1 - 5,000**

Using less intense laser conditions and two matrices CHCA and SA, mass spectra were collected in the m/z range 1-5000 on phases P2 and P3. These four data sets were cross-correlated and only the peaks that appeared in all four (20 in number) were included in a first list of candidate species that is reproduced in Table 2. This selection approach mostly ensured that the peaks were not matrix artifacts, i.e. polymers of matrix alone or complexes of matrix with the molecule under study, because the different matrix masses (189 for CHCA and 224 for SA) would give rise to different complexes, and these would not cross-correlate between the spectra. For continuity with our previous analyses [1,2] we show in Table 2 the amino acid composition of glycine, hydroxy-glycine and alanine that we would have assigned if this had been the total of information available. The average residue count implied by this assignment was 23, a large increase above the previous average of 11 residues that we had observed [2].

**Table 2. List of peaks that appeared in all four tests, with the (now outdated) assignment in terms of three amino acids and an aqueous termination.**

| Integer name of peak | Average m/z (four traces) | Historical assignment according to [1] in terms of Gly, $Gly_{OH}$, Ala, and a termination. | | | | Calc. mass based on assignment |
|---|---|---|---|---|---|---|
| | | Gly | $Gly_{OH}$ | Ala | termination | |
| 693 | 693.256 | 8 | 3 | 0 | 18 | 693.231 |
| 947 | 947.309 | 15 | 1 | 0 | 18+H | 947.357 |
| 963 | 963.274 | 14 | 2 | 0 | 18+H | 963.352 |
| 1001 | 1001.502 | 16 | 0 | 1 | 18 | 1001.391 |
| 1011 | 1011.567 | 11 | 5 | 0 | 18+H | 1011.336 |
| 1027 | 1027.542 | 10 | 6 | 0 | 18+H | 1027.331 |
| 1033 | 1033.537 | 14 | 2 | 1 | 18 | 1033.381 |
| 1417 | 1417.513 | 22 | 1 | 1 | 18+H | 1417.544 |
| 1433 | 1433.486 | 21 | 2 | 1 | 18+H | 1433.539 |
| 1449 | 1449.532 | 20 | 3 | 1 | 18+H | 1449.534 |
| 1483 | 1483.864 | 18 | 6 | 0 | 18+H | 1483.503 |
| 1503 | 1503.741 | 21 | 2 | 2 | 18 | 1503.568 |
| 1519 | 1519.697 | 20 | 3 | 2 | 18 | 1519.563 |
| 1535 | 1535.683 | 19 | 4 | 2 | 18 | 1535.558 |
| 1551 | 1551.650 | 18 | 5 | 2 | 18 | 1551.553 |
| 1567 | 1567.729 | 17 | 6 | 2 | 18 | 1567.548 |
| 1639 | 1639.630 | 17 | 6 | 3 | 18+H | 1639.593 |
| 1995 | 1995.993 | 27 | 6 | 0 | 18 | 1995.688 |
| 2012 | 2013.019 | 26 | 7 | 0 | 18+H | 2012.691 |
| 2124 | 2124.882 | 28 | 6 | 1 | 18+H | 2124.755 |
| | sum | 362 | 76 | 19 | | |

It was remarkable that all 20 peaks could be immediately assigned in this manner, which reinforced the certainty that the prior molecular type was again being observed, this time at higher signal-to-noise ratio. The assignments of Table 2, however, are now incorrect in light of the data to be presented below.

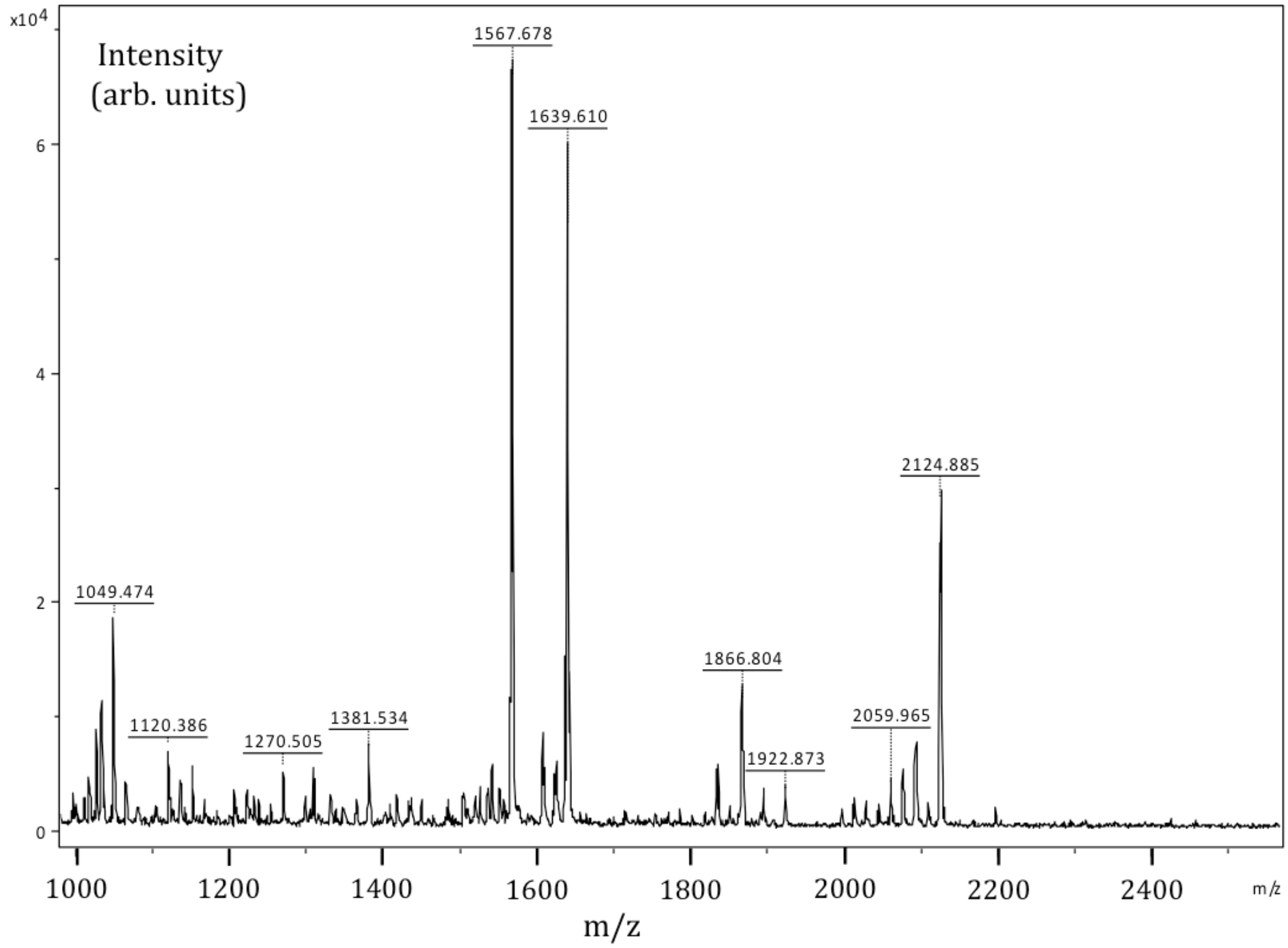

**Figure 5. Major m/z peaks in the P2 mass spectrum with matrix CHCA. The spectrum stops at m/z 2124 and in this range of m/z with matrix CHCA most peaks are related to the molecule under study.**

The fundamental mass spectrum from P2 is shown in Figure 5. Using matrix CHCA no major peaks are seen above m/z 2124.885, which we shall refer to by its integer part as just “2124”. Below this there are dominant peaks at 1866, 1639, 1567, and 1049. The highest intensity peak in this spectrum is at m/z 730, shown in detail, on the same intensity scale, in Figure 6. It was chosen in order to illustrate the isotopologues in a simple low molecular weight species with only one iron atom.

In conventional terminology m/z 730 is named the “monoisotopic” peak, defined as the peak within a cluster whose mass is derived using the dominant isotope of each element in the molecule. In organic molecules the dominant isotope has lower mass than other isotopes so the “monoisotopic” peak is indeed monoisotopic with respect to each of the constituent elements. With metals such as Fe and Li involved the “monoisotopic” peak can contain many additional combinations, for example the m/z 730 peak contains (in part) $^{54}$Fe + two H->D substitutions to recover the mass of $^{56}$Fe, however we continue to use that designation, and additionally designate the “monoisotopic” peak as the (0) component, surrounded by negative “isotopologues” (-4), (-3), (-2), (-1) to the low mass side, and by positive isotopologues (1), (2), (3) etc. to the high mass side.

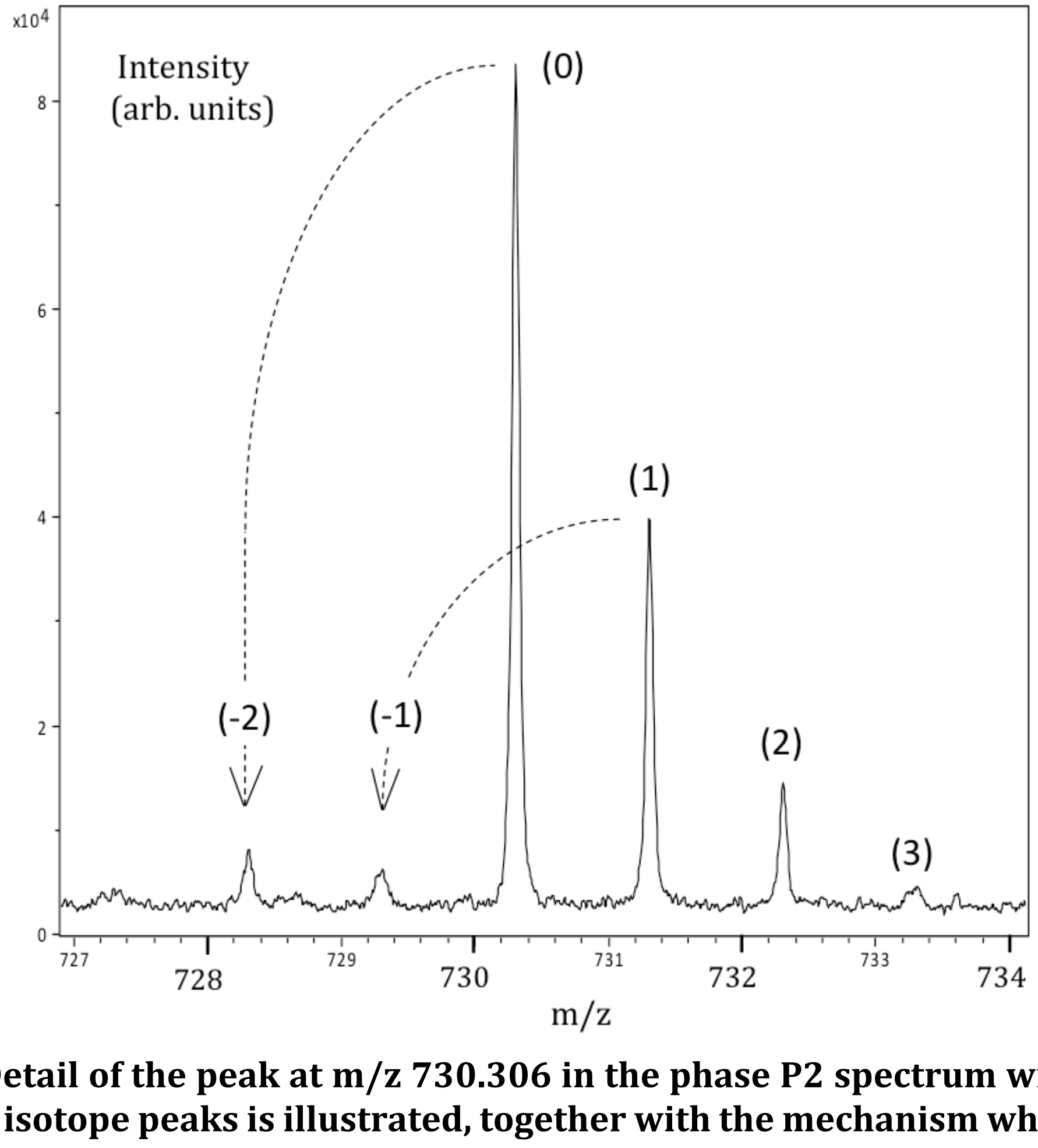

**Figure 6. Detail of the peak at m/z 730.306 in the phase P2 spectrum with CHCA. The labeling of isotope peaks is illustrated, together with the mechanism whereby Fe with majority $^{56}$Fe in the molecular peaks (0), (1), (2) etc. contributes to peaks 2 units to the left via its minority $^{54}$Fe isotope. Isotope fitting for 730 is listed in S3.**

Iron can be shown to be present in the ion at m/z 730 by its (-2) and (-1) isotopologues that relate respectively to the (0) m/z 730 peak and the (+1) peak in the abundance ratio 0.05845 : 0.91754 of iron isotopes $^{54}$Fe and $^{56}$Fe. There are in addition two less abundant stable iron isotopes that contribute to peaks on the high mass side, $^{57}$Fe with abundance 0.0212 and $^{58}$Fe at 0.0028, but these are not separately visible above the many other heavy isotope contributions at locations (1), (2), etc. due to D, $^{13}$C, $^{15}$N, and $^{18}$O.

The same characteristic (-2), (-1), (0), (1) pattern is seen throughout the present study, with varying isotopologue amplitudes indicating the presence of from 1 to 4 iron atoms in a species. For additional depth, 14 mass spectrum peaks with this pattern are shown in S5. As an example of the pattern the high intensity '1567' peak at S/N = 135 is shown in Figure 7. The 1567 isotope fit shown in Figure 8, with details in Table 3, shows that it has 3 iron atoms. The calculation of an isotope pattern from a given trial composition is described in S1. The signal-to-noise level around a main peak has to be high enough to distinguish between different numbers of iron atoms. The 54/56 abundance ratio is 0.064 [8], so a single iron atom gives a (-2)/(0) ratio of approximately 6.4% and the signal-to-noise ratio therefore has to exceed about 30 to distinguish between, say, one and two iron atoms, or three and four of them.

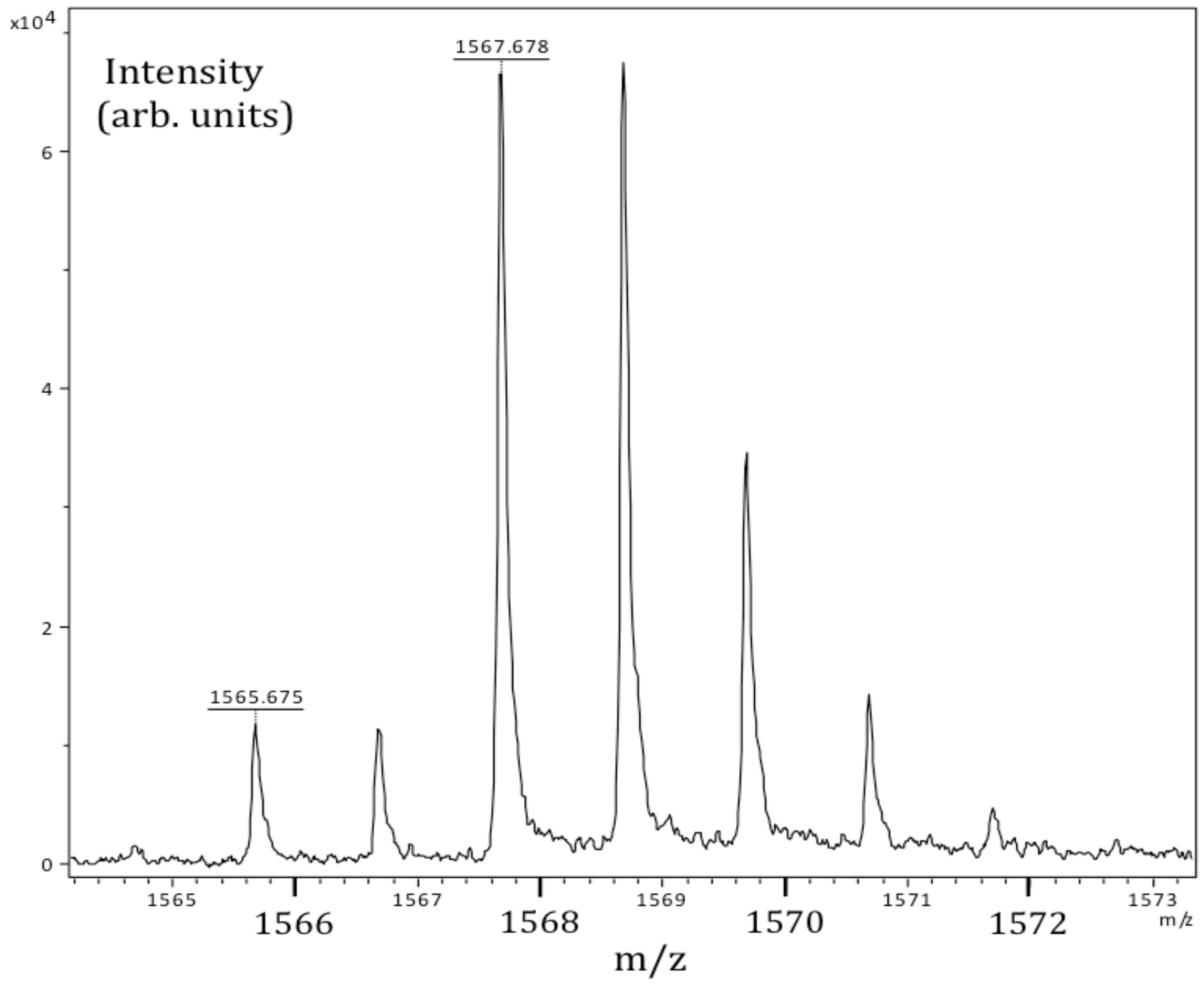

**Figure 7 The 1567.678 m/z peak from phase P2 with matrix CHCA**

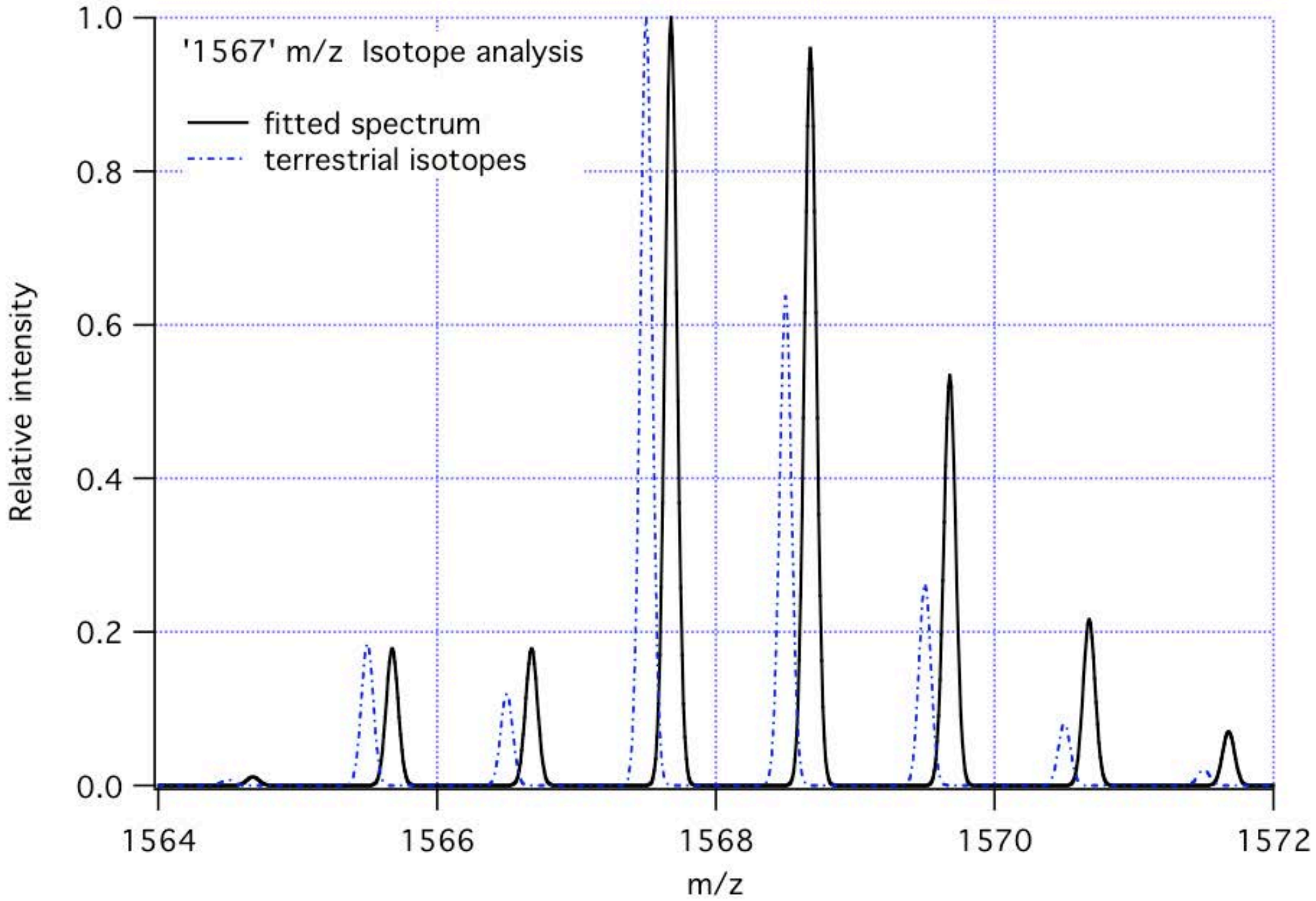

**Figure 8. Fitted isotope spectrum (solid line) to 1567 peak and simulation of the same structure with terrestrial isotope values (dotted line, offset for clarity). Details of the fit are given in Table 3.**

In Figure 8 we show that terrestrial isotope values (S1) do not at all represent the relative isotopologue amplitudes around an m/z peak of known composition derived via our analysis of MS data. The solid curve in Figure 8, which closely represents the data in Figure 7, was generated using enhanced D/H and $^{15}N/^{14}N$ [9] ratios while keeping exact terrestrial ratios for the isotopes of elements C, O and Fe that comprise the remaining constituents of the molecular entity at m/z 1567. This may seem like an arbitrary choice when analyzing extra-terrestrial material that could in principle be from pre-solar interstellar space, but the isotopes of C, O, Fe and Li are relatively close to terrestrial in every measurement of solar system material and also in astronomical observations beyond the solar system, as reviewed with references in S1.2. Because we could in principle be measuring pre-solar material we have constantly revisited this isotope fitting assumption but not seen significant anomalies. The enhanced D/H ratios of solar system and interstellar material are discussed below and in S1.2.

**Table 3. Isotope analysis for peak at m/z 1567 of Figure 7. S/N = 135.**
**Modeled structure = (18Gly, 4Gly$_{OH}$, 3Fe) + 5'O' + H.**

| Satellite | (-3) | (-2) | (-1) | (0) | (1) | (2) | (3) |
|---|---|---|---|---|---|---|---|
| Data | 1,427 | 11,721 | 10,794 | 66,519 | 67,450 | 34,637 | 14,141 |
| Fitted $^{2}H$ 27,500 ‰ Preset $^{15}N$ 1,015 ‰ | .004 | .060 | .060 | .314 | .301 | .168 | .068 |
| Normalized data | .007 | .055 | .051 | .314 | .318 | .164 | .067 |
| Calc. terrestrial | .003 | .058 | .037 | .314 | .200 | .083 | .026 |

**Reproducibility of isotopologue intensities and presence of Fe**

In this section the proof of iron content is explored in more detail. Supporting data is as follows:

**a)** If the (-2) peak is derived from (0) via substitution of $^{54}Fe$ in place of $^{56}Fe$, then the mass difference ought to be 1.9953 units. Measurement gives 1.998±0.004 mass units ($1\sigma$, n=23). The alternative possibility of the loss of 2 hydrogen atoms is associated with a mass difference of 2.0157, which would be 4 standard deviations different from measurement.

**b)** The (-2):(0) amplitude ratio of m/z 730 is measured as 0.067±0.006, in agreement with the calculated amplitude from a single Fe in the molecule of 0.064, at terrestrial $^{54}Fe$:$^{56}Fe$, which is expected to be the case (discussion in S1.2).

**c)** There is no change to the isotopologue intensity ratios for any given m/z fragment when we change from matrix CHCA to SA, even though they have protonation rates that differ by orders of magnitude [10]. Data that illustrates this is presented for multiple m/z cases in S5 This again rules out any side peak generation process that involves differential hydrogen atom gain or loss.

**d)** Similarly, there is no change with MALDI laser intensity across different runs whereas the protonation rate changes at approximately the $6^{th}$ power of intensity.

**e)** Both (-4) and (-3) isotopologues due to multiple $^{54}Fe$ substitutions also appear consistently in the m/z 1639 data given in S1 and S5. One of these m/z 1639 traces is compared in fine detail with modeled isotope amplitudes in Section S1.3, Figure S1.2, confirming beyond doubt that the (-4) to (-1) satellites are due to iron and agreeing with the terrestrial $^{54}Fe$:$^{56}Fe$ ratio to within better than 100 parts per mil (100‰). There does not appear to be any way to create this specific pattern via hydrogen atom variation.

**The 1567 fragment**

One of the highest intensity species present in the fundamental spectrum (Figure 5) is at m/z 1567. In Figure 8 the 1567 data has been fitted using the composition {(18Gly, 4$Gly_{OH}$, 3Fe) + 5'O' + H} that emerged from the analysis to be discussed below. Only two parameters were varied in order to match the experimental isotope spectrum: a) the number of Fe atoms, and b) the hydrogen isotope excess above terrestrial (method in S1). With terrestrial $^{2}H$ content the spectrum would have been very different at the (1), (2), (3) etc. levels, as illustrated alongside the fitted spectrum in the dotted trace of Figure 8. The level of $^{2}H$ enhancement above terrestrial to fit the 1567 spectrum was 27,500‰ ± 2,500‰ (the ‰ symbol represents parts per thousand above terrestrial). The fit was performed with a simultaneous $^{15}N$ preset enhancement of 1,015‰ that had been measured by a different method [9]. The modeled structure for m/z 1567 was mainly derived as described below by assembling smaller fragment species such as m/z 304 and m/z 730, and confirmed by MS/MS analysis (S7). A similarly high or occasionally very much higher $^{2}H$ enhancement was observed in all peaks to be analyzed in this way, data summarized in S3. It indicates extra-terrestrial origin for the molecules under study.

**The 2069 species and others with lithium.**

In addition to the analysis of iron content mainly via the (-2) isotopologue, a measure of lithium content was obtained mainly from the amplitude of the (-1) peak. This arises from the $^{6}Li$ : $^{7}Li$ abundance ratio of 0.0759 : 0.9241. In a previous analysis there had been evidence for the lithium mass of 7Da in the fitting of polymers to m/z observations [2], but there had not been a strong technique to differentiate between the presence of the pair ($^{16}O$ + $^{7}Li$) and the presence of $^{23}Na$. With higher signal-to-noise ratios the (-1) : (0) ratio allows us to confirm definitely the presence of lithium and to differentiate between one and two lithium atoms at a spectrum signal-to-noise ratio of greater than about 50. The peak at m/z 2069 with S/N = 63 shows the presence of two lithium atoms upon isotope analysis. The data is shown in Figure 9, and Figure 10 shows the fitted spectrum together with a simulated spectrum based on the expectation for the same structure if the isotopes had been terrestrial. The quality of the 2069 isotope fit in Table 4 is excellent, with the positive side peaks matching data to better than 2%. The negative ones, being closer to noise, achieve a fit to better than 10% in amplitude.

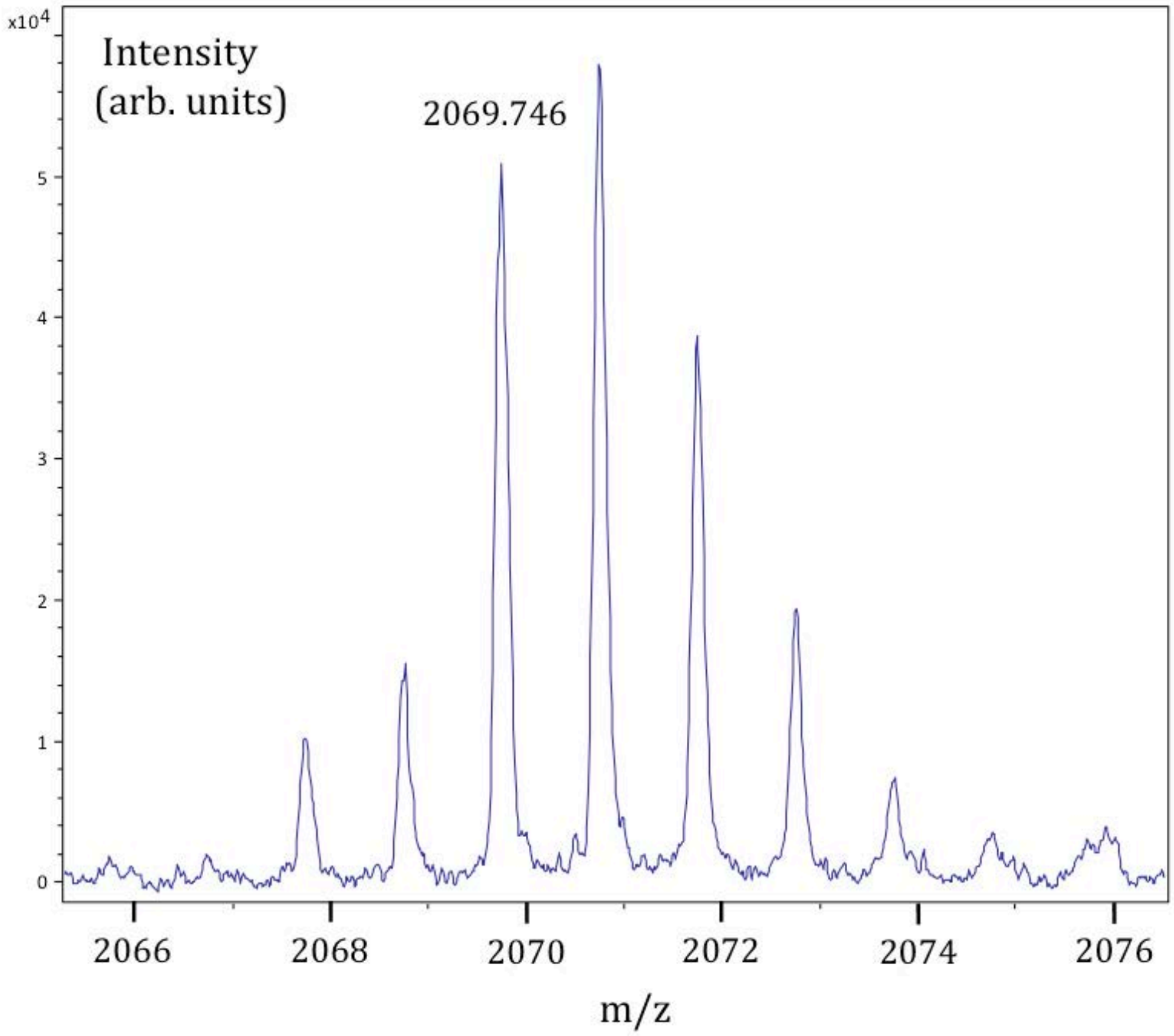

**Figure 9. Spectrum in the region of m/z 2069. Phase P2, matrix SA.**

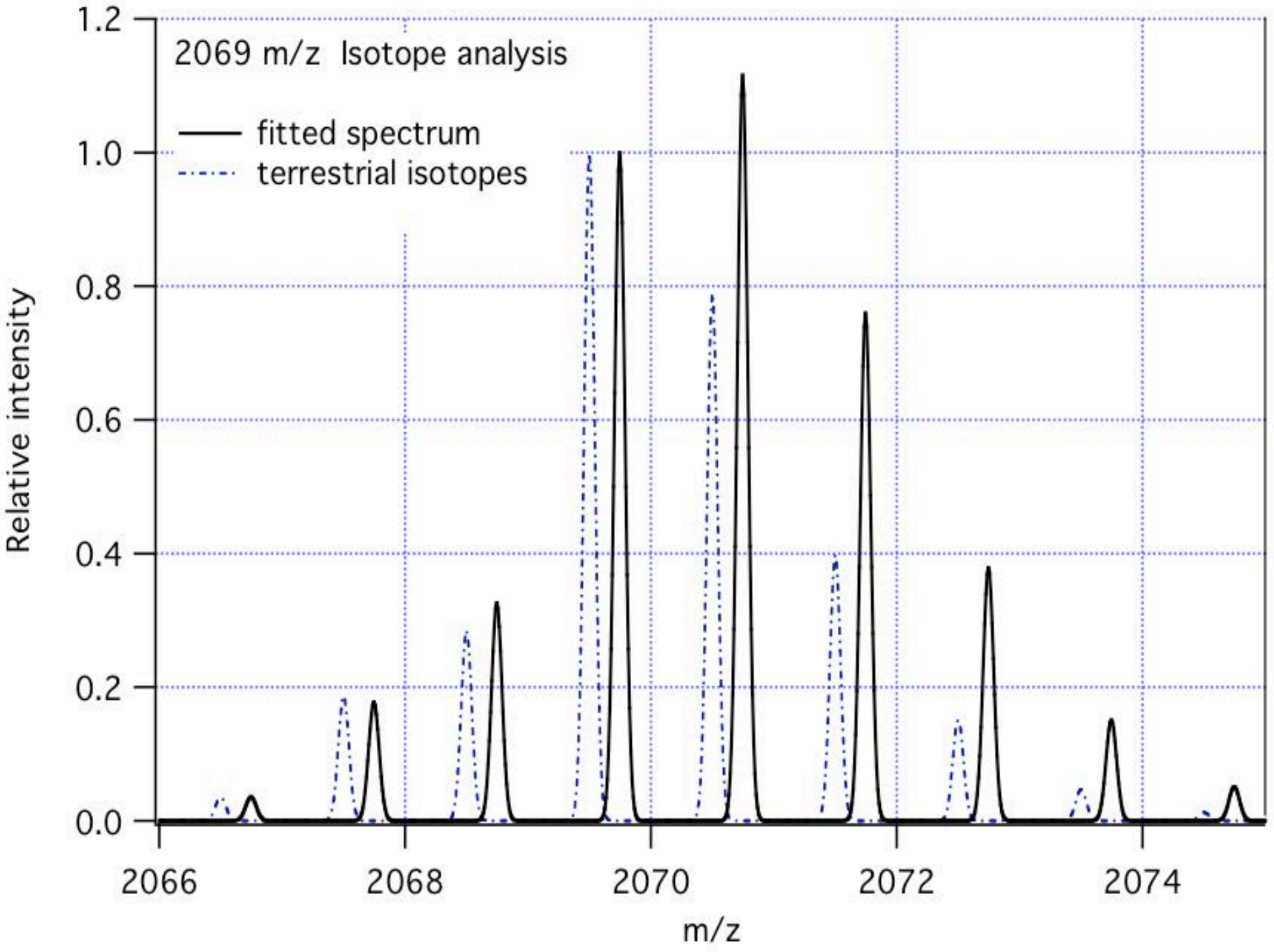

**Figure 10. Fitted isotope spectrum (solid line) to 2069 peak and simulation of the same structure with terrestrial isotope values (dotted line, offset for clarity). Details of the fit are given in Table 4.**

**Table 4. Isotope analysis for the m/z 2069 complex of Figure 9. S/N = 63.**
**Modeled structure = (26Gly, 4Gly$_{OH}$, 3Fe) + 7'O' + 2Li + H.**

| Satellite | (-3) | (-2) | (-1) | (0) | (1) | (2) | (3) |
|---|---|---|---|---|---|---|---|
| Data | 1,992 | 10,151 | 15,465 | 50,924 | 57,957 | 38,699 | 19,170 |
| Fitted $^{2}H$ 25,000 ‰ Preset $^{15}N$ 1,015 ‰ | .009 | .044 | .081 | .248 | .277 | .189 | .094 |
| Normalized data | .010 | .049 | .075 | .248 | .282 | .188 | .093 |
| Calc. terrestrial | .009 | .046 | .070 | .248 | .196 | .099 | .038 |

**Achieving a unified structure via synthesis**

Key MALDI fragment species were identified at m/z 304, 730, 1331 and 1417 that pointed toward the structure of higher polymers.

**a) m/z 304**

In the low m/z data of Table 1 there is an unidentified peak at m/z 304 that is strongly represented in the spectra for each of SA, DHB and CHCA matrices (Figure 11a). It has (-2) and (-1) isotopologues in the typical iron pattern, and their intensities correspond to the presence of two iron atoms within the entity (Figure 11b). The three traces were re-scaled to have the same peak intensity, averaged, and fitted with variable $\delta$D, keeping C, N and O at terrestrial levels. The m/z 304 formula is (0Gly, 2Gly$_{OH}$, 2Fe) + 3'O' -2H and its probable structure is shown in Figure 12, part A.

When fitted for deuterium enhancement the astonishing value of δD = 200,000 ± 15,000‰ was obtained, placing the likely origin of this molecular fragment in interstellar space. It appears to be a primary component within a number of the higher m/z structures to be discussed below.

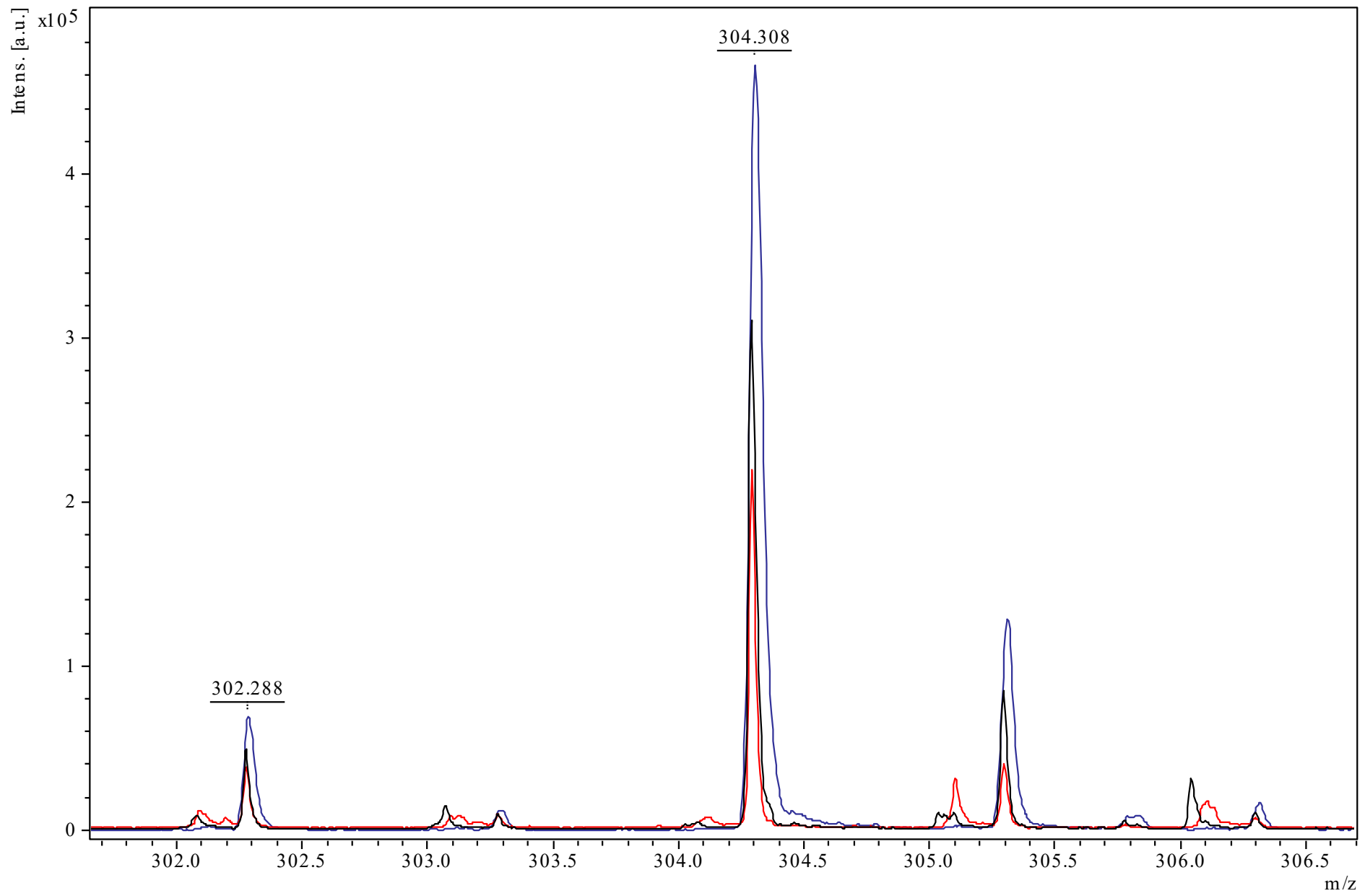

**Figure 11a. The m/z 304 complex with each of SA (blue), DHB (black) and CHCA(red) matrices, on the same intensity scale. Sample P2A.**

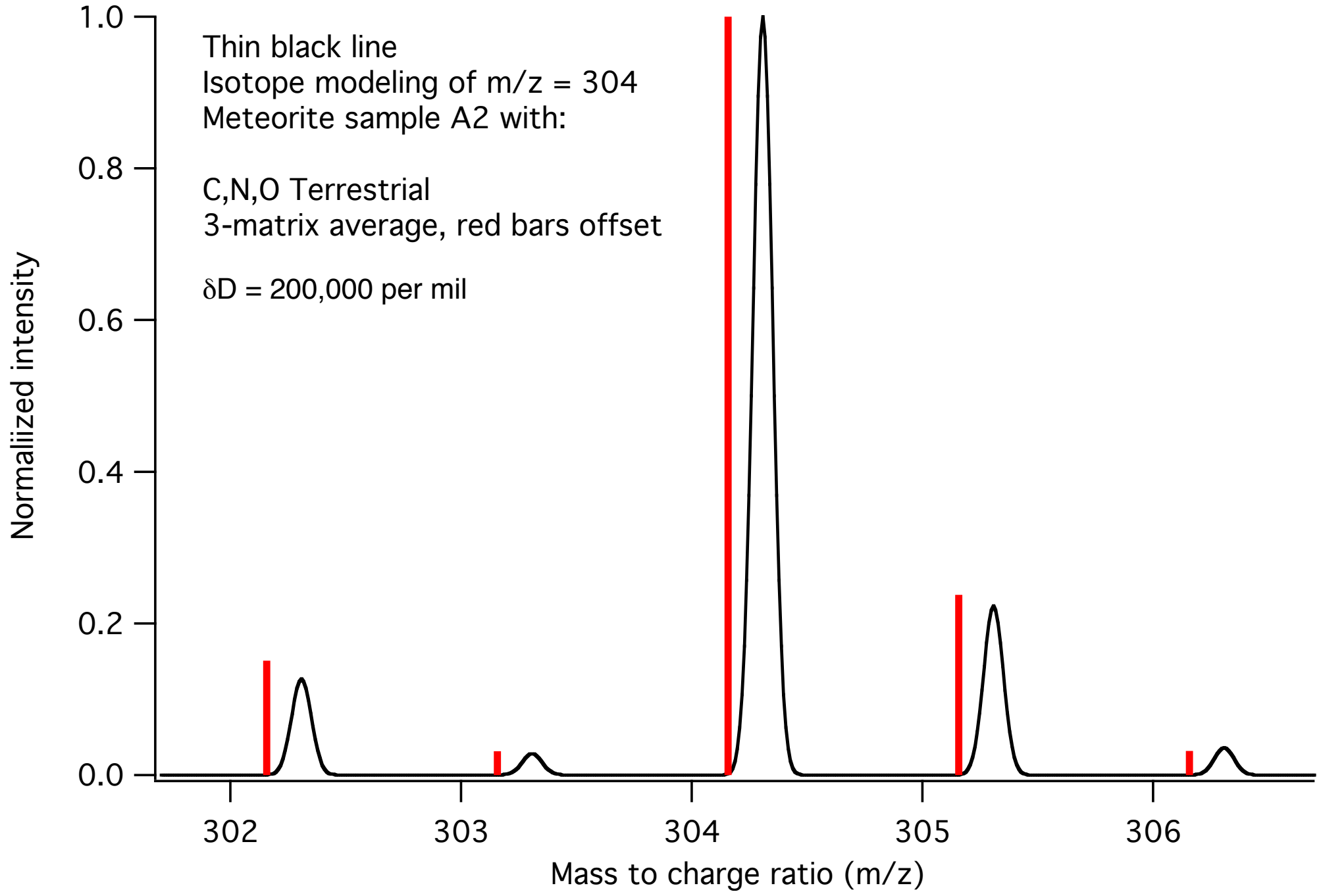

**Figure 11b. Fitted isotope distribution to an equal-weight average of the 3 data traces of Figure 11a. Data shown by red bars, offset for clarity. The isotope fit to formula (0Gly, 2Gly$_{OH}$, 2Fe) + 3'O' -2H is shown by a thin black line.**

**b) m/z 730**

This entity, shown to contain 1 Fe atom in the discussion above, is the highest peak in the fundamental spectrum with CHCA shown in Figure 5. Its formula is proposed to be (9Gly, $2Gly_{OH}$, 1Fe) + ‘O’ – H and isotope fitting is listed in Table 5. A possible structure is presented in Figure 12, part B.

**Table 5. Data and isotope fit for m/z 730**

| 730.306 | (-3) | (-2) | (-1) | 0 | (1) | (2) | (3) |
|---|---|---|---|---|---|---|---|
| Data | .008 | .038 | .025 | .572 | .262 | .083 | .014 |
| Fit | 0 | .036 | .017 | .572 | .273 | .080 | .018 |
| **730** | (9,2,1) + ‘O’ –H $\delta\,^2H‰$ / $^{15}N‰$ = 25,000 / 1,015 | | | | | | |

**c) m/z 1331 and 1417**

These do not have iron or lithium side-peaks, but have the compositions shown in Figure 12, parts C and D. There are surrounding (±16Da) peaks (S2) that have the expected mass difference of 15.974 for Na – K interchange. The formulae given in Figure 12 are derived by taking into account all of the (±16Da) substitutions seen in the data (Figure S2.2). Each of m/z 1331 and 1417 has the very high isotope enhancement of $\delta D$ = 75,000 ± 7,000‰, details in S2.

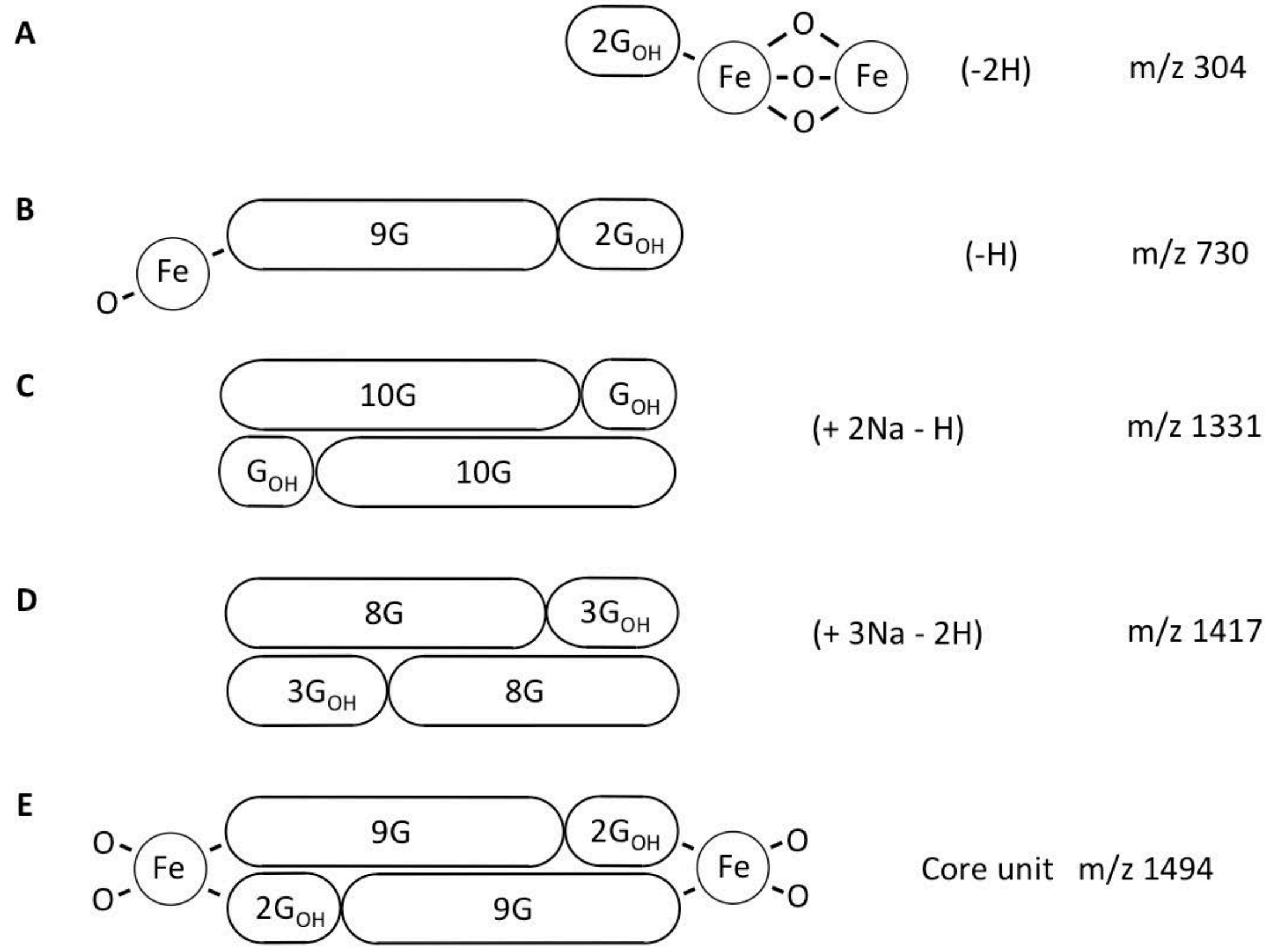

**Figure 12. Dominant molecular fragments at increasing m/z: assembly of a core unit at mass 1494.**

**Synthesis of a core unit at 1494Da and the observed m/z 1567 peak**

The MALDI fragments in Figure 12 parts B, C and D suggest a core unit at 1494Da (Figure 12 E), which has nominal end-for-end symmetry. This together with the m/z 304 motif can be assembled into the m/z 1567 molecule shown in Figure 13, parts A and B, which is one of the most intense species with matrix CHCA, but relatively much weaker in matrix SA. In contrast, a (+16) peak at m/z 1583 is usually stronger in SA but weaker in CHCA. There are several possibilities for the oxygen locations, as shown in Figure 13, parts A thru D. In the SA data the difference between m/z 1583 and m/z 1600 fits an oxygen mass but not a Na –> K substitution, and there are not (-16) intervals that match the reverse (K -> Na) substitution so the suggested m/z formula in Figure 13 part C for m/z 1583 does not contain K, Na, or Li and in SA the difference between m/z 1583 and m/z 1600 is an OH group. Additionally, extensive MS/MS data has been used for m/z 1567 to show that the proposed version in Figure 13A produces consistent fragments (S7).

Above m/z 1567 there are peaks principally at m/z 1639 and m/z 2124 with matrix CHCA, and peaks such as m/z 2069 with matrix SA that do not have related peaks at (M-H) lesser mass that would arise from having partial matrix content. However, SA peaks such as m/z 1862 are related by ($M_{SA}$ –H = 223Da) to m/z 1639 and all such peaks such as m/z 1862, with a "root" mass lower by one or more matrix masses are not included as viable candidates. Peaks with a fair certainty of belonging to the glycine–iron chemistry are summarized in supplementary S3, together with tentative assignments and isotope analyses.

**The prior m/z 4641 entity as a triskelion**

A higher mass complex centered at 4641Da [2] has been reported previously in meteorites Allende and Acfer 086, with details presented again in S4. It is now possible to demonstrate a convincing structure for m/z 4641 as a triskelion containing three identical subunits based upon the 1494Da core, bonded to each other by silicon, with additional sodium adducts. The array of lesser peaks around m/z 4641 is now readily assigned to different silicon fragments, as listed in S4. In a prior version of the present paper we had proposed that m/z 2320 hemolithin could dimerize to form the 4641 set, but an additional m/z 2402 entity was necessary to complete the match, and in any case the measured peak was at m/z 2313. With 4641 as a triskelion a single core unit at 1494Da suffices to build all of the set around m/z 4641, which is much more elegant. The triskelion is also consistent with the filaments, vesicles and angled structures observed (consider carbon nanotubes, buckyballs, etc.). Lastly, multiples of 4641 are observed (Table S4.1) that are consistent with silicon bonding of triskelia, one to another, in an extended sheet.

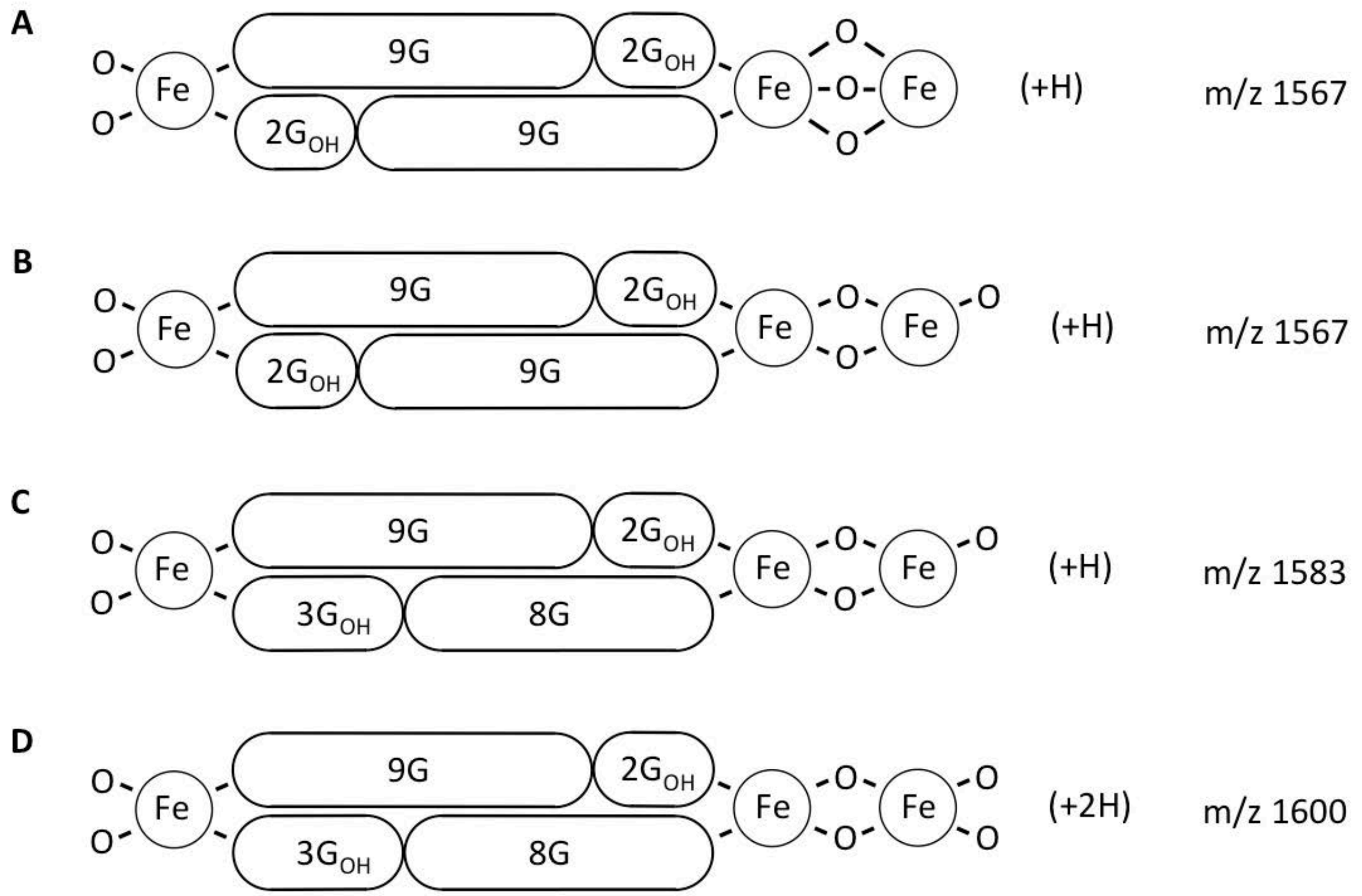

**Figure 13. Structures for the main hemoglycin fragments in the region of m/z 1567 .**

## Structure of the triskelion

The hemoglycin triskelion is proposed to assemble as shown in Figure 14.

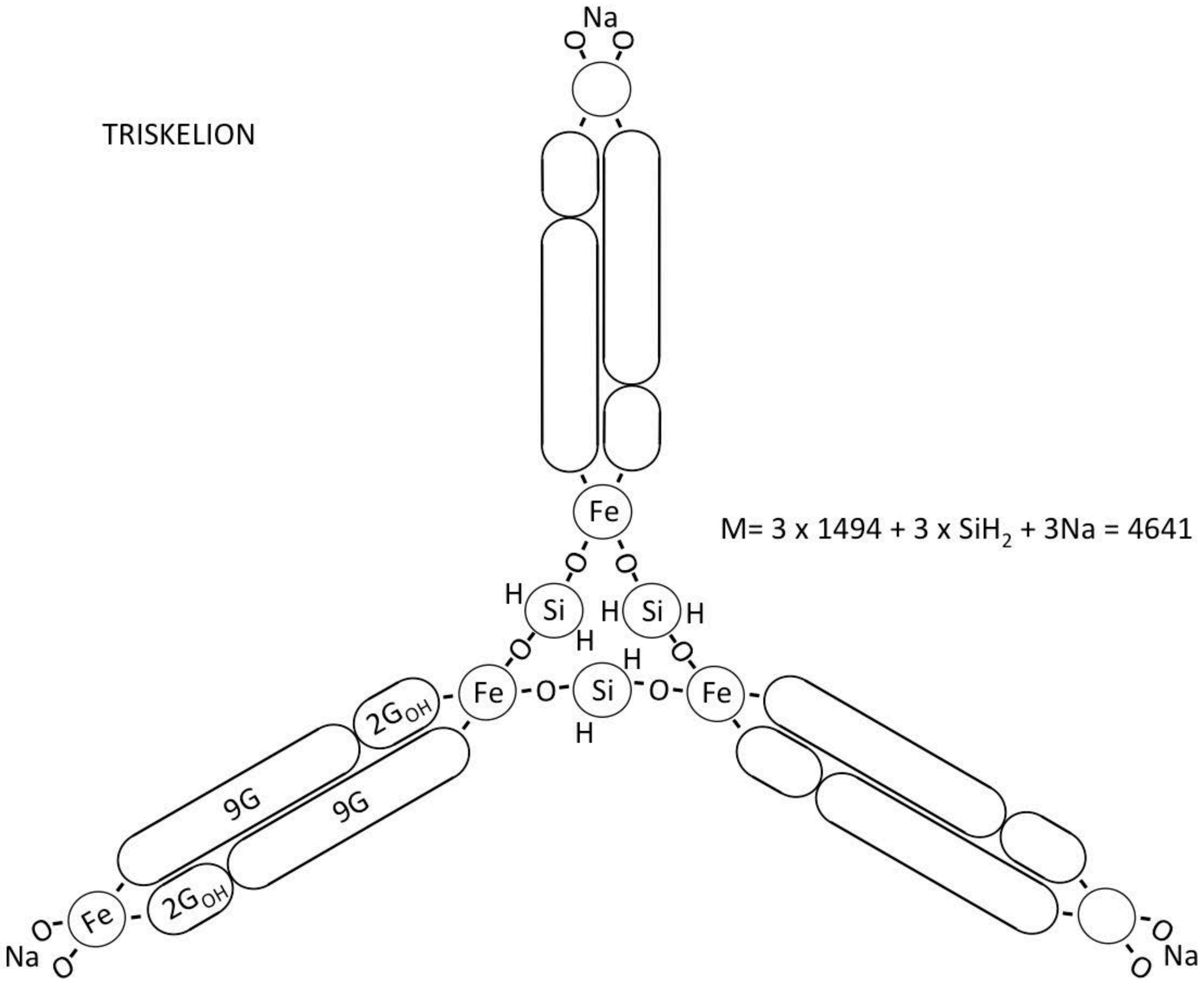

**Figure 14. Schematic of the m/z 4641 hemoglycin triskelion. When assembled as an extended sheet, the sodium adducts observed in mass spectrometry are replaced by silicon groupings, as at the center.**

The "legs" of such a structure need end-for-end symmetry in order to extend in a uniform sheet, and this is provided by the 1494Da entity. It is covalently bonded in the extended sheet. There is some evidence that the sheet could have magnetic properties. Every feature of this structure, including silicon-oxygen bonding, and multimer assembly into sheets, is supported by the prior 4641Da observations [2], with full details in Supplementary S4. Figure 15 illustrates the triskelion of Figure 14 following MMFF energy minimization in Spartan software [11,12], and juxtaposes 7 of these without full silicon bonding in order to simulate sheet assembly with its endless structural possibilities.

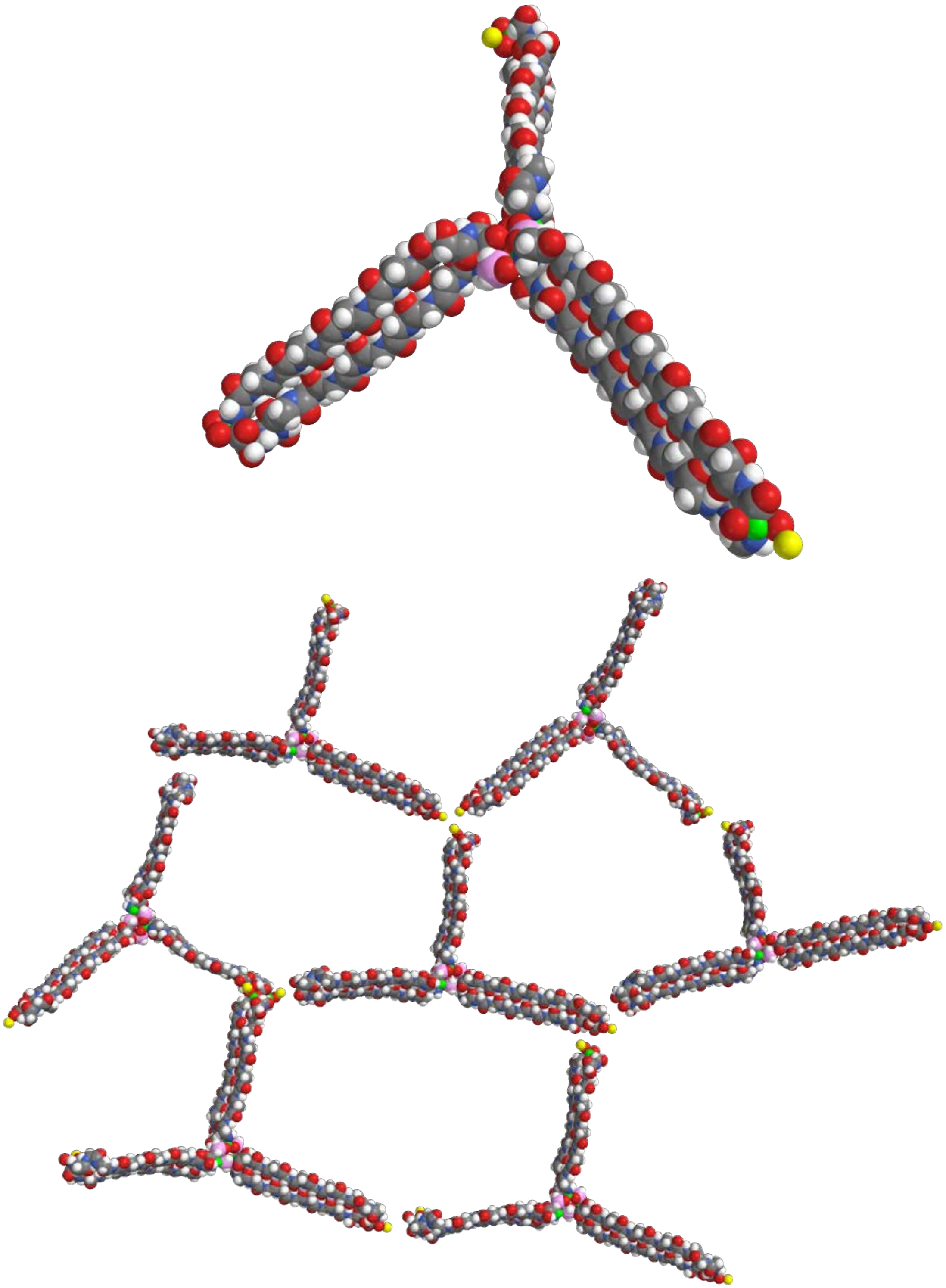

**Figure 15. Top: Space-filling MMFF model of the 4641Da triskelion. Bottom: 7 triskelia juxtaposed without full silicon bonding in order to simulate sheet assembly with its endless structural possibilities.**

## DISCUSSION

From this research we have two potential molecular structures, termed hemolithin and hemoglycin. Additionally, three-dimensional visible structures form in every Folch extraction. Some are of a fibrous protein nature and others are crystals of salt. The salt is expected as we are always working with everything that is present in micron-scale meteorite particles. The protein fibers, however, diffuse out of the particles, find one another and accrete to structures such as a lattice of triskelia that can entrap gas.

Progress toward the molecular structures has only been possible due to higher signal-to-noise data that allowed us to quantify the number of iron and lithium atoms associated with a mass spectrum peak, via their side peaks in the (-1) or (-2) positions relative to the intensity of the (0) and (+1) peaks. A program was written (S1) to include all the stable isotopes of H, Li, C, N, O and Fe and calculate the full isotope spectrum for any trial compound containing them. Complex isotope spectra could be fitted to within a few percent in many cases, data summarized in S3. Starting with the clear presence of polymers of glycine and hydroxy-glycine both in the present spectra and in the prior results [1,2], and adding the newly-determined numbers of iron and/or lithium, it was found that only additional oxygen atoms, sometimes in large numbers, were needed to match an unknown m/z peak. The m/z 304 fragment and details of the smaller fragments in the 1567 MS/MS spectra (S7) enabled us to piece together the $FeO_3Fe$ motif so prevalent throughout the data. Trial structures were simulated using the "Spartan" software from Wavefunction [11,12] in MMFF and at higher levels. It was found that anti-parallel peptide chains had significantly lower energy than parallel owing to better hydrogen bonding. In a conceptual breakthrough iron, with its four or five-fold coordination radiating outward, was considered for the role of terminating the strands, with a peptide 'C' terminus and an 'N' terminus joining onto one iron atom. The remaining two or three coordinations could carry oxygen atoms that themselves bonded to a second iron atom.

In a review of the known binding geometries for iron within proteins such as hemoglobin or hemerythrin, this proposed new motif was not found. Its overwhelming simplicity and the new association with beta sheet proteins make its future study very important.

In agreement with two prior MALDI studies of polymer amide in meteorites [1,2] the spectral peaks had higher (1,2,3 etc.) isotopologue intensities than would be calculated for the same structures using only terrestrial isotope values. In fact, from fitting the spectra, the average enhancement of $^{2}H$ above terrestrial was found to be 25,700‰, with a standard deviation of 3,500‰, n=15 (S3). Three cases (m/z 304, 1331 and 1417) were excluded from this average owing to exceptionally high $^{2}H$ levels that appear to be from a different origin. The 25,700‰ enhancement may also be expressed as a ratio D/H = (4.1 ± 0.5) x $10^{-3}$. It was not found necessary to vary Li, C, O or Fe isotope ratios away from terrestrial in order to fit the main series of data. These elements have relatively constant isotope ratios within all solar system material, as reviewed in S1.2. Only $^{15}N$ is known to be raised in this same Acfer 086 meteorite sample [9], by a completely different technique, and its enhancement value of 1,015 ± 220 ‰ was incorporated as a baseline. The $^{2}H$ enhancement quoted here is calculated with this $^{15}N$ enhancement as a preset condition of the fit to data. If desired, the $^{15}N$ value can be changed in the future with predictable effect on the $^{2}H$ determination. If we consider only polymer glycine with its H:N ratio of 3:1, a change to N(‰) of -1,000 would be equivalent to a change to H(‰) of +7,700, approximately, in the fit to data.

Such high $^{2}H$ (D) enhancements are well documented in molecular clouds [13] where low-temperature isotopic selection has operated for timespans of more than 10 million years. D/H fractions in simple molecules within the clouds are generally in the region of 0.01 to 0.1, to be compared with terrestrial D/H = $1.55 \times 10^{-4}$. Our measured average enhancement of 25,700‰ is equivalent to a D/H ratio of 0.0041, which would be marginally consistent with an interstellar origin for the hemolithin molecule. The exceptional isotope levels in m/z 304, 1331 and 1417 are more definitely consistent with an interstellar origin. We have estimated [14] that simple amino acids should be able to slowly polymerize in the conditions of "warm dense molecular clouds", without necessarily requiring surfaces for the reaction. Furthermore, the elements H, Li, C, N, O and Fe comprising this molecule were initially the most abundant when the first massive stars released them about 13B years ago.

Deuterium is also enriched in molecules of the proto-planetary disc. Aikawa and Herbst [15] have applied complete molecular cloud chemistry models to the disc and one of their predictions relates to the amount of DCN/HCN enrichment expected at 30AU radius, which is the likely region for the origin of comets. There they found that DCN/HCN converged in calculations to a value of 0.004, only slightly larger than the ratio measured in comet Hale-Bopp [16,17] of $(2.3 \pm 0.4) \times 10^{-3}$. The D/H ratio of 0.0041 that we find in polymer amide is therefore consistent with both calculated and measured DCN/HCN ratios for comets and outer solar system material. Further to this, the $^{15}N$ enhancement that we measured by a different technique is also consistent with cometary values [9]. Despite this consistency, the presumed source of CV3 class meteorites containing hemoglycin and hemolithin is a parent body in the asteroid belt that is only between 2.2 and 3.3AU from the sun.

High deuterium enhancement, comparable to the present readings, has been found in ultra-carbonaceous micro-meteorites from Antarctic snow [18]. The method of detection was secondary ion mass spectrometry (SIMS). There was no sign of these particles having experienced high temperature and because found in snow, rather than ice, there was less chance of aqueous alteration. The highest D/H reading was $(4.6 \pm 0.5) \times 10^{-3}$, comparable to our present average of $(4.1 \pm 0.5) \times 10^{-3}$ for meteoritic proteins. The polyglycine chains of the present work have C/H = 0.66, which is rather lower than reported in [18] where C/H ratios were generally greater than 1 and indeed the organic carrier of the deuterium was not identified. In [18] the presence of crystalline silicates as opposed to amorphous silicates in small inclusions was cited as evidence in favor of a protoplanetary rather than an interstellar origin, although different ultimate origins could have applied to the organic and mineral regions. As in the present work there was minimal processing before measurement, which could be important in the determination of deuterium.

A small fraction of the particles of acid-insoluble organic matter (IOM) from Antarctic CR2 chondrite Yamato-793495 has displayed via SIMS a local (sub-micron) correlation of enhanced D (up to 8,957‰) together with enhanced $^{15}N$ [19]. This is not a general rule, but within one particle there was a linear correlation in which $\delta^{15}N = (0.031 \pm 0.017) \times \delta D$ + (53

± 25). This correlation is consistent, within the error bars, with our measurement [9] of $\delta^{15}N$ = 1,015 ± 220‰ in Acfer 086 particles together with our present main value for $\delta D$ in hemolithin and hemoglycin. It should therefore be kept in mind that hemolithin and hemoglycin could be components within the so-called IOM, that may only have been released in the present experiment as the result of etching into very fine particles with high surface area.

It appears that there is a basic building block of hemoglycin with chain of length 11 glycine residues, arranged in antiparallel with a second similar chain. Several of these residues are modified to hydroxy-glycine via addition of OH in place of H on the alpha carbon. If this molecular type grows spontaneously when its glycine, Fe, O, Li and H components are present together in a particular environment then it is difficult to imagine such a tight length distribution being the result. A function yet to be identified could determine the length and maintain such a tight distribution. If successful in that function, energy could be harvested, which is a thermodynamic requirement [20], to aid in the creation of molecular copies of identical length.

Although not directly related to the present finding, polymers of up to 16 glycine residues have been catalytically assembled on a $TiO_2$ surface [21], out of the gas phase. The distribution of polymers was “diffusion-like” with a maximum intensity in mass spectrometry at around 5 glycine residues in length. In [21] there also was evidence of a degree of beta sheet organization after hydration. In our own work the possibility of “in vitro” glycine polymerization can be discounted because:

a) The typical polymer appears to have a core length of exactly 11 residues.
b) Extended polymer amide -C-C-N- backbone chains existed within the intact dry meteorite prior to solvent extraction as evidenced by the exact 2:1 ratio of CN to $C_2$ negative ions in focused ion beam (FIB) SIMS milling [9], associated with high $^{15}N$ levels in the CN ions.
c) The solvent extraction time was only 24 hours at room temperature (approximately 293K), whereas 403K had been used in the surface-catalyzed case [21].
d) Terrestrial contamination glycine was not involved because of the very high extra-terrestrial D/H ratios.

The $FeO_3Fe$ motif at the ends of hemolithin has been studied in relation to the photo-induced splitting of water [22] on hematite surfaces, pointing to a direction for further investigation. As a footnote, the outdated assignments in Table 2 are explained by the mass equivalence of (2 x alanine + t=18) to $FeO_3Fe$, each having mass 160, and similar items. In the prior work [1,2] the signal-to-noise ratios were too low to identify Fe and Li via their light isotopes, although Li was deduced to be present simply via its dominant mass 7 isotope [2].

## CONCLUSIONS

One-step room temperature solvent extraction from micron scale particles of the meteorite Acfer 086 (class CV3), has yielded via MALDI mass spectrometry a relatively simple spectrum in the m/z range 1-2,000 that is dominated by a single class of proteins. This appears to be composed of anti-parallel beta strands of glycine, with about 20% oxidation to hydroxy-glycine, and termination at each end with an iron atom directly bonded to C and N terminals of the peptide strands. There can be additional oxygen/iron groups at each end, and lithium adducts. The method of MALDI mass spectrometry yielded both intact molecules and fragments, with 11-residue antiparallel chains (22 residues in all) being clearly dominant. These molecules have extra-terrestrial origin as proved by the extreme raised D/H ratio in their isotopic satellite peaks. The average molecular deuterium excess above terrestrial is (25,700 ± 3,500)‰, or a D/H ratio of (4.1 ± 0.5) x$10^{-3}$, comparable to cometary levels, interstellar levels and also equal to the highest prior report in micro-meteorites. One component at m/z 304 shows an interstellar $^{2}H$ enhancement of (200,000 ± 15,000)‰. A triskelion of hemoglycin, built from the 1494Da hemoglycin core residue derived here, is now identified firmly as the source of the 4641Da entities previously reported [2]. Observations are reported of novel triskelion macromolecular structures consistent with formation of an extended sheet, and MS data of polymers of the 4641 triskelion shows fragments of up to 5 units of the sheet.

**ACKNOWLEDGEMENTS**
Authors MWM and JEMM wish to thank Professor Guido Guidotti of Harvard for encouragement and advice and Brian Stall at Bruker for allowing the use of their instrument for this analysis. We thank Charles H. Langmuir and Zhongxing Chen of the Department of Earth and Planetary Science, Harvard for use of their Hoffman clean room facilities. Dr. Shao-Liang Zheng is thanked for his X-ray analyses at the Harvard Department of Chemistry and Chemical Biology. The KABA meteorite sample was made available at Harvard to J E M McG by the director, Professor, Dr. Béla Baráth and the deputy director, Dr Teofil Kovács of the Museum of the Debreceni Reformatus Kollegium, Kalvin ter 16, H4026, Debrecen, Hungary.

## REFERENCES

1. McGeoch JEM and McGeoch MW. Polymer amide in the Allende and Murchison meteorites. *Meteorit and Planet Sci.* 2015; **50**: 1971-1983.

2. McGeoch JEM and McGeoch MW. A 4641Da. Polymer of Amino Acids in Allende and Acfer 086 Meteorites. arXiv:1707.09080 (28th July 2017).

3. Shimoyama A and Ogasawara R. Dipeptides and deketopiperazines in the Yamato-791198 and Murchison carbonaceous chondrites. *Orig. Life Evol. Biosph.* 2002 **32**, 165-179.

4. Bruker rapifleX MALDI-TOF/TOF system. Bruker Daltonik GmbH, Bremen, Germany and Bruker Scientific LLC, Billerica, MA , USA. Holle A, Haase A, Hoehndorf J. Laser spot control in MALDI mass spectrometers. US Patent 8,431,890 (2013). Haase A and Hoehndorf J. Mass spectrometer with MALDI laser system. US Patent 8,872,103 (2014).

5. Folch J, Lees M and Sloane Stanley GH. A simple method for the isolation and purification of total lipids from animal tissues. *J Biol Chem.* 1957; **226**: 497-509.

6. Zhang J and Zenobi R. Matrix-dependent cationization in MALDI mass spectrometry. *J Mass Spectrometry*. 2004; **39**: 808-816.

7. Harris WA, Janecki DJ and Reilly JP. The use of matrix clusters and trypsin autolysis fragments as mass calibrants in matrix-assisted laser desorption/ ionization time-of-flight mass spectrometry. *Rapid Commun. Mass Spectrom.* 2002; **16**: 1714-1722. doi: 10.1002/rcm.775.

8. Atomic Weights and Isotopic Compositions for All Elements. https://physics.nist.gov/cgi-bin/Compositions/stand_alone.pl

9. McGeoch MW, Samoril T, Zapotok D and McGeoch JEM. Polymer amide as a carrier of $^{15}N$ in Allende and Acfer 086 meteorites. www.ArXiv.org/abs/1811.06578

10. Lu I-C, Chu KY, Lin CY et al. Ion-to-Neutral Ratios and Thermal Proton Transfer in Matrix-Assisted Laser Desorption/Ionization. *J. Am. Soc. Mass Spectrom.,* (2015) **26**:1242-1251. DOI: 10.1007/s13361-015-1112-3

11. Spartan 16 for Macintosh and updates, Wavefunction Inc, Irvine, CA.

12. Halgren TA. Merck molecular force field. I. Basis, form, scope, parameterization, and performance of MMFF94. *J Comp. Chem.* 1996; **17**: 490–519. doi:10.1002/(SICI)1096-987X(199604)17:5/6,490::AID-JCC1.3.0.CO;2-P.

13. Roueff E and Gerin M. Deuterium in molecules of the interstellar medium. *Space Science Reviews* 2003; **106**: 61-72.

14. McGeoch JEM and McGeoch MW. Polymer Amide as an Early Topology. *PLoS ONE* **9**(7) e103036 (2014). doi:10.1371/journal.pone.0103306

15. Aikawa Y and Herbst E. Deuterium Fractionation in Protoplanetary disks. *The Astrophysical J.* 1999; **526**: 314-326.

16. Meier R et al. Deuterium in Comet C/1995 01 (Hale-Bopp): Detection of DCN. *Science* 1998; **279**: 1707-1710.

17. Meier R et al. A Determination of the HDO/H2O ratio in Comet C/1995 01 (Hale-Bopp). *Science* 1998; **279**: 842-844.

18. Duprat J et al. Extreme Deuterium Excesses in Ultracarbonaceous Micrometeorites from Central Antarctic Snow. *Science* 2010; **328**: 742-745.

19. Hashizume K, Takahata N, Naraoka H and Sano Y. Extreme oxygen isotope anomaly with a solar origin detected in meteoritic organics. *Nature Geosci.* 2011; **4**: 165-168.

20. England JL. Statistical physics of self-replication. *J Chem. Phys.* 2013; **139**: 121923.

21. Martra G, Deiana C, Sakhno Y, Barberis I, Fabbiani M, Pazzi M and Vincenti M. The Formation and Self-Assembly of Long Prebiotic Oligomers Produced by the Condensation of Unactivated Amino Acids on Oxide Surfaces. *Angewandte Chemie International Edition*; 2014; **53**: 4671-4674.

22. Hellman A and Pala RGS. A first principles study of photo-induced water-splitting on $Fe_2O_3$. *J. Phys. Chem C* 2011; **115** :12901-12907. DOI: 10.1021/jp200751j

23. Reference and inter-comparison materials for stable isotopes of light elements. IAEA, Vienna, 1995. IAEA-TECDOC-825, ISSN 1011-4289.

24. Sephton MA, James RH, Zolensky ME, The origin of dark inclusions in Allende: New evidence from lithium isotopes. *Meteorit and Planet Sci* 2006; **41**: 1039-1043.

25. Seitz H-M, Brey GP, Zipfel J, Ott U, Weyer S, Durali S, et al. Lithium isotope composition of ordinary and carbonaceous chondrites, and differentiated planetary bodies: Bulk solar system and solar reservoirs. *Earth and Planet Sci Lett.* 2007; **260**: 582-596.

26. Dauphas N, Cook DL, Sacarabany A, Frohlich C et al. Iron 60 evidence for early injection and efficient mixing of stellar debris in the protosolar nebula. *Astrophys J* 2008; **686**: 560-569.

27. Dauphas N, Janney PE, Mendybaev RA, Wadhwa M et al. Chromatographic separation and multicollection ICPMS analysis of iron. Investigating mass-dependent and -independent isotope effects. *Anal Chem* 2004; **76**: 5855-5863.

28. Clayton RN, Mayeda TK. Oxygen isotope studies of carbonaceous chondrites. *Geochim et Cosmochim Acta* 1999; **63**: 2089-2104.

29. Pearson VK, Sephton MA, Franchi IA, Gibson JM and Gilmour I. 2006. Carbon and nitrogen in carbonaceous chondrites: Elemental abundances and stable isotope compositions. *Meteorit and Planet Sci.* 2006; **41**:1899-1918.

30. Pizzarello S, Huang Y and Fuller M. The carbon isotopic distribution of Murchison amino acids. *Geochim et Cosmochim Acta*. 2004; **68**: 4963 – 4969.

31. Prombo CA and Clayton RN. A Striking Nitrogen Isotope Anomaly in the Bencubbin and Weatherford Meteorites. *Science* 1985; **230**: 935-937.

32. Pizzarello S and Holmes W. Nitrogen-containing compounds in two CR2 meteorites: $^{15}N$ composition, molecular distribution and precursor molecules. *Geochim et Cosmochim Acta*. 2009; **73**: 2150-2162.

33. Arpigny C, Jehin E, Manfroid J, Hutsemekers D, Schulz R, Stuwe JA, Zucconi JM and Ilyin I. Anomalous Nitrogen Isotope Ratio in Comets. *Science* 2003; **301**:1522-1524.

34. Wirstrom ES, Charnley SB, Cordiner MA, Milam SN. Isotopic Anomalies in Primitive Solar System Matter: Spin-state Dependent Fractionation of Nitrogen and Deuterium in Interstellar Clouds. *Astrophys J.* 2012; **757**: 1208.0192

35. Pizzarello S and Huang Y. The deuterium enrichment of individual amino acids in carbonaceous meteorites: A case for the presolar distribution of biomolecule precursors. *Geochim et Cosmochim Acta*. 2005; **69**: 599 – 605.

36. Hily-Blant P, Bonal L, Faure A, Quirico E, The $^{15}N$-enrichment in dark clouds and Solar System objects, *Icarus* 2013; **223**: 582-590.

**Meteoritic Proteins with Glycine, Iron and Lithium.**

Malcolm W. McGeoch[1], Sergei Dikler[2]
and Julie E. M. McGeoch[3*].

[1] PLEX Corporation, 275 Martine St., Suite 100, Fall River, MA 02723, USA.
[2] Bruker Scientific LLC, 40 Manning Rd, Billerica MA 01821.
[3] Department of Molecular and Cellular Biology, Harvard University, 52 Oxford St., Cambridge MA 02138, USA.
*Corresponding author. E-mail: mcgeoch@fas.harvard.edu

# Supplementary Information

## S1 Isotope Spectrum Calculations

### S 1.1 Analysis via isotopologue amplitudes

The less commonly observed (-1) and (-2) side peaks to a principal m/z peak are used to find the number of iron and lithium atoms in any given m/z fragment of the protein under study. Conventional amino acid polymers that only contain H, C, N, O and S atoms, all display "heavy isotope" peaks to the higher mass side of the principal peak giving predictable (+1), (+2) etc. amplitudes once the composition of a molecule is known. The "light" iron and lithium isotopes that contribute to the low mass side of a (0) peak fortunately differ from each other, with $^{54}Fe$ being at (-2) and $^{6}Li$ at (-1). The (-1) and (-2) amplitudes are therefore clear indicators of iron and lithium content, although there are higher order interactions that have to be included in the fitting. The object here is to test the fidelity of an isotopologue spectrum against the terrestrial expectation and if necessary adjust iron, lithium and one or more of the heavy isotope ratios (particularly $^{2}H$) to trial extra-terrestrial levels until a spectrum is accurately reproduced. In this section the mechanism for this calculation is described.

We start with terrestrial heavy isotope fractions ([23] Vienna) as reference values:

| | |
|---|---|
| VSMOW water | $R_H = {}^2H/{}^1H = 155.76 \pm 0.05 \times 10^{-6}$ |
| VSMOW water | $R_O = {}^{18}O/{}^{16}O = 2{,}005.20 \pm 0.45 \times 10^{-6}$ |
| V-PDB | $R_C = {}^{13}C/{}^{12}C = 11{,}237.2 \times 10^{-6}$ |
| Atmospheric Nitrogen | $R_N = {}^{15}N/{}^{14}N = 3{,}612 \pm 7 \times 10^{-6}$ , |

to which we add the iron and lithium terrestrial values [8]:

| | |
|---|---|
| Iron | $R_{F54} = {}^{54}Fe/{}^{56}Fe = 0.06371$ |
| | $R_{F57} = {}^{57}Fe/{}^{56}Fe = 0.02309$ |
| Lithium | $R_{L6} = {}^{6}Li/{}^{7}Li = 0.08217$ |

$^{58}Fe$ is at the 0.2% level and is neglected here.

We first illustrate the calculation of isotope probabilities with reference to $^1H$ and $^2H$ and then extend to the full array. In a molecule with $n$ hydrogen ($^1H + {}^2H$) atoms in its formula we calculate the probability of obtaining $k$ $^2H$ atoms from the binomial distribution as

$$P_H(k) = \frac{n!}{k!(n-k)!} p_H^k (1-p_H)^{n-k}$$

where $p_H = \frac{[^2H]}{[^1H]+[^2H]}$ is the probability of a heavy isotope substitution. For example, if the ratio of heavy to light isotopes is $R_H = {}^2H/{}^1H$ then $p_H = R_H/(1+R_H)$ = 1.5574 x $10^{-4}$ for the terrestrial hydrogen standard. Similar considerations lead to values for $p_C$, $p_N$, $p_O$, $p_{F54}$, $p_{F57}$ and $p_{L6}$ denoted in general by $p_X$ in Table S1.1.

We illustrate the calculation with a terrestrial matrix cluster [7] at m/z=861 that is frequently observed in MALDI spectra with matrix CHCA. The 861 molecule is due to ($M_4KNa_3 - 3H$) in which $M$ is the CHCA molecule of mass 189.043 with formula $M = C_{10}H_7NO_3$. We will calculate its isotopologues and compare them with the observed matrix cluster (Figure S1.1, detail from [1], Fig. 2) which has isotopic (Δm = +1 and Δm = +2) peaks of amplitude 47% and 17% relative to the main peak.

The composition of the '861' cluster is $H_{25}C_{40}N_4O_{12}K_1Na_3$. Of these constituents only $^{23}Na$ is naturally mono-isotopic. In the following table its heavy isotope probability is zero. Entries in the table are obtained from the binomial distribution above for $k$ = 0,1,2,3 heavy substitutions in the $n$ atoms of the species concerned, with $n$=25 for hydrogen, etc.

**Table S1.1 Isotope satellite calculation for the '861' matrix cluster peak**

| | $H_{25}$ | $C_{40}$ | $N_4$ | $O_{12}$ | $K_1$ | $Na_3$ |
|---|---|---|---|---|---|---|
| $p_X$ | 1.5574 x $10^{-4}$ | 0.011112 | 3.599 x $10^{-3}$ | 2.0012 x $10^{-3}$ | 0.06731 | 0 |
| $P_X(0)$ | 0.996 | 0.640 | 0.986 | 0.976 | 0.933 | 1 |
| $P_X(1)$ | 0.0039 | 0.288 | 0.0036 | 0.023 (Δm = +2) | 0.067 (Δm = +2 | 0 |
| $P_X(2)$ | 7 x $10^{-6}$ | 0.063 | - | - | - | - |
| $P_X(3)$ | - | 0.009 | - | - | - | - |

The isotope ratios are presumed uncorrelated, so the probability of having zero mass addition is the product of the individual $P_X(0)$ terms:

$\Pi(0) = P_H(0)P_C(0)P_N(0)P_O(0)P_K(0)P_{Na}(0) = 0.572$

There are three different ways to have Δm = +1, involving $H$, $C$ and $N$ respectively. Oxygen and potassium jump to Δm = +2 immediately. The probability of having Δm = +1 is therefore:

$$\Sigma\,\Pi(1) = P_H(1)P_C(0)P_N(0)P_O(0)P_K(0)P_{Na}(0) + P_H(0)P_C(1)P_N(0)P_O(0)P_K(0)P_{Na}(0) + P_H(0)P_C(0)P_N(1)P_O(0)P_K(0)P_{Na}(0) = 0.262$$

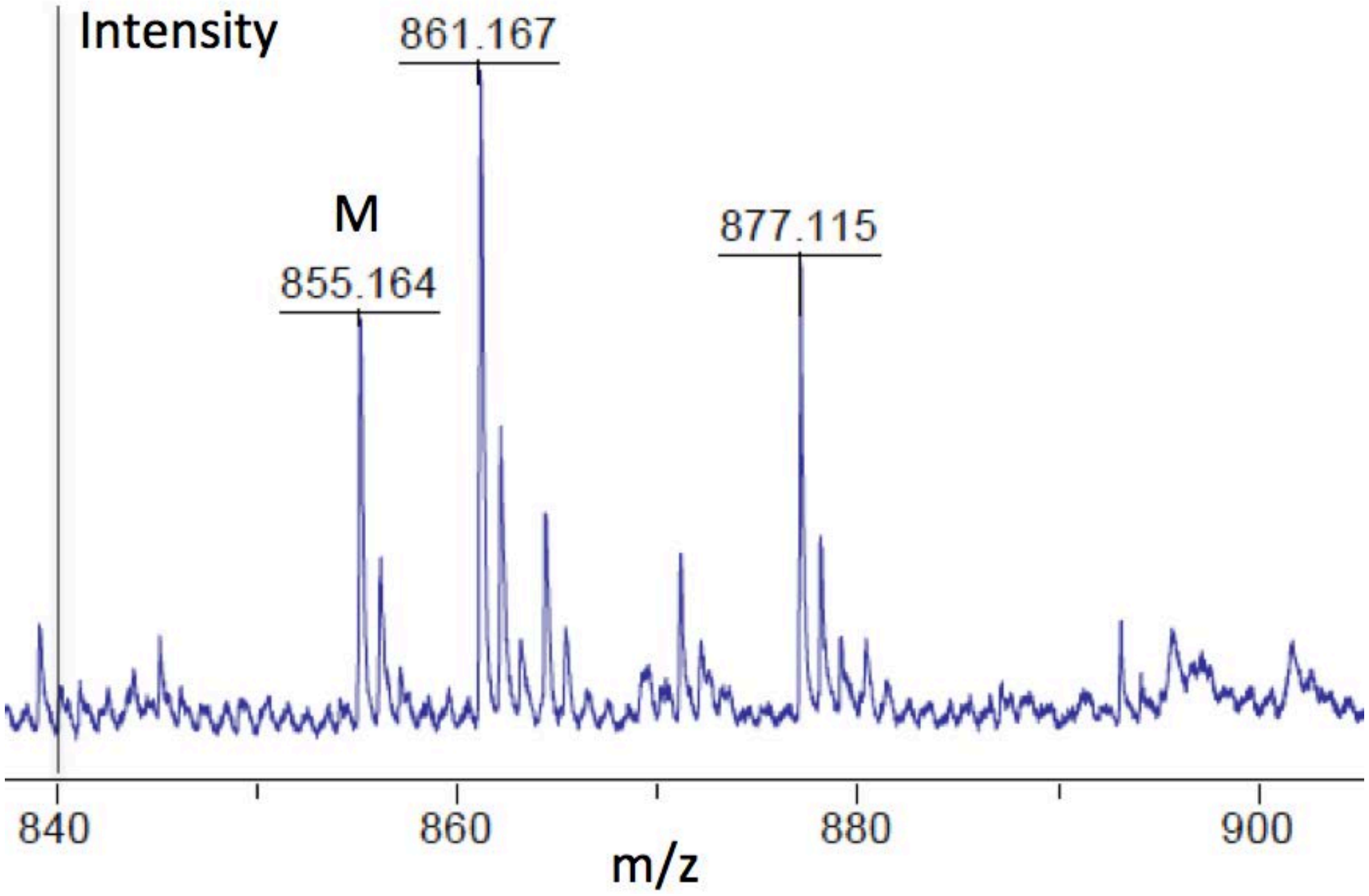

**Fig S1.1. Matrix cluster peaks at m/z 855, 861 and 877 reproduced from [1], Fig. 2.**

Similarly, the possible combinations that give $\Delta m = +2$ sum to give $\Sigma\, \Pi(2) = 0.113$ and the $\Delta m = +3$ sum is $\Sigma\, \Pi(3) = 0.033$. By now, we have accounted for 0.98 of the collective isotope amplitude, with the remaining 0.02 assigned to $\Delta m = +4$ and higher species. The total of probabilities should sum to unity, when correctly carried to completion.

Continuing with our terrestrial example at m/z 861 we normalize the $\Delta m = 0$ peak to 100%, as listed in Table S1.2. In the calculated satellites of the 861 matrix peak there is good agreement with the $\Delta m = +1$ and $\Delta m = +2$ measured intensities but the $\Delta m = +3$ location appears to also carry another peak that obscures the true reading.

**Table S1.2 Calculated and experimental isotope satellite ratios for the 861 matrix cluster peak.**

| | $\Delta m = 0$ | $\Delta m = +1$ | $\Delta m = +2$ | $\Delta m = +3$ |
|---|---|---|---|---|
| Calc. | 100% | 45.8% | 19.8% | 5.8% |
| Expt. | 100% | 47% | 17% | - |

We handle the complexity of the present calculations via a computer program that inputs a trial molecular composition and trial values for the isotope ratios, whether heavy or light. The output comprises peak amplitudes from (-8) to (+8) that can then be compared with data. The calculation is accurate to better than 1 part in 10,000. We found that it was efficient to start the comparison with a trial $^{2}H$ enhancement of 25,000‰, on top of the preset $^{15}N$ enhancement of 1,015‰ discussed in the main text. Depending upon the closeness of the fit the $^{2}$H enhancement was adjusted in increments of ± 2,500‰ until the various side peaks were collectively fitted. If the (-2) and (-1) peaks did not clearly indicate a number of iron or lithium atoms, either the formula under test was adjusted, or the fitting attempt was abandoned for that peak because of inadequate signal-to-noise ratio. After a formula adjustment the whole set of isotopologues has to be fitted again because of the numerous "knock-on" effects from one to another.

The (0) "monoisotopic" peak contains the dominant isotope of each molecular constituent. However, when elements such as Fe and Li are involved, with (X-2) and (X-1) isotopes $^{54}Fe$ and $^{6}Li$, the (0) peak carries many isotopic variations. For example, if a single molecule contains one atom of $^{54}Fe$ and has in addition two $^{1}H$ -> $^{2}H$ substitutions, then that molecule sits on the (0) peak. A centroid "monoisotopic" mass is calculated via the weighted addition in the above program of the product probabilities $P_H(a)P_{Li}(b)P_C(c)P_N(d)$.. etc. times the associated mass.

## S1.2 The Terrestrial Isotope Assumption for Li, C, O, Fe

In this section we discuss the baseline assumption made during isotope spectrum fitting that the isotopes of Li, C, O and Fe, but not H or N, are terrestrial.

In regard to meteorites, lithium and iron isotope ratios are *always* very close to terrestrial in the carbonaceous chondrites, whereas the D/H ratio is highly variable. Also, oxygen and carbon isotope ratios remain relatively close to terrestrial. However, $^{15}N/^{14}N$ is subject to considerable variation, but on a lesser scale than D/H. The following brief review of prior work supports our variation of D/H and $^{15}N/^{14}N$ in the spectrum fits and, initially at least, the use of static terrestrial ratios for Li, C, O and Fe.

**Lithium**

Multiple measurements of the $^{7}Li/^{6}Li$ ratio in meteorites have yielded results very close to the NIST terrestrial standard "L-SVEC" which has $^{7}L/^{6}Li$ = 12.02. Within the CV3 class of carbonaceous chondrites that includes Acfer 086, the Allende meteorite contains slightly raised $^{7}Li$ relative to the L-SVEC standard at $\delta^{7}Li$ = + 2.0‰ (i.e. 2 parts in 1,000), when measured in the "whole rock" [24] (Sephton et al. 2006}. Additionally $\delta^{7}Li$ = + 2.8‰ was measured by [25] (Seitz et al. 2007). Sephton studied the "dark inclusions", an unidentified interior morphology of Allende, and saw a very similar lithium result to the bulk rock.

**Iron**

The most abundant three of the four stable iron isotopes ($^{54}Fe$, $^{56}Fe$ and $^{57}Fe$) are believed to be formed in stars by a statistical nuclear equilibrium and exhibit remarkably consistent ratios wherever they are measured [26] (Dauphas et al. 2008). The minority 0.2% abundant $^{58}Fe$ isotope is thought to form in Type IA or Type II supernovae. In regard to meteorites: "… the Allende CV3.2 and Orgueil CI1 chondrites (and etc.) are normal within uncertainties" [27] (Dauphas et al. 2004). In that work the uncertainties are less than one part per ten thousand. Also from Dauphas et al. 2004, "This suggests that Fe isotopes were thoroughly homogenized in the solar system on the scale of planets and planetismals". For $^{54}Fe$, $^{56}Fe$ and $^{57}Fe$, therefore, we adopted these "normal" and universal isotope ratios, which also apply to terrestrial iron. We neglected to include in our fitting the slightly more variable $^{58}Fe$, as it is present at only the 0.2% level.

**Oxygen**

Oxygen shows relatively minor isotopic variation across meteorites, our planet's oceans, and terrestrial and lunar minerals. In an extensive study [28] (Clayton and Mayeda 1999] reported $\delta^{18}O$ = + 3.76‰ for "whole rock" Acfer 086 and $\delta^{18}O$ = + 1.51‰ for "whole rock" Allende. Within meteoritic refractory inclusions is there a significant departure to $\delta^{18}O$ = -40‰ [Clayton and Mayeda 1999] but such inclusions would not have entered solution in our method of Folch extraction. There has been a measurement of enhanced $\delta^{18}O$ up to an extreme of + 530‰ [19], discussed in the main text, but this would raise $^{18}O$ from its terrestrial 2% to 3%, which would not add significantly to the +2 satellite in comparison with the D effects that are seen. An upper bound of about 500‰ can be inferred from our present data.

**Carbon**

$^{13}C$ has been measured across a wide range of carbonaceous chondrites [29] (Pearson et al 2006) in "whole rock" samples and found to range between $\delta^{13}C$ = - 24‰ and + 13‰ relative to the terrestrial (PDB) standard. In studies of Murchison fractions containing amino acids [30] (Pizzarello et al 2004) measured a range between $\delta^{13}C$ = - 28‰ and +41‰. The highest of these is 20 times smaller than the $^{15}N$ enhancements found in several meteorites.

**Nitrogen**

Unlike in the case of Li, Fe, O and C isotopes, which are consistently terrestrial, the $^{15}N/^{14}N$ nitrogen ratio has been reported to be highly elevated in many meteorites, reaching a level of up to $\delta^{15}N$ = +973‰ (973 parts per 1,000) in bulk material from Bencubbin [31] (Prombo and Clayton 1985). Chemical separation has indicated that individual amino acids can have $\delta^{15}N$ > +100‰ [32] (Pizzarello and Holmes 2009). Also, cometary CN carries an average $\delta^{15}N$ = +873‰ [33] (Arpigny et al 2003). In our prior work with Xe ion milling and TOF/SIMS [9] (McGeoch et al. 2018) we measured $\delta^{15}N$ = +410‰ in Allende, and $\delta^{15}N$ = +1,015‰ in Acfer 086, in relation to CN ions from a polymer amide entity that is very probably Hemolithin (an updated version of that manuscript is available from the authors). This result for Acfer 086 gave us the nitrogen baseline when we fitted isotope spectra in the present paper. The only remaining variable was the D/H ratio.

**Hydrogen**

The D/H ratio of molecules in molecular clouds, pre-stellar cores, meteorites, interplanetary dust particles, and comets, is often very high compared to terrestrial. An extreme case is that of Antarctic micro-meteorites that have regions with $\delta D$ = +29,000‰ [18] (Duprat et al 2010). An overview is given [34] by (Wirstrom et al 2012) in which theory for $^{15}N$ fractionation along with D fractionation is presented for the cold conditions in interstellar clouds. Isolated meteoritic amino acids exhibit $\delta D$ up to +3,600‰ [35] (Pizzarello and Huang 2005).

**Hydrogen correlated with Nitrogen?**

In our binomial probabilistic approach to calculation of the isotope spectrum it is necessary to assume statistical independence of D and $^{15}N$ enhancements <u>within any given molecule</u>.

Although some observations in meteorites via SIMS have shown "hot spots" which simultaneously contain enhanced D and $^{15}N$ (without any molecular identification and at a spatial resolution of microns), this correlation has not proved to be widespread [19], and usually such enhancements are not spatially correlated in SIMS. Possible reasons for this are discussed [34] by (Wirstrom et al 2012). It is now believed that D and $^{15}N$ fractionation may in general follow different histories [36] (Hily-Blant et al 2013) and the latter authors conclude from astronomical observations that apart from within molecular clouds considerable nitrogen isotopic fractionation can occur at low temperature in the gas phase of pre-stellar cores. The very high interstellar D fractionation that is observed possibly feeds into different pre-solar reservoirs, so it is not obvious that D and $^{15}N$ will be correlated beyond chance statistics within individual molecules of a given type.

In our analysis of the spectra we made the assumption that D and $^{15}N$ are un-correlated within Hemoglycin and Hemolithin. This does not much affect the lithium and iron determinations, because D and $^{15}N$ mainly affect the relative height of the (+1) and (+2) isotopologues. In regard to that ratio, the fits do not indicate significant anomalies that would be a signature for D and $^{15}N$ correlation within a molecule, at the present precision of a few percent in peak amplitudes.

## S1.3 Precision of Isotope Fitting

In Figure S1.2 (below) we compare an m/z 1639 complex of peaks with peak amplitudes calculated under the assumption of terrestrial C, O and Fe but elevated $^{15}N$ (+1,015‰) and D(+27,500‰). A composition of (18Gly, 4$Gly_{OH}$, 4Fe) + 6'O' + H has been proposed for this entity which has 4Fe atoms and therefore relatively high (-2) and (-1) satellite amplitudes. The (-4) satellite is almost entirely due to the presence of 2 $^{54}Fe$ atoms because the m/z 1639 composition does not contain Li, the only other contributor to negative satellites. The calculated ratio for (-4):(0) is 0.021, in agreement with data, to within the noise.

When we vary $^{54}Fe$ above terrestrial by as little as +100‰ (100 parts per thousand) in the calculation, the (-4):(0) amplitude ratio rises to 0.026, which begins to deviate from the observed height, even allowing for noise. In S5 we give examples of spectra at m/z 1567 and 1639 with each of matrices CHCA and SA that have orders of magnitude different protonation rates [10] but yield nearly identical ratios of the negative side peaks to the (0) peak. We believe that the exact agreement of the four observed negative isotopologue amplitudes with the present model based on iron content is impossible to explain via hydrogen loss from peaks (0), (1) to generate peaks (-4) to (-1).

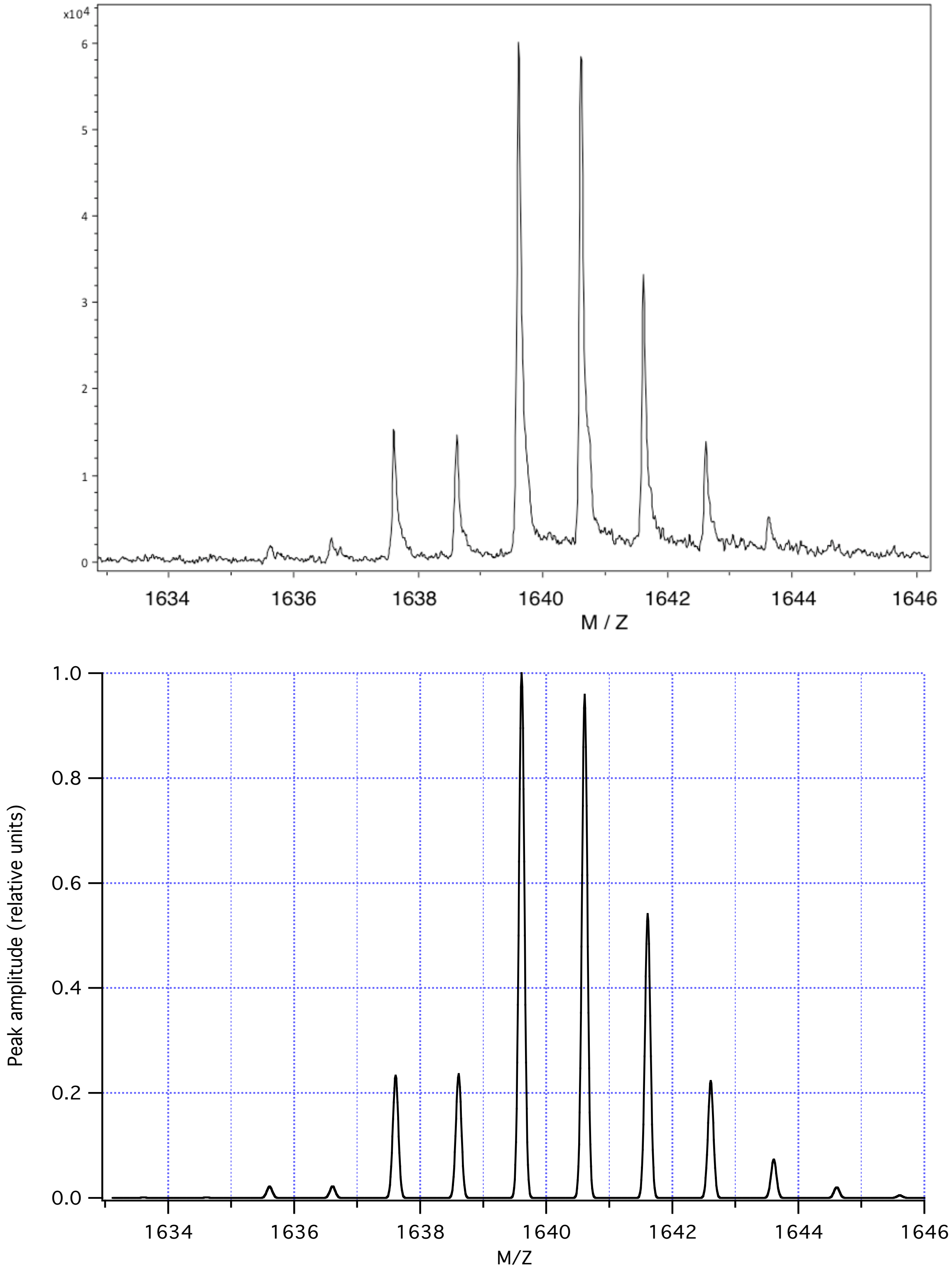

**Figure S1.2 Top: Data file P2_REF_CHCA_2019 0919_bb\0_K4\1 M/Z 1639.610 Bottom: Isotope fitting to composition (18Gly, 4Gly$_{OH}$, 4Fe) + 6'O' + H. $\delta$D = 27,500‰ $\delta^{15}$N = 1,015‰ and C, O, Fe terrestrial.**

# S2. Primary Fragment m/z Analyses

In this section a number of core spectral peaks are shown together with their isotope analysis.

## m/z 1331

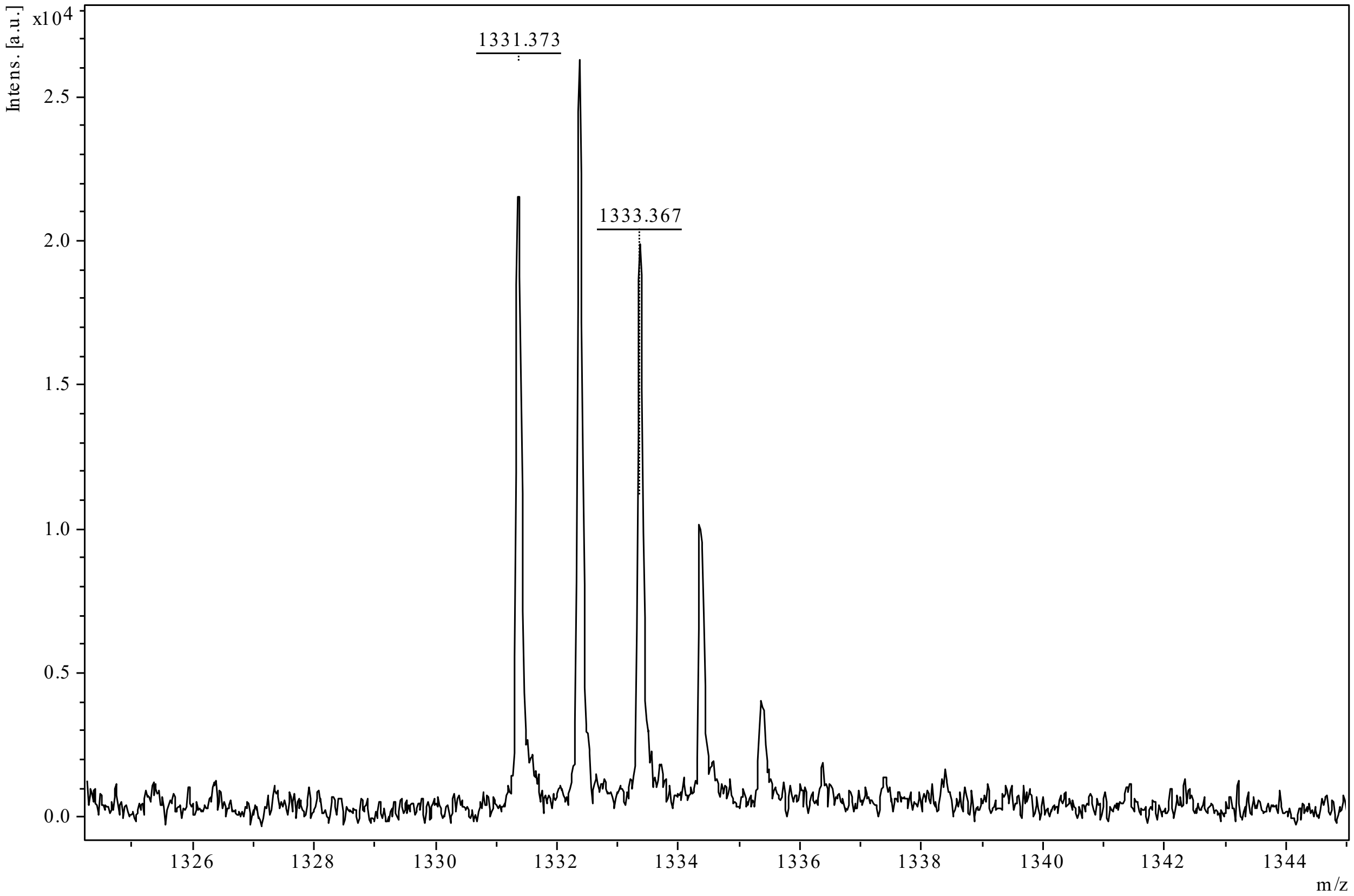

**Figure S 2.1 Spectrum from sample P3 with matrix CHCA around m/z = 1331.**

M/z 1331 is an example of a peak without Fe or Li. It does, however, show high isotope enhancement, as in Table S2.1 (δD = 75,000 ± 7,000‰).. Here we have set the $^{15}N$ to terrestrial as the $^{2}H$ enhancement greatly dominates the $\delta^{15}N$ enhancement of 1015‰ that we have routinely applied to most polymer glycine peaks based on a separate measurement [9]. M/z 1331 is seen in a group ranging from m/z 1315, 1331, 1347 and 1363 (Figure S2.3). These are similar to the set of corresponding entities around m/z 1417 that arise from Na – K substitutions.

**Table S2.1 Isotope analysis for m/z 1331. Notation is (Gly, $Gly_{OH}$, Fe).**

| 1331.373 | (-1) | (0) | (1) | (2) | (3) | (4) | (5) |
|---|---|---|---|---|---|---|---|
| Data | 0 | .255 | .336 | .233 | .113 | .043 | .013 |
| Fit | 0 | .255 | .311 | .226 | .120 | .048 | .023 |
| **1331** | (20,2,0) + 2Na –H  $\delta\,^{2}H$‰ = 75,000 ± 7,000 | | | | | | |

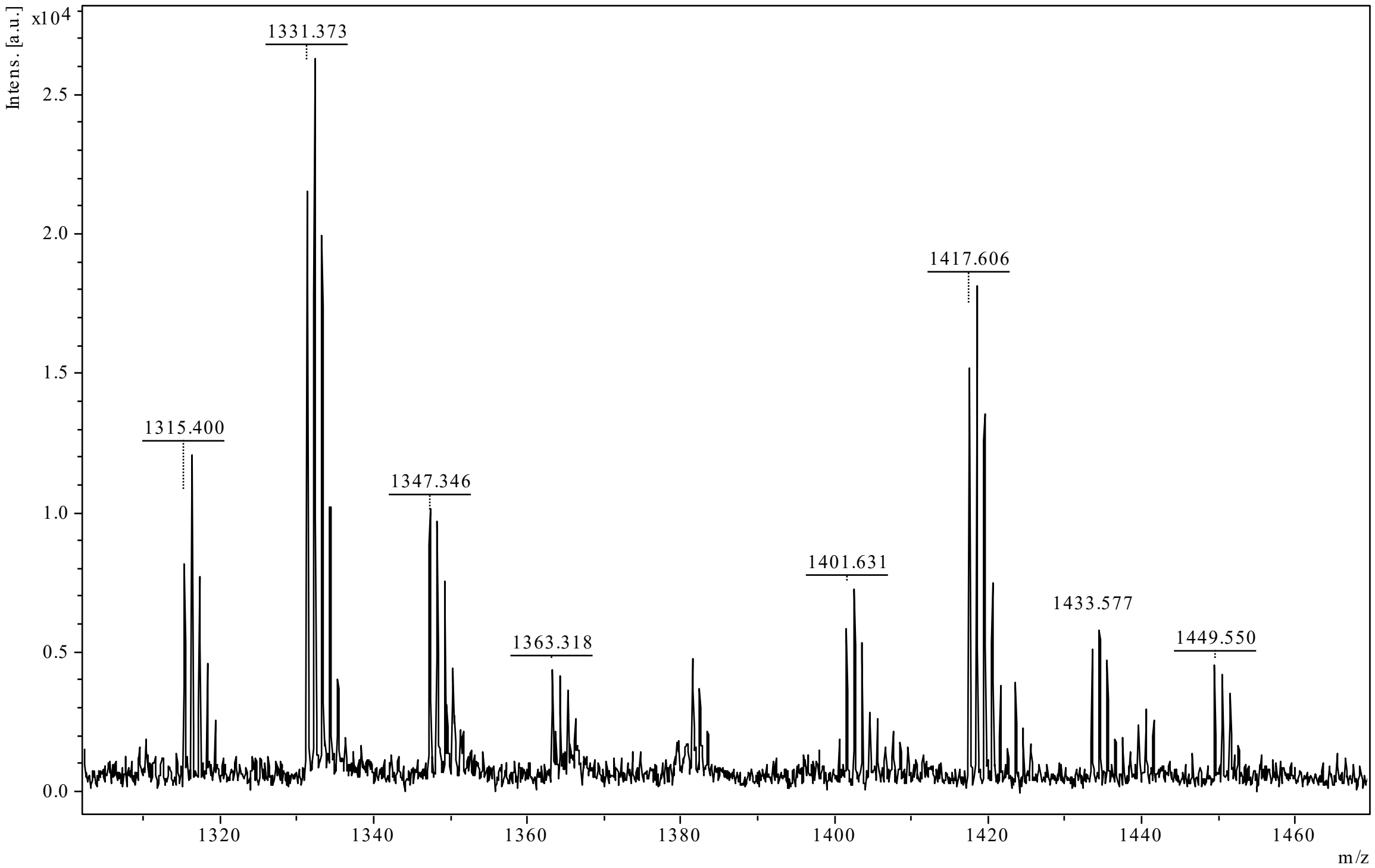

**Figure S2.2 Sample P3, matrix CHCA, related m/z series around m/z 1331 and m/z 1417.**

## m/z 1417

This peak, like 1331, does not contain Fe or Li, and also is seen within a group, this time with peaks at m/z 1401, 1417, 1433 and 1449 (Figure S2.2). The separation of these peaks by 16Da is superficially like an oxygen gain or loss, but in this case it is due to Li substitution in place of Na (-16) and K for Na (+16). The measured spacing of the complexes around m/z 1417 is 15.973 ± 0.010 Da, equal to the 15.974 difference between the masses of Na and K, whereas the difference expected for 'O' is 15.995Da. The composition of the 1417 peak is therefore proposed to be (16Gly, 6$Gly_{OH}$, 0Fe) + 3Na – 2H. This peak is shown in finer detail in Figure S.2.3 and its isotope analysis is summarized in Table S2.2. It has an estimated $\delta^2$D = 75,000 ± 7,000 ‰ , identical to that for m/z 1331.

**Table S2.2 Isotope analysis for m/z 1417. Notation is (Gly, $Gly_{OH}$, Fe).**

| 1417.606 | (-1) | (0) | (1) | (2) | (3) | (4) | (5) |
|---|---|---|---|---|---|---|---|
| Data | 0 | .252 | .336 | .235 | .115 | .044 | .014 |
| Fit | 0 | .255 | .300 | .225 | .124 | .058 | .025 |
| **1417** | (16,6,0) + 3Na -2H $\delta$ $^2$H‰ = 75,000 ± 7,000 | | | | | | |

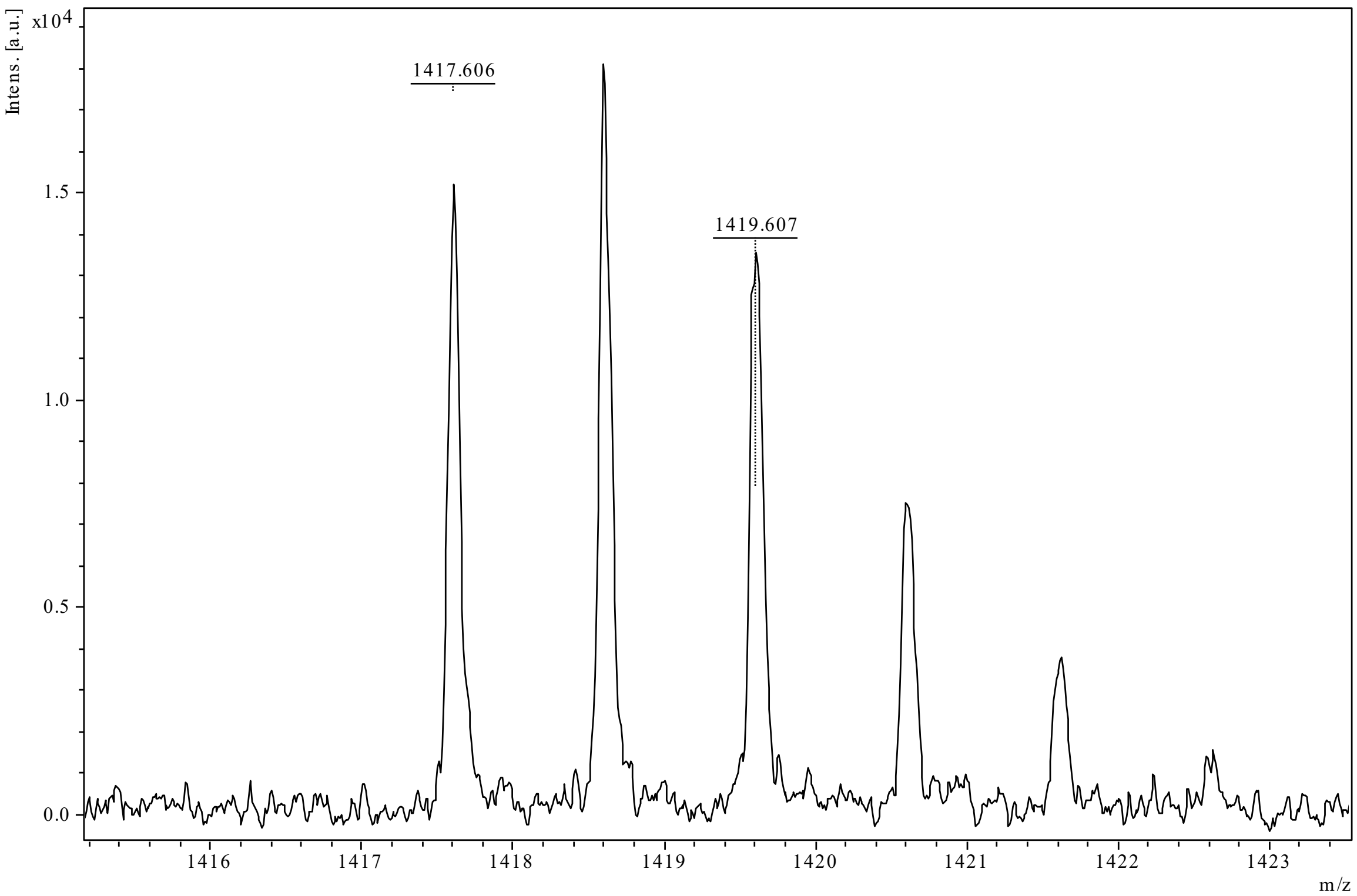

**Figure S2.3 Spectrum from sample P3 with matrix CHCA at m/z = 1417.**

# S3. Data Compilation.

**Table S3.1 Isotope fitting to peaks for given structures in notation (i,j,k) = (Gly, $Gly_{OH}$, Fe); *m*'O' represents oxygen atoms associated with iron. $^{2}H/^{15}N$ per mil enhancements from fit to data with preset $\delta^{15}N$ = 1,015‰ [9]. Average $^{2}H$ enhancement = 25,700‰ (σ = 3,500‰, n = 15).**

| **Obs. m/z Integer m/z** | **Satellite** | | | | | | | **S/N 0 peak** |
|---|---|---|---|---|---|---|---|---|
| 2364.385 | (-3) | (-2) | (-1) | 0 | (1) | (2) | (3) | |
| Data | .012 | .053 | .074 | .220 | .228 | .195 | .100 | 16 |
| Fit | .009 | .048 | .076 | .220 | .264 | .197 | .108 | |
| **2364** | (24,8,4) + 11'O' + Li + 5H 22,500/1,015 | | | | | | | |
| | | | | | | | | |
| 2124.885 | (-3) | (-2) | (-1) | 0 | (1) | (2) | (3) | |
| Data | .008 | .045 | .065 | .237 | .279 | .202 | .105 | 135 |
| Fit | .006 | .053 | .067 | .237 | .278 | .195 | .099 | |
| **2124** | (26,4,4) + 8'O' -2H 25,000/1,015 | | | | | | | |
| | | | | | | | | |
| 2069.746 | (-3) | (-2) | (-1) | 0 | (1) | (2) | (3) | |
| Data | .010 | .049 | .075 | .248 | .282 | .188 | .093 | 63 |
| Fit | .009 | .044 | .081 | .248 | .277 | .189 | .094 | |
| **2069** | (26,4,3) + 7'O' +2Li + H 25,000/1,015 | | | | | | | |
| | | | | | | | | |
| 1866.804 | (-3) | (-2) | (-1) | 0 | (1) | (2) | (3) | |
| Data | .030 | .053 | .093 | .274 | .332 | .185 | .101 | 27 |
| Fit | .006 | .049 | .072 | .274 | .293 | .187 | .087 | |
| **1866** | (24,4,2) + 5'O' + 2Li 25,000/1,015 | | | | | | | |
| | | | | | | | | |
| 1639.610 | (-3) | (-2) | (-1) | 0 | (1) | (2) | (3) | |
| Data | .013 | .075 | .071 | .300 | .291 | .165 | .069 | 112 |
| Fit | .007 | .070 | .071 | .300 | .287 | .162 | .067 | |
| **1639** | (18,4,4) + 6'O' + H 27,500/1,015 | | | | | | | |
| | | | | | | | | |
| 1567.678 | (-3) | (-2) | (-1) | 0 | (1) | (2) | (3) | |
| Data | .007 | .055 | .051 | .314 | .318 | .164 | .067 | 135 |
| Fit | .004 | .060 | .060 | .314 | .301 | .168 | .068 | |
| **1567** | (18,4,3) + 5'O' + H 27,500/1,015 | | | | | | | |
| | | | | | | | | |
| 1381.534 | (-3) | (-2) | (-1) | 0 | (1) | (2) | (3) | |
| Data | .017 | .076 | .107 | .335 | .249 | .147 | .073 | 13 |
| Fit | .012 | .062 | .098 | .335 | .279 | .139 | .051 | |
| **1381** | (17,2,3) + 5'O' + 2Li + 4H 30,000/1,015 | | | | | | | |
| | | | | | | | | |
| 1226.527 | (-3) | (-2) | (-1) | 0 | (1) | (2) | (3) | |
| Data | .001 | .055 | .034 | .402 | .309 | .147 | .039 | 36 |
| Fit | .001 | .049 | .038 | .402 | .307 | .138 | .046 | |
| **1226** | (11,6,2) + 3'O' + H 27,500/1,015 | | | | | | | |
| | | | | | | | | |
| 1049.474 | (-3) | (-2) | (-1) | 0 | (1) | (2) | (3) | |
| Data | .011 | .054 | .056 | .432 | .310 | .137 | .051 | 31 |
| Fit | .005 | .054 | .068 | .432 | .284 | .113 | .033 | |
| **1049** | (13,2,2) + 3'O' + 2H 30,000/1,015 | | | | | | | |
| | | | | | | | | |
| 1043.344 | (-3) | (-2) | (-1) | 0 | (1) | (2) | (3) | |
| Data | .016 | .070 | .097 | .423 | .275 | .114 | .037 | 114 |
| Fit | .009 | .055 | .096 | .423 | .272 | .104 | .030 | |
| **1043** | (15,0,2) + 4'O' + 2Li – 2H 27,500/1,015 | | | | | | | |
| | | | | | | | | |
| 1027.253 | (-3) | (-2) | (-1) | 0 | (1) | (2) | (3) | |
| Data | - | .032 | .023 | .470 | .320 | .125 | .035 | 111 |
| Fit | 0 | .029 | .020 | .470 | .311 | .123 | .036 | |
| **1027** | (11,4,1) + 3'O' + 4H 25,000/1,015 | | | | | | | |
| | | | | | | | | |
| 765.350 | (-3) | (-2) | (-1) | 0 | (1) | (2) | (3) | |
| Data | - | .048 | .152 | .517 | .225 | .069 | .020 | 24 |
| Fit | - | .042 | .132 | .517 | .222 | .063 | .013 | |
| **765** | (10,1,1) + 3'O' + 3Li -3H 20,000/1,015 | | | | | | | |
| | | | | | | | | |
| 730.306 | (-3) | (-2) | (-1) | 0 | (1) | (2) | (3) | |
| Data | .008 | .038 | .025 | .572 | .262 | .083 | .014 | 84 |
| Fit | 0 | .036 | .017 | .572 | .273 | .080 | .018 | |
| **730** | (9,2,1) + 'O' –H 25,000/1,015 | | | | | | | |
| | | | | | | | | |
| 531.335 | (-3) | (-2) | (-1) | 0 | (1) | (2) | (3) | 33 |
| Data | - | - | .003 | .691 | .243 | .054 | .098 | |
| Fit | 0 | 0 | 0 | .691 | .243 | .055 | .095 | |
| **531** | (9,0,0) + 'O' + 2H 17,500/1,015 | | | | | | | |
| | | | | | | | | |
| 401.086 | (-3) | (-2) | (-1) | 0 | (1) | (2) | (3) | |
| Data | - | .010 | .005 | .717 | .217 | .061 | .015 | 182 |
| Fit | 0 | 0 | 0 | .717 | .229 | .046 | .007 | |
| **401** | (7,0,0) + 2H 30,000/1,015 | | | | | | | |

# S4. The 4641Da polymer peaks revisited.

The prior observation of a relatively high mass isolated system of peaks around m/z 4641 [2] is now explained by the hemoglycin triskelion, with its MS structure illustrated in Figure 14 of the main text. There are close surrounding variants (Table S4.1), especially with added {SiO} groups, as would be expected if the 4641 entity had previously been part of a sheet (Figure 15). The original spectrum from [2] is reproduced in Figure S4.1 below. A control of volcanic rock yielded a flat spectrum in this region.

Multimers of m/z 4641 were observed at m/z 9289, 13,935, 18,591 and 23,228, which is consistent with an origin in a polymer sheet. A very similar spectrum was observed in Allende, another CV3 type meteorite [2].

The observed MS variant of the hemoglycin triskelion has the following composition:

54Gly + 12 $Gly_{OH}$ + 6Fe + 3Si + 12 'O' + 3Na + 6H = 4641.851

If connected in a large area sheet, the triskelion has (0) component unit mass 4662.859Da, having 3Na replaced by 3 $SiH_2$ units.

**Table S4.1 Molecular assignments to the previously observed 4641Da complex.**

| Assignment | Observed m/z | Intensity | Calculated m/z |
|---|---|---|---|
| Hemolithin? (prior obs.) | 2313.284 | 82,221 | 2313.377 |
| | | | |
| Hemoglycin triskelion | | | |
| 4641 – 2Na -H | 4595.116 | 49,287 | 4594.864 |
| 4641 – $SiH_2$ | 4611.582 | 49,804 | 4611.858 |
| 4641 | 4641.582 | 317,999 | 4641.851 |
| 4641 + 'O' | 4657.215 | 260,378 | 4657.846 |
| 4641 + SiO + 2'O' - H | 4716.357 | 48,394 | 4716.805 |
| 4641 + 2{SiO} + 'O' - H | 4744.880 | 42,350 | 4744.782 |
| 4641 + 3{SiO} + 2'O' | 4805.988 | 42,823 | 4805.756 |
| 4641 + 3{SiO} + 3'O' | 4821.152 | 49,821 | 4821.752 |
| 2 x 4641 + {Si – Na} | 9,289.418 | 75,426 | 9,288.796 |
| 3 x 4641 + 2{Si – Na} | 13,935.257 | 7,687 | 13,935.741 |
| 4 x 4641 + 3{Si – Na} | 18,591.267 | 794 | 18,582.686 |
| 5 x 4641 + 4{Si – Na} | 23,228.506 | 124 | 23,229.631 |

In the first version of the present MS it was proposed that a 2320Da entity with iron and lithium (hemolithin) could be responsible for the 4641 peak, as a dimer. However, the surrounding peaks required an additional structural variant with an incremental glycine count, making it much less likely that hemolithin itself was the source of the 4641 peak. The prior m/z 2313 observation in Table S4.1 is still a candidate (minus 1Li) for hemolithin, which has the proposed complete structure:

2320 = (26,6,4) + 10'O' + 2Li + 2H

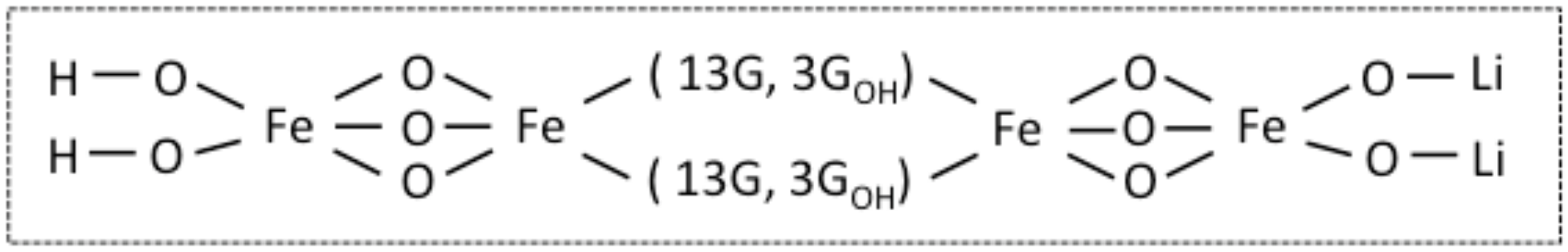

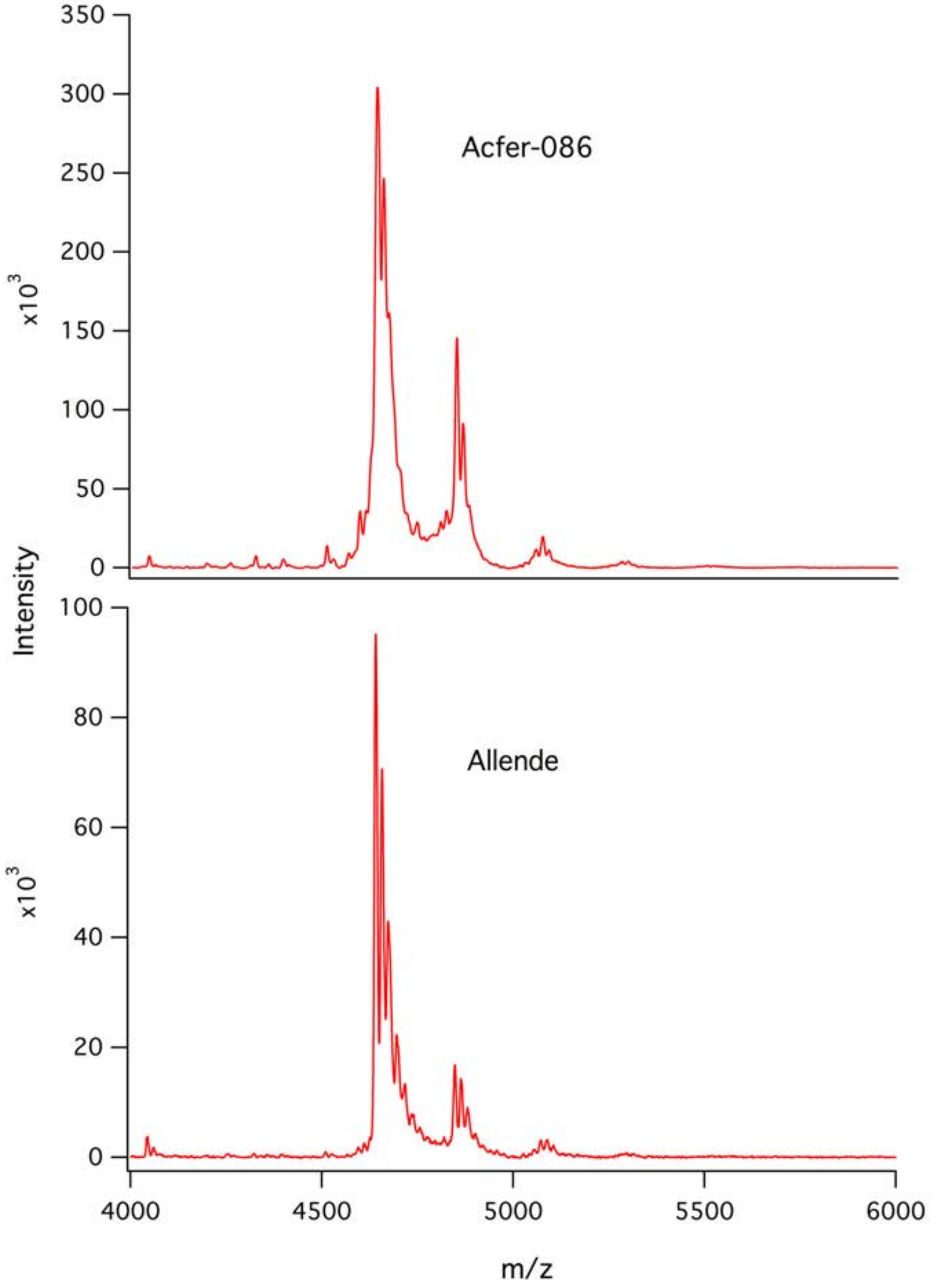

**Figure S4.1 Reproduced from [2]. Phase P2 mass spectra in the 4000 < m/z < 6000 range of Acfer 086 and Allende using sinapinic acid (SA) matrix. A control trace from volcanic rock presented in [2] is flat in this region of m/z. The "shadow" peaks at 223Da higher mass are matrix additions to the primary 4641Da complex.**

## S5. Partial raw data for isotope spectra at m/z 730, 1270, 1639, 2069, and 1567.

A sample of raw data is presented to illustrate the reliability of isotopologue relative amplitudes across:
Different sample phases (Phases P1, P2 and P4)
Different MALDI matrices CHCA and SA
Different spots and laser conditions.

**Notes**

**1.** Peak m/z 730.306 in Figure S5.1 is also shown in Figure 6 of the main text.

**2.** Peak m/z 1270 in Figure S5.3 is shown as an intermediate m/z case with (-2) and (-1) amplitudes that indicate the formula (13,2,2) + 3'O' + 2H in which the bracket notation stands for (Gly, $Gly_{OH}$, Fe).

**3.** Peaks m/z 1639.646, 1639.610 and 1639.632 (Figs. S5.4 – S5.6) correspond to formula (18,4,4) + 6'O' + H in which the bracket notation stands for (18Gly, 4$Gly_{OH}$, 4Fe). This specie had the highest Fe content among those analyzed with the result that isotopologues at masses 1635.6 and 1636.6, in the (-4) and (-3) positions respectively, are clearly visible in Figs. S5.4 and S5.5. The (-4) satellite can only be formed via the presence of two $^{54}$Fe isotopes within the total of 4 Fe atoms in the formula. An example of m/z 1639 with matrix SA is also given (Fig. S5.6), although the S/N ratio was not so good as with CHCA.

**4.** m/z 2069.746 (Fig. S5.7) is also shown in Figure 9 of the main text. It corresponds to formula (26,4,3) + 7'O' + 2Li + H, as does m/z 2069.718 (Fig S7.8).

**5.** m/z 1567.678 (Fig. S5.10) is also shown in Figure 7 of the main text. It corresponds, as do the five additional m/z 1567 spectra presented below, to the formula (18,4,3) + 5'O' + H. (Figs. S5.9 – S5.14). Two of these spectra are taken with matrix SA.

**6.** It is found via MS/MS that m/z 1567 is contained within the m/z 1639 peak. The difference of 72 mass units corresponds to the loss of FeO.

### Interpretation

The side peak amplitudes to the low mass side of a principal peak are very consistent across measurements with the two matrices CHCA and SA, that have orders of magnitude different protonation rates ([10], Lu et al.). This, together with the precise (-2) -> (0) mass difference (main text) and the amplitude consistency from sample to sample and from spot to spot at changing laser fluences, rules out hydrogen atom loss as the source of the negative satellites, leaving the $^{54}$Fe iron and $^{6}$Li lithium isotopes as the only viable explanation.

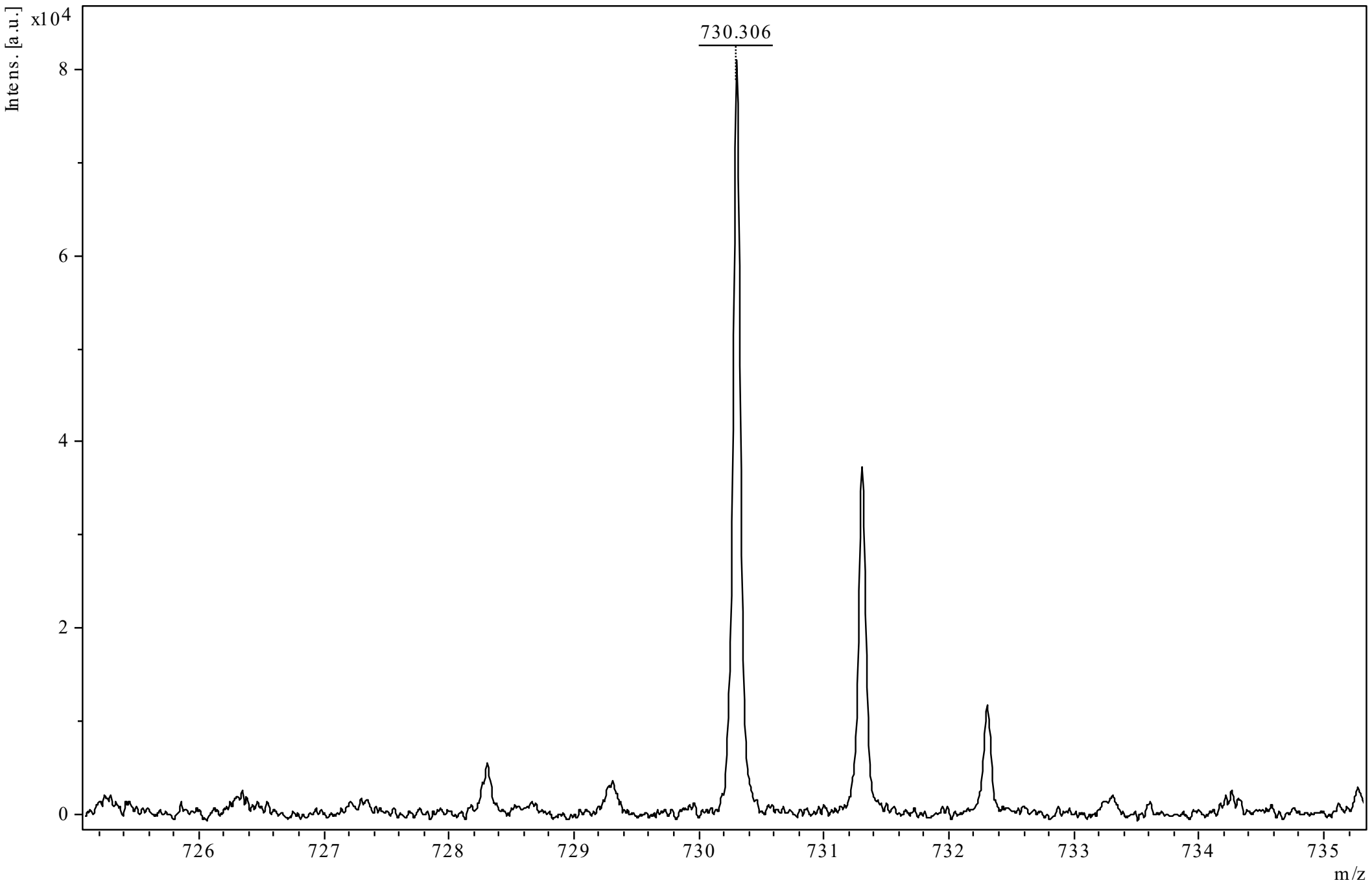

**Fig. S5.1 m/z 730.306 sample P2, matrix CHCA, S/N = 84**

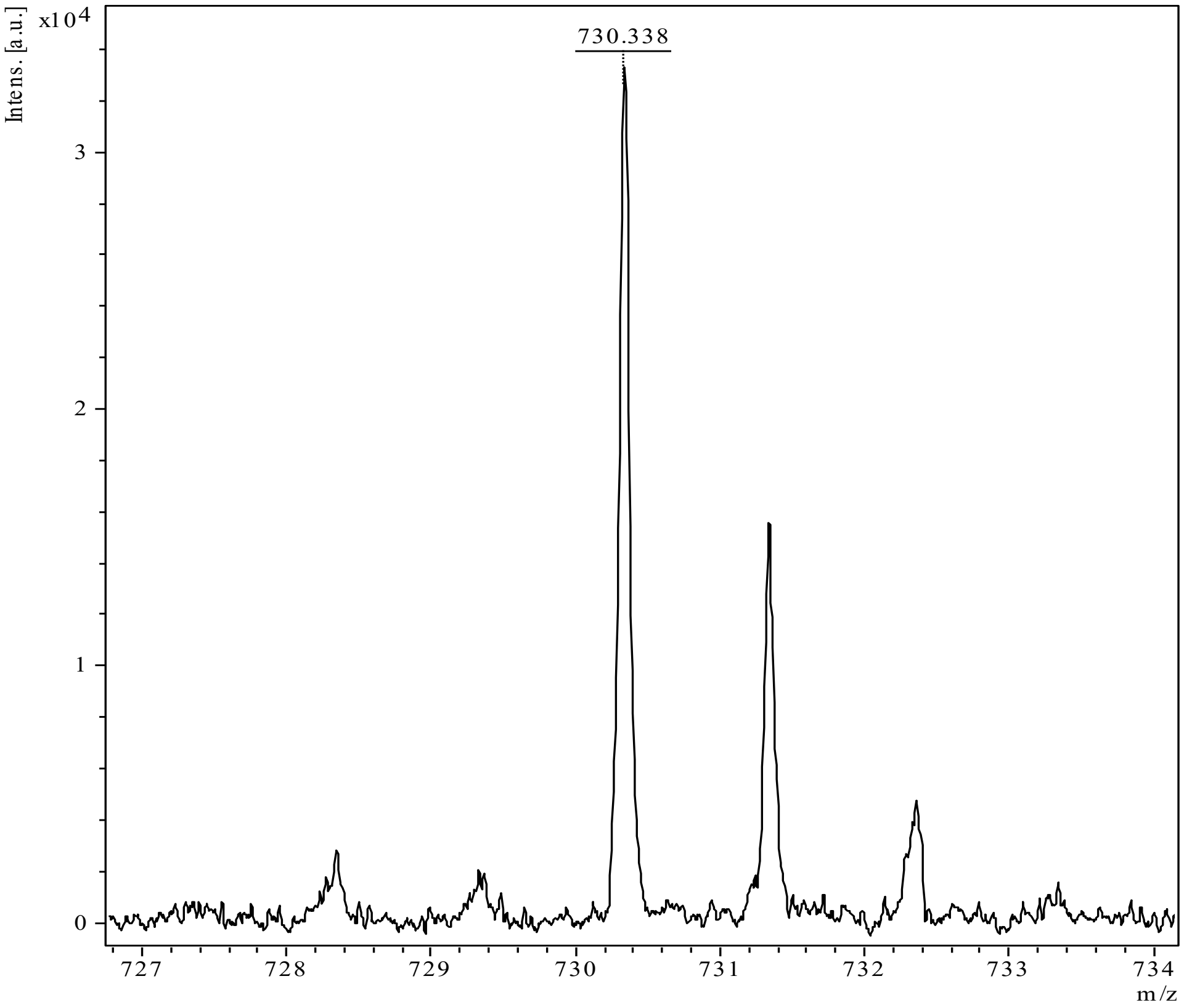

**Fig. S5.2 m/z 730.338 sample P2, matrix CHCA, S/N = 51**

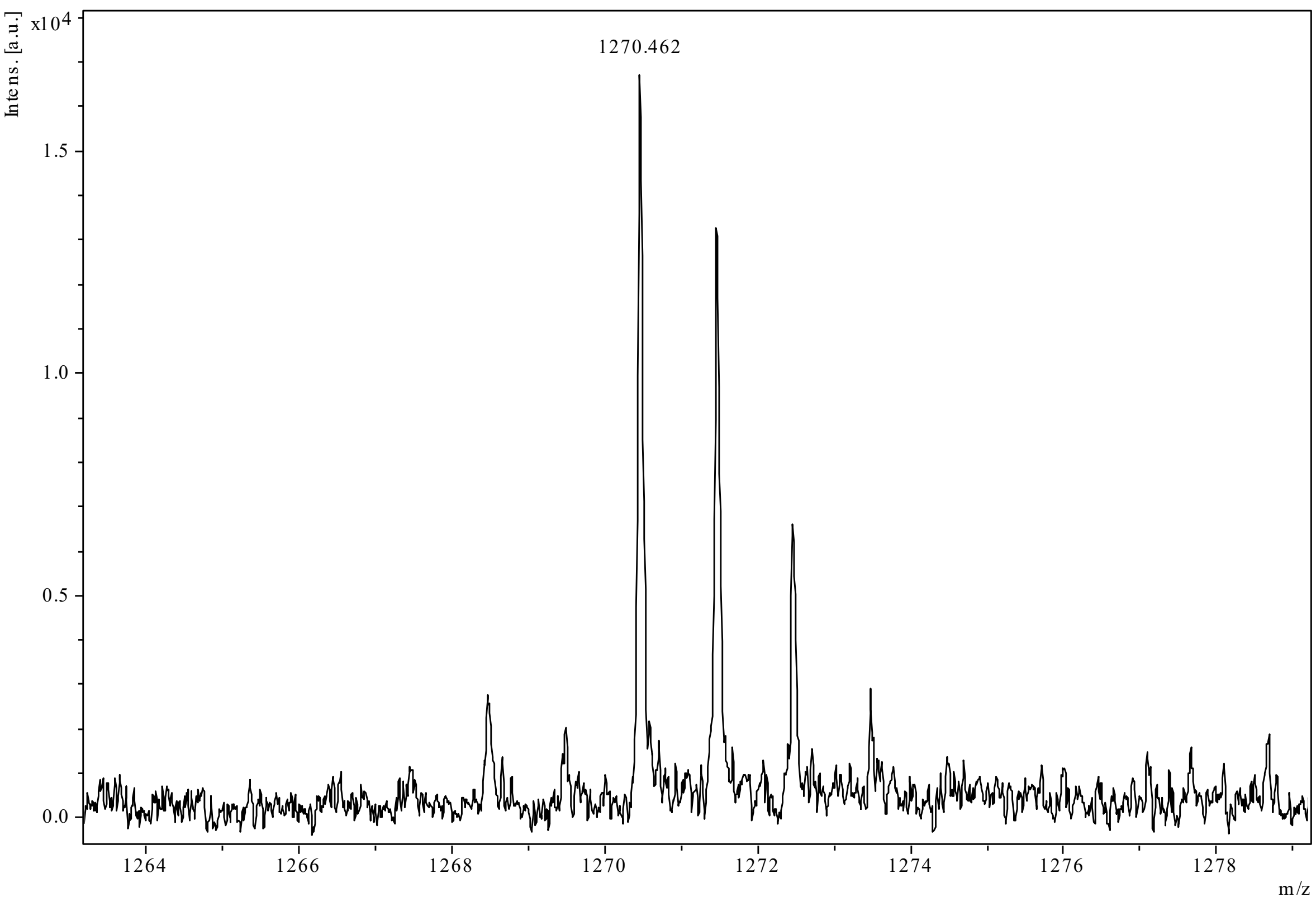

**Fig S5.3 m/z 1270.462 Sample P2, matrix CHCA, S/N = 22**

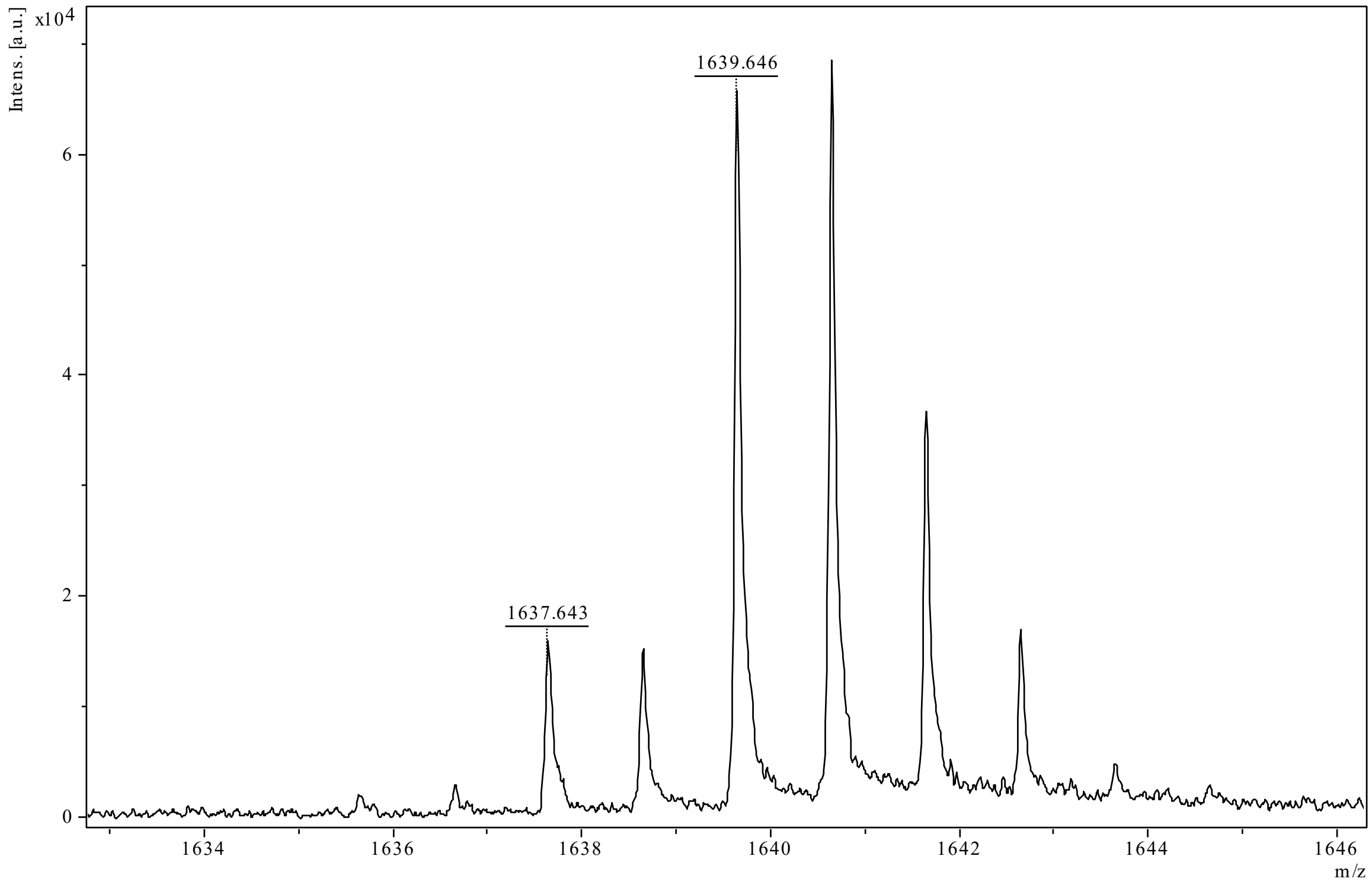

**Fig. S5.4 m/z 1639.646 Sample P2, matrix CHCA, S/N = 132**

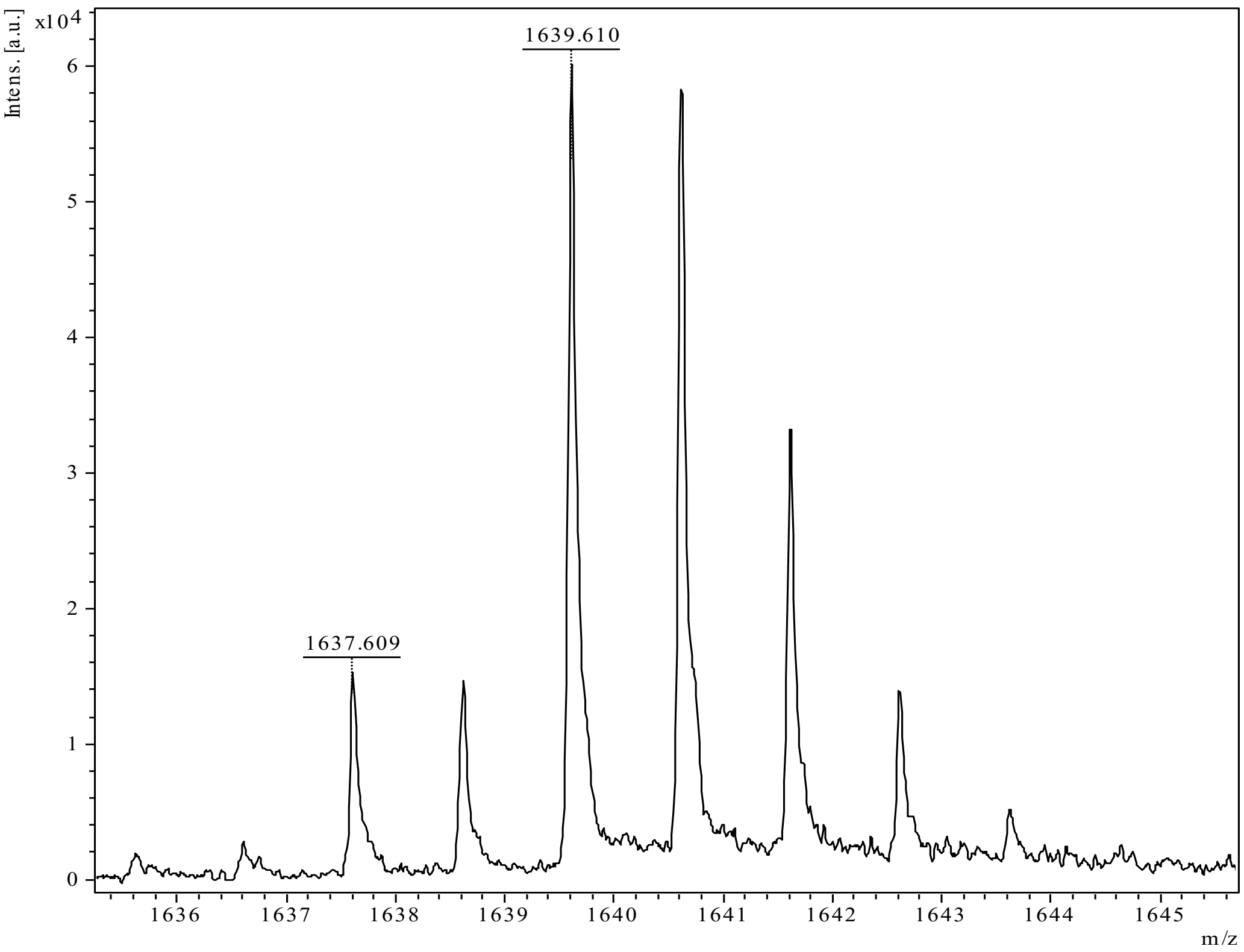

**Fig. S5.5   m/z 1639.610   Sample P2, matrix CHCA,  S/N = 112**

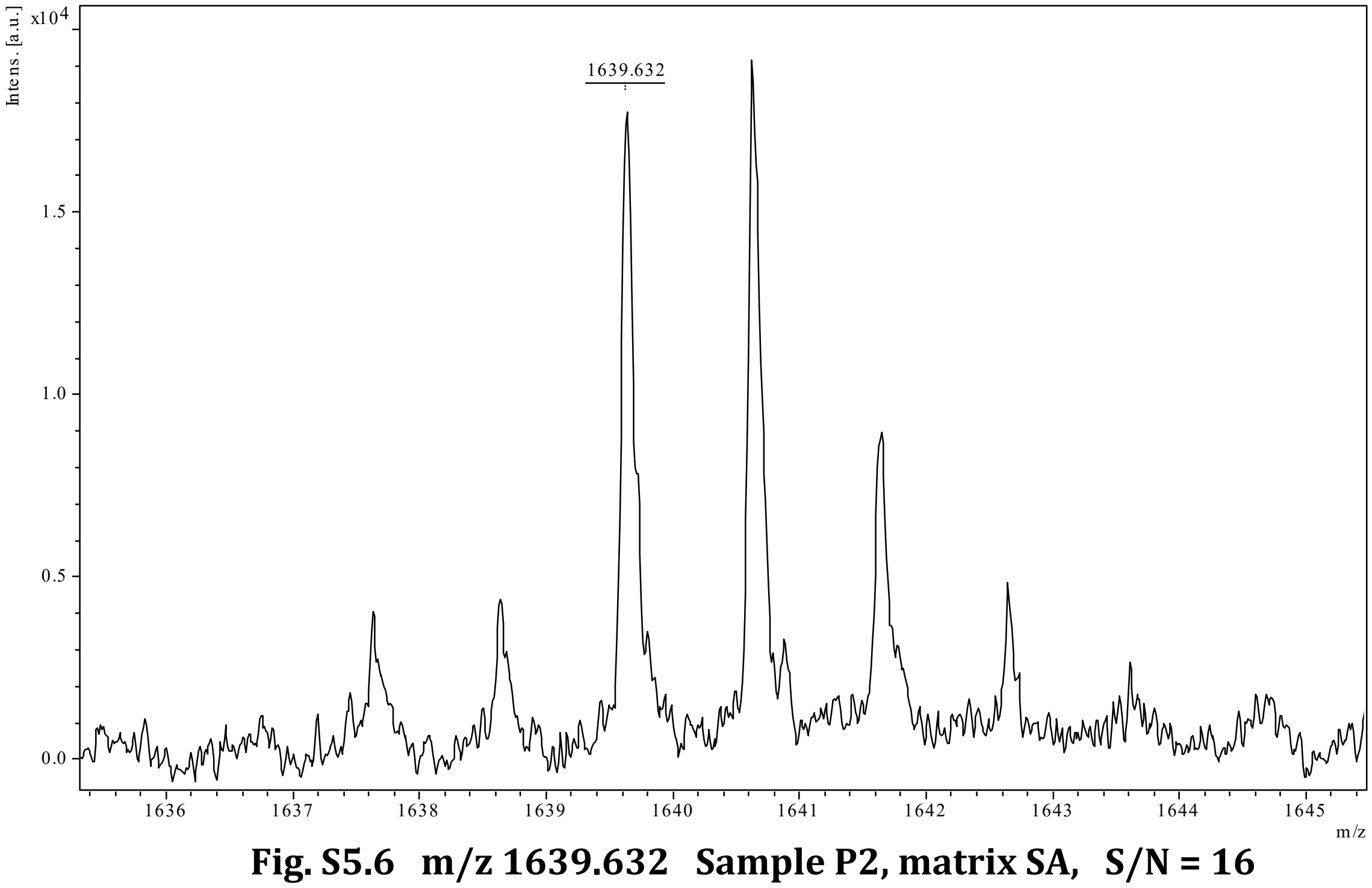

**Fig. S5.6   m/z 1639.632   Sample P2, matrix SA,   S/N = 16**

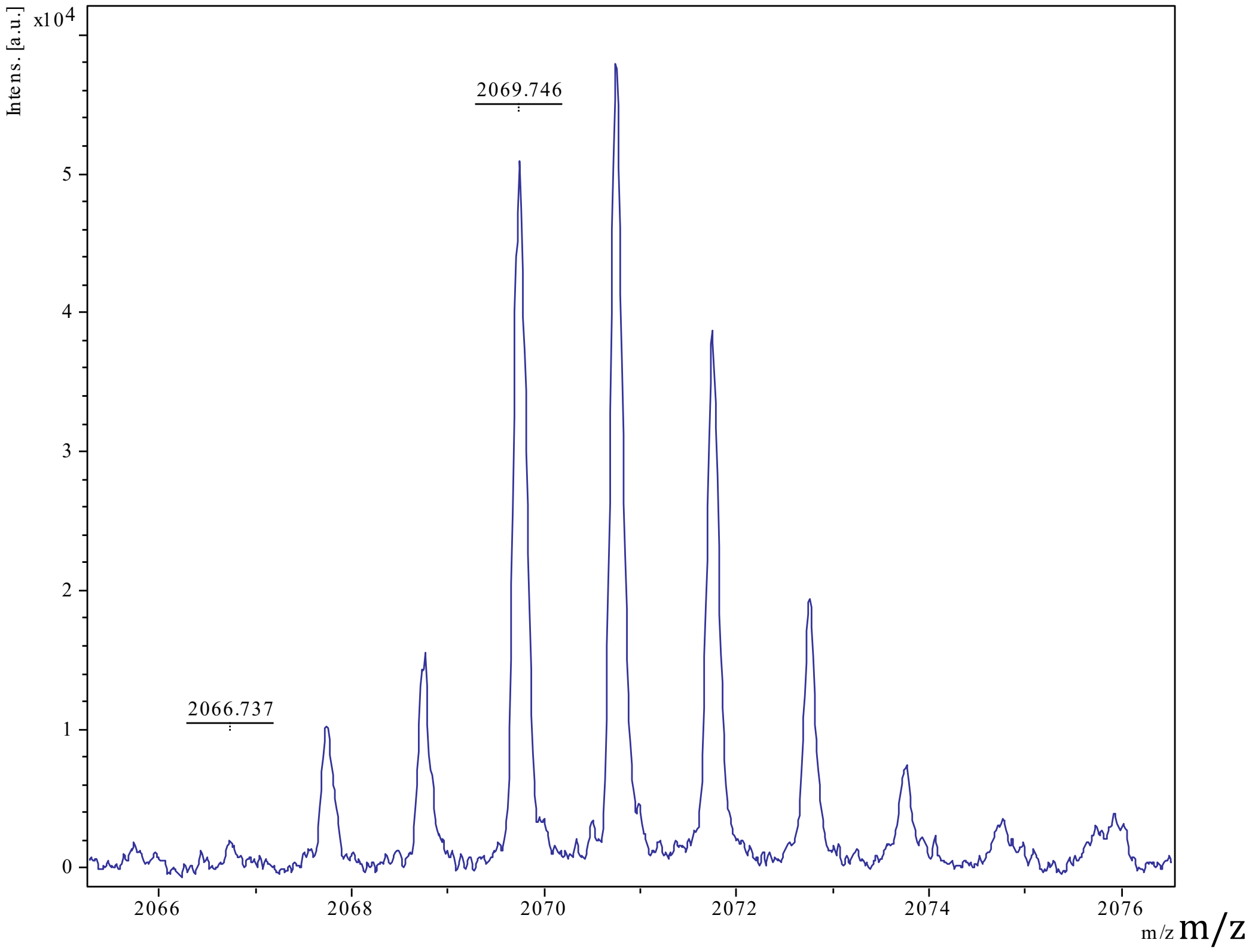

**Fig S5.7 m/z 2069.746 Sample P2, matrix SA, S/N = 63**

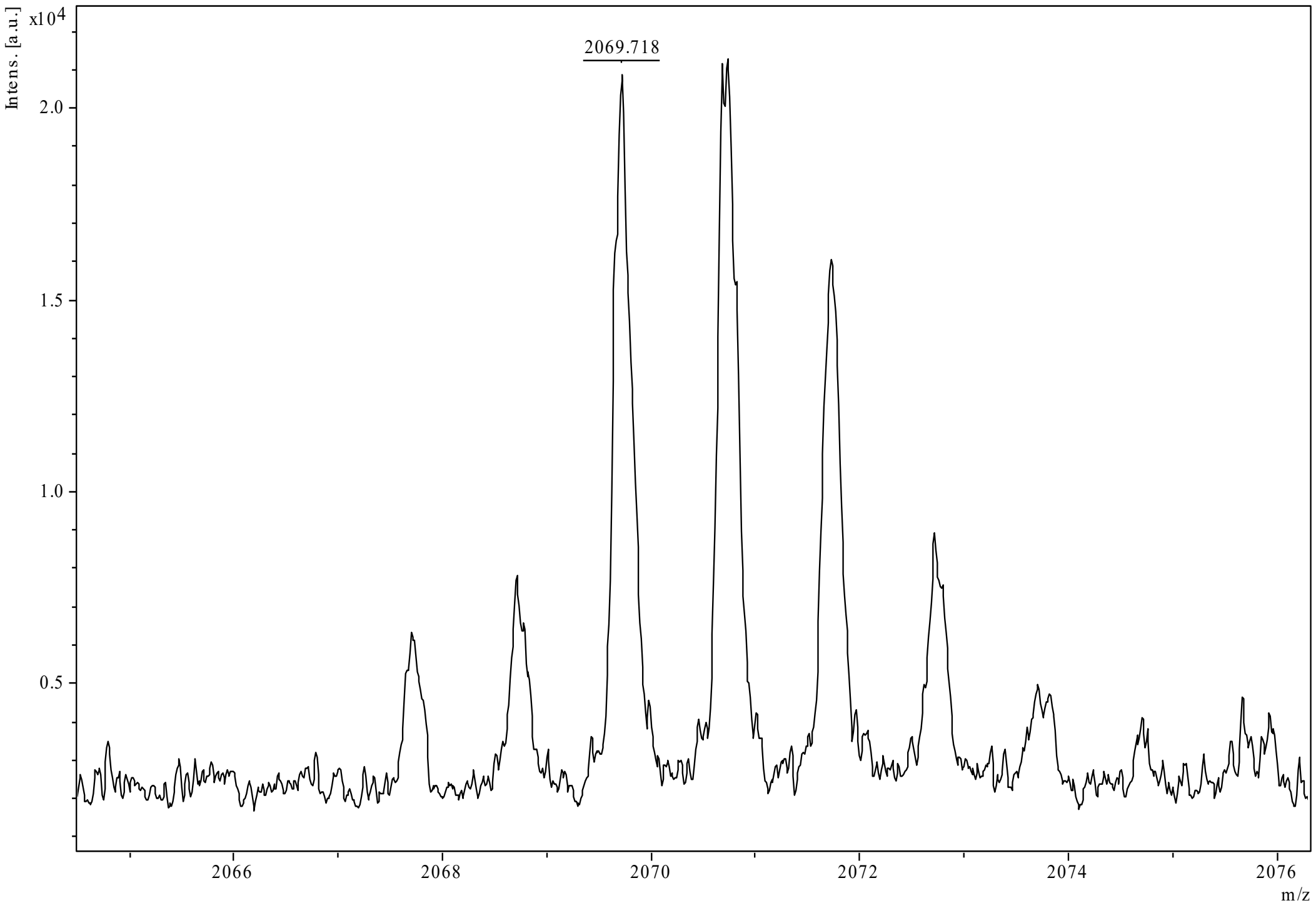

**Fig. S5.8 m/z 2069.718 Sample P1, matrix SA, S/N = 25**

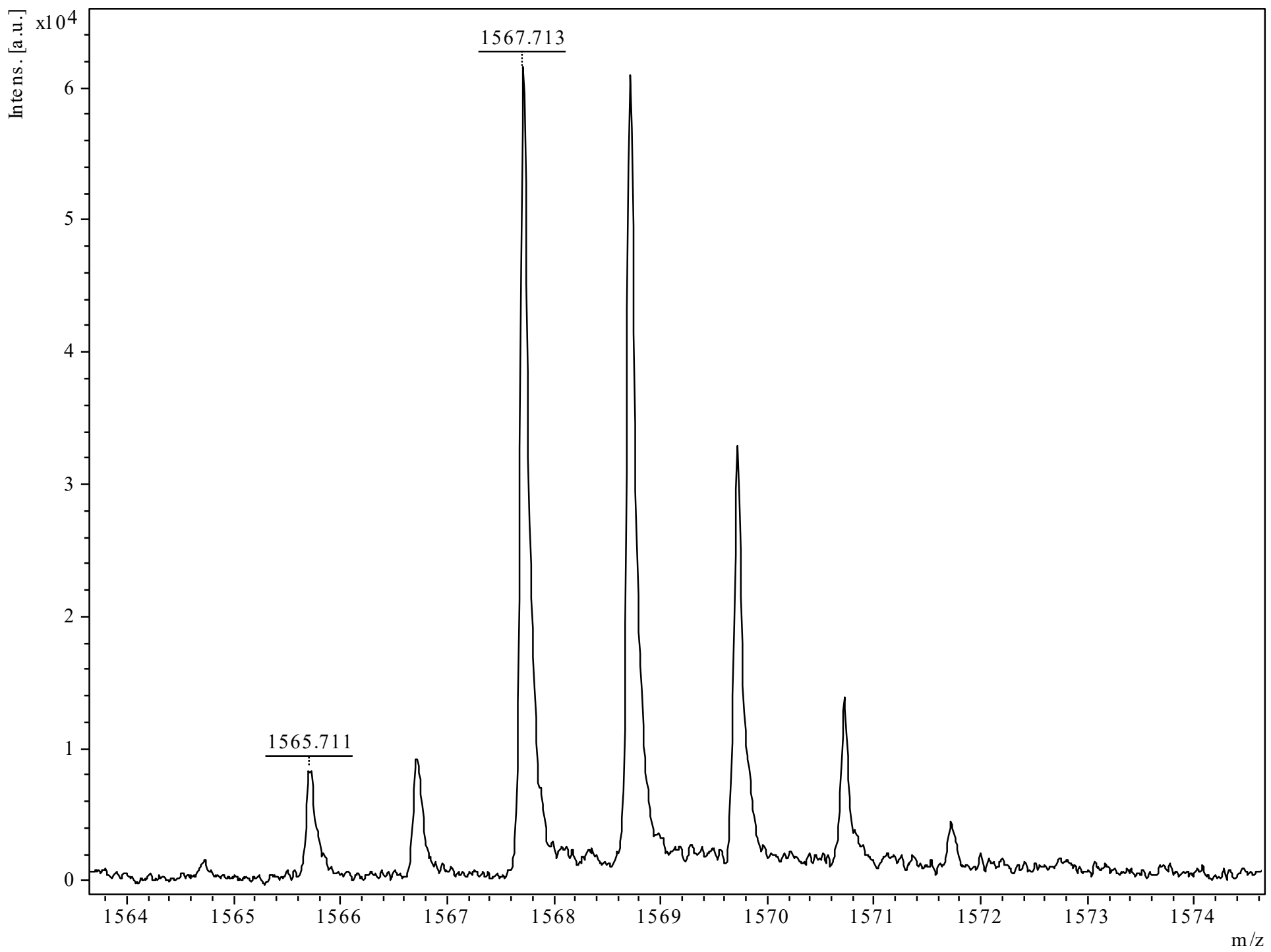

**Fig. S5.9 m/z 1567.713 Sample P2, matrix CHCA, S/N = 132**

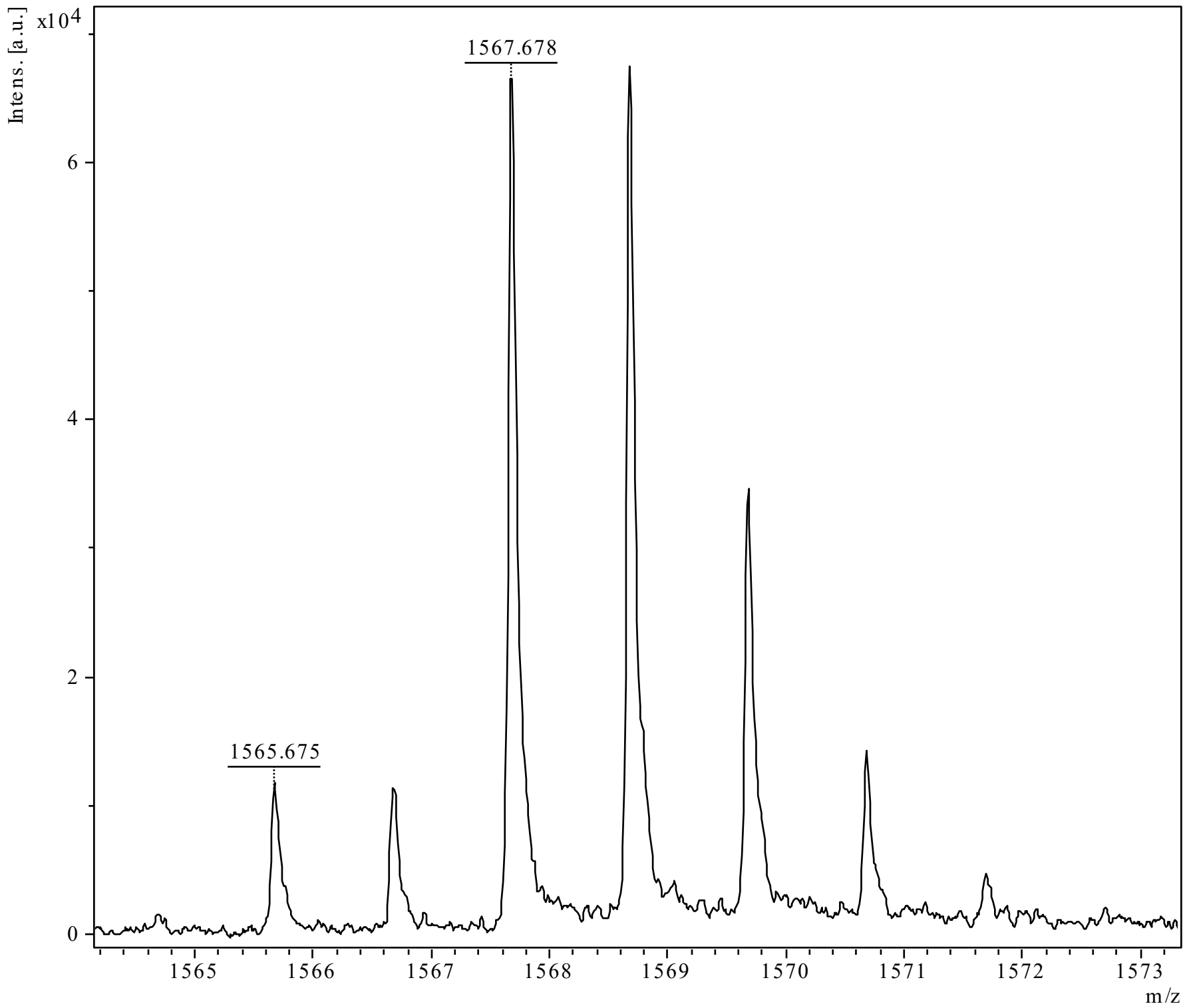

**Fig. S5.10 m/z 1567.678 Sample P2, matrix CHCA, S/N = 135**

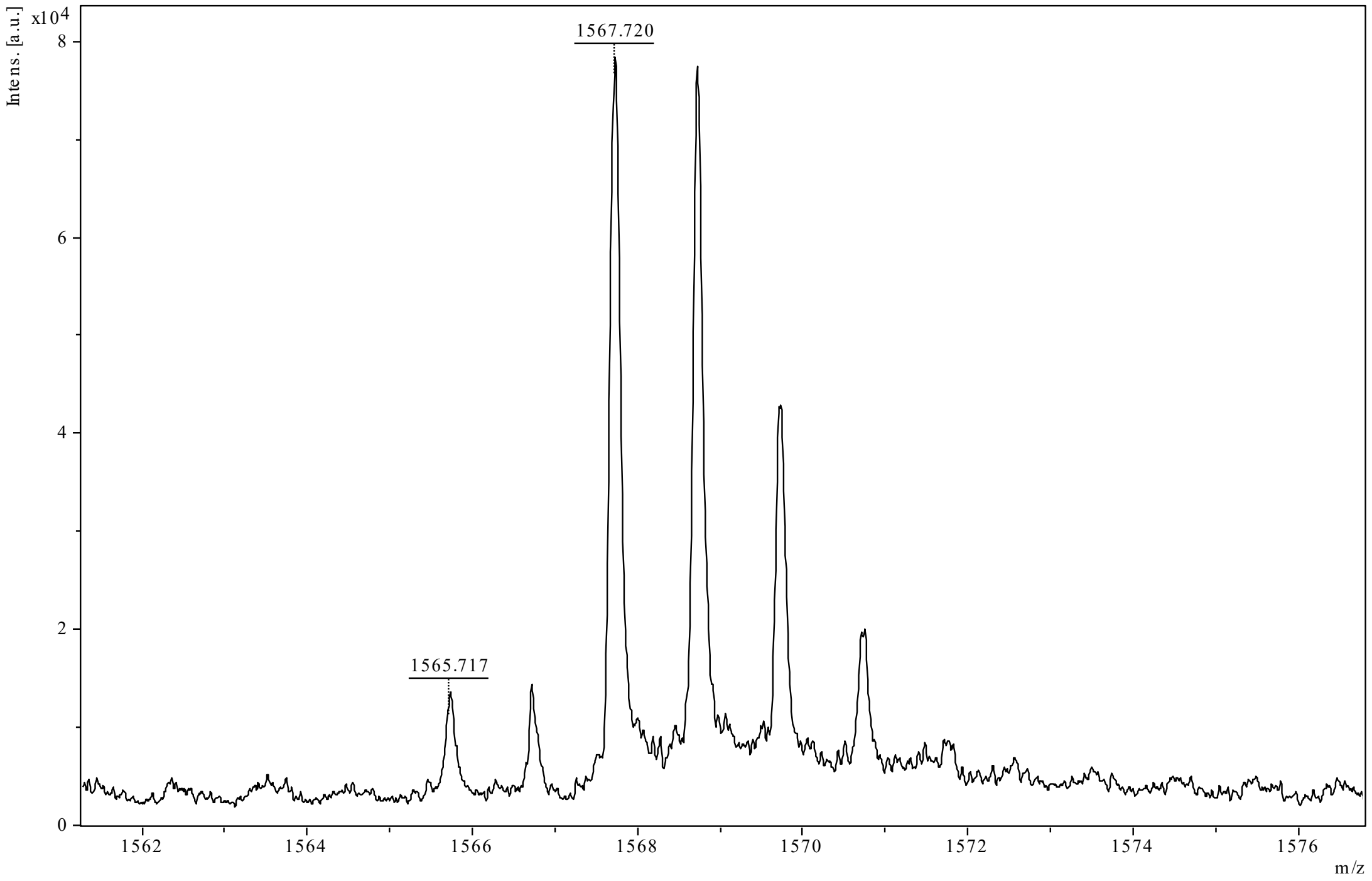

**Fig. S5.11  m/z 1567.720  Sample P1, matrix SA,  S/N = 67**

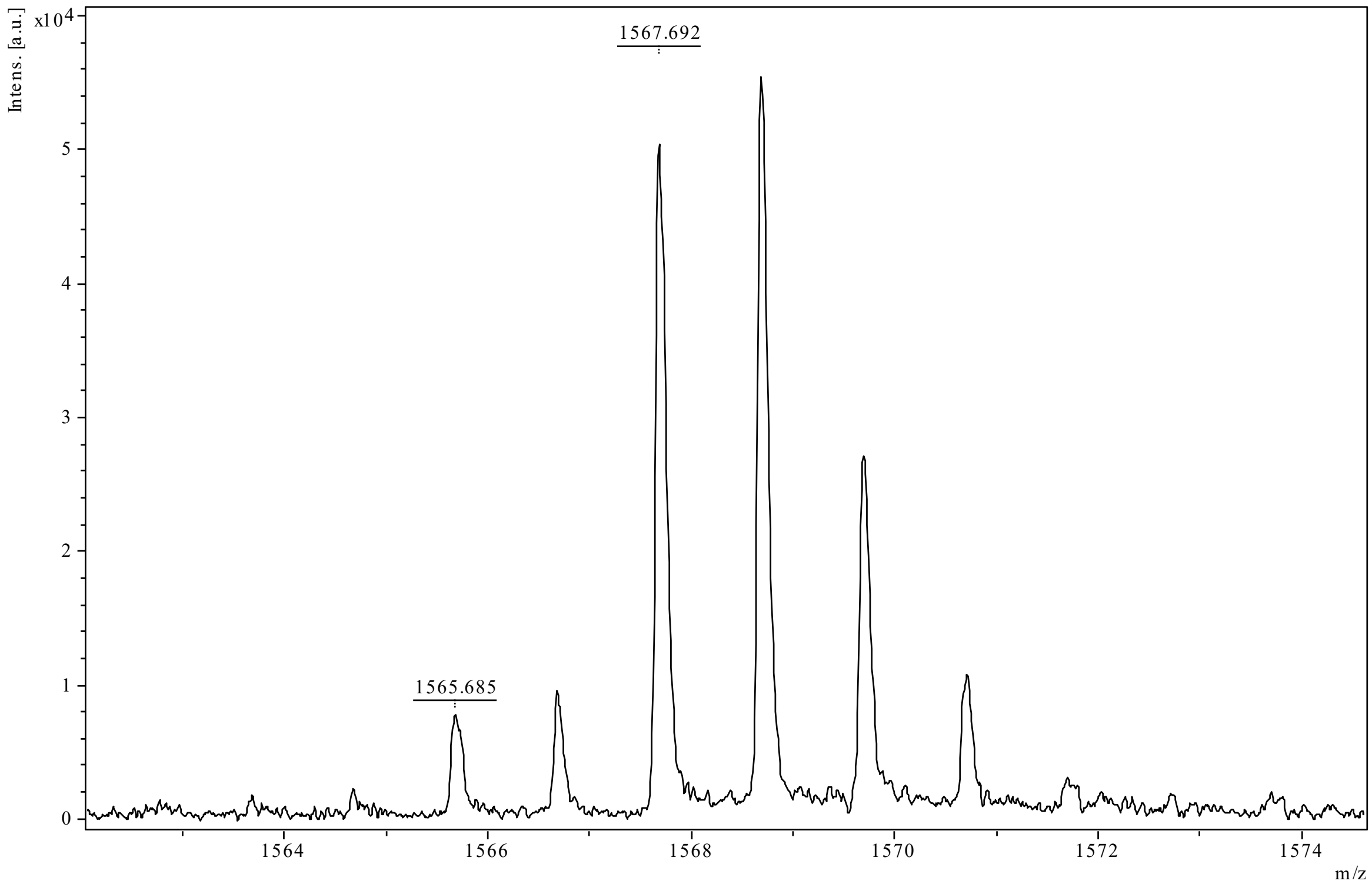

**Fig S5.12  m/z 1567.692  Sample P2, matrix CHCA,  S/N = 87**

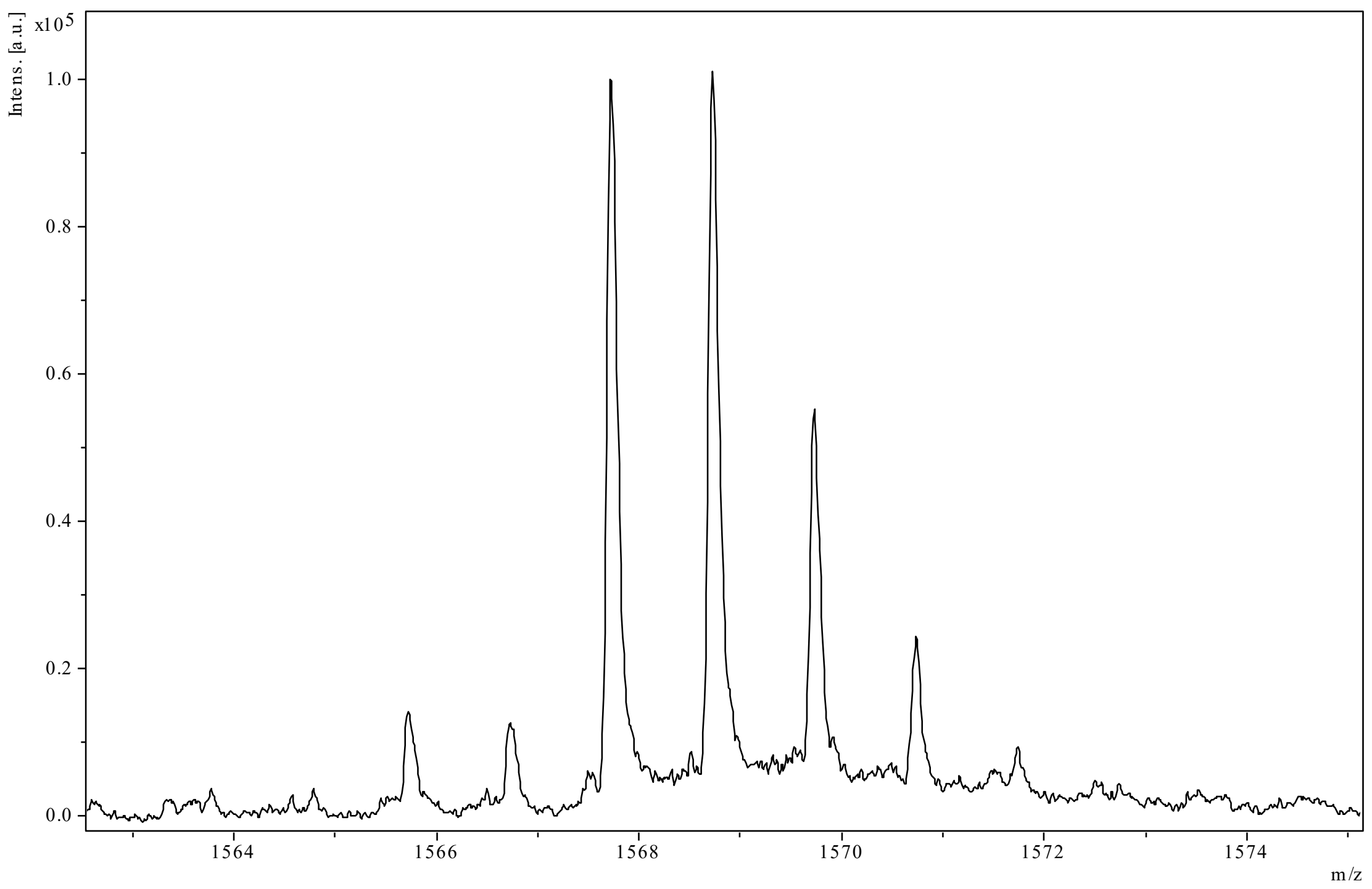

**Fig. S5.13 m/z 1567.725 Sample P2, matrix SA, S/N = 88**

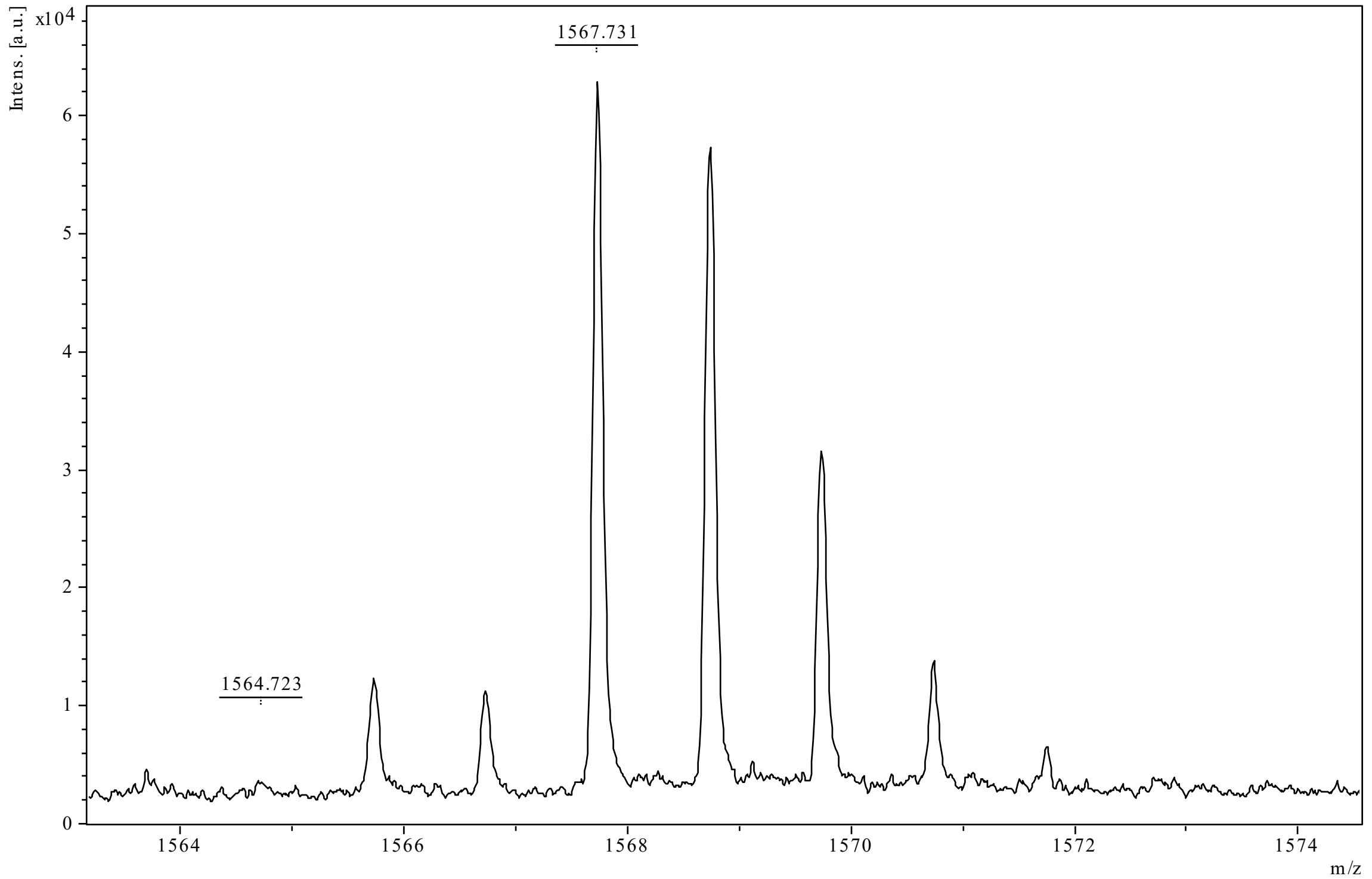

**Fig. S5.14 m/z 1567.731 Sample P4, matrix CHCA, S/N = 91**

## S6 Blank and control spectra

Figure S6.1 compares the blank from matrix alone with sample trace P2.

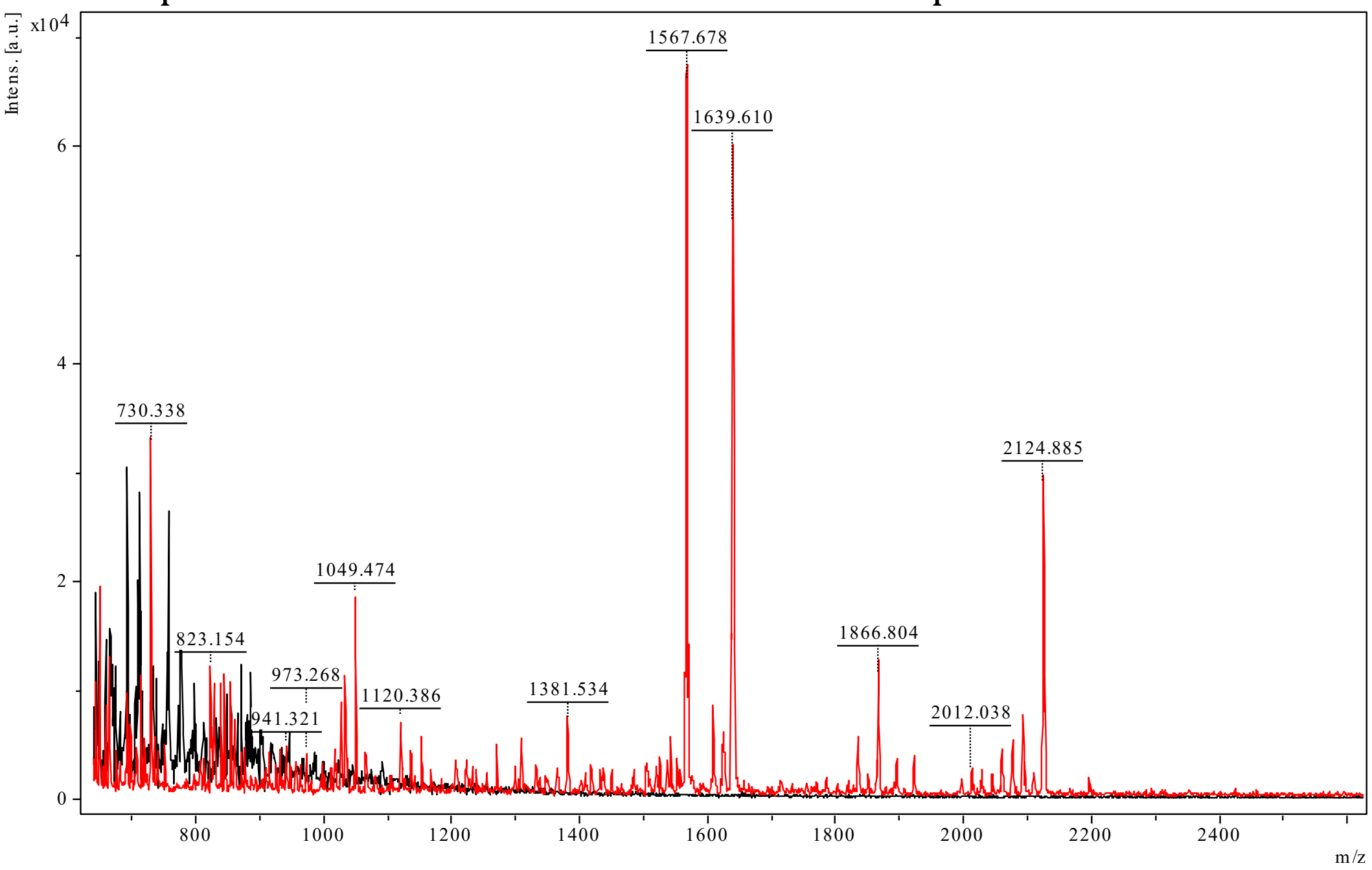

**Fig. S6.1 (Black trace) Matrix blank, CHCA**
**(Red trace) Sample P2, matrix CHCA**

Fig. S6.2 compares the top phase of the extract (P1) with the interphase (P2) that carries the highest hemoglycin and hemolithin signals.

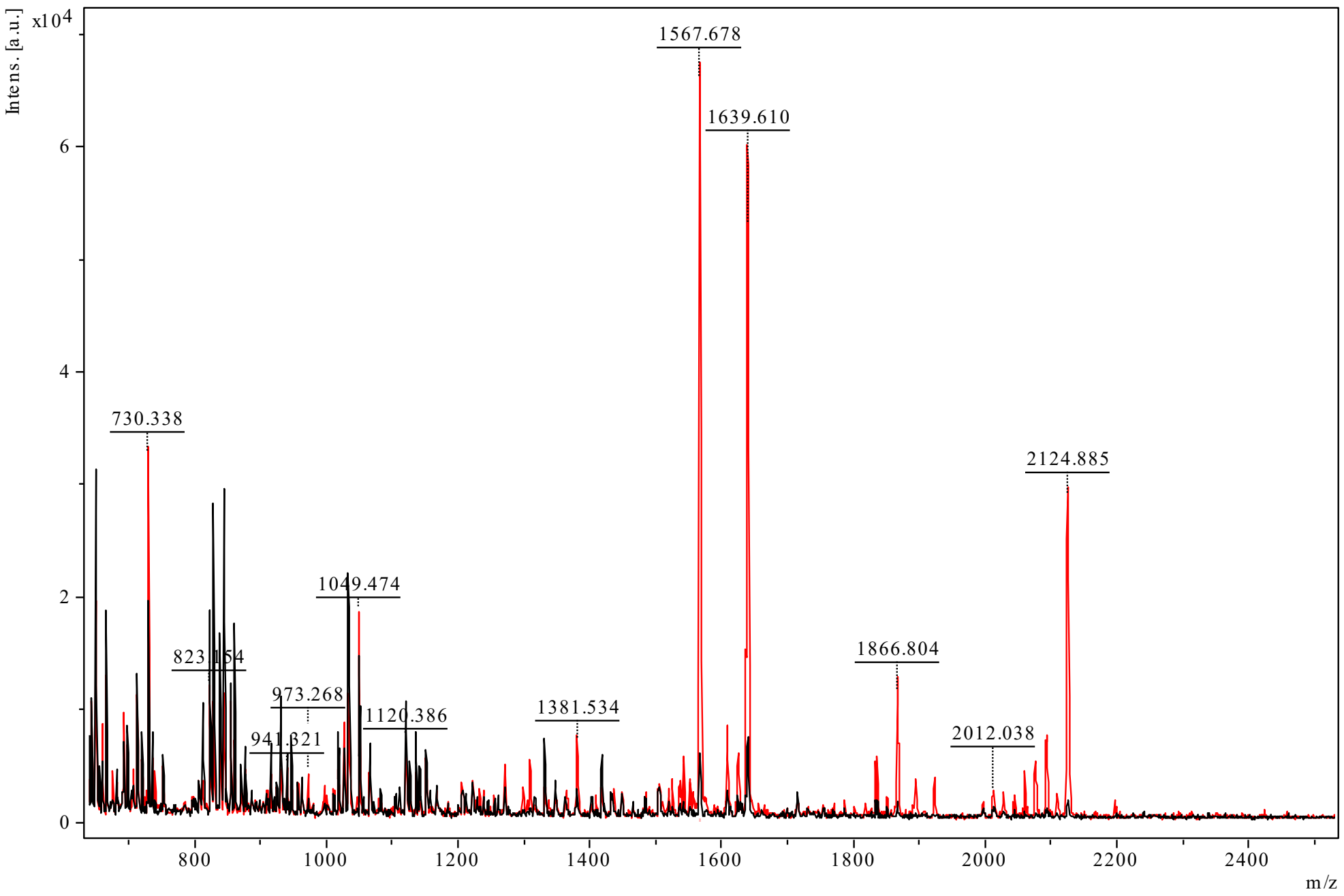

**Fig. S6.2 (Black trace) Sample P1, matrix CHCA. (Red trace) Sample P2, matrix CHCA**

## S7 Structural Analysis via MS/MS.

### MS/MS of the m/z 1567 peak

There is strong MS/MS support for the synthesized structure of the m/z 1567 entity of Figure 13 in the main text.

The 1567 peak that appeared prominently in the full spectrum (Figure 1, main paper) contained upon MS/MS analysis (method in main text) more than 30 peaks at signal-to-noise ratio between 3 and 35, as shown in Figures S7.1 and S7.2. This was by far the most informative MS/MS spectrum that we could obtain. MS/MS analysis selects via an applied voltage pulse the small region around m/z 1567, extending over isotopologues between m/z 1565 and 1570. There can be some under-recording at the limits of this range, which in consequence renders isotope analysis of the MS/MS fragments less accurate in terms of measurement of the number of iron atoms. However, approximate iron measurement is available at S/N levels exceeding 20. The molecule, once selected, is irradiated by laser to induce a high degree of internal energy and induce dissociation at many different bond locations, producing the MS/MS spectrum of fragments.

The fragments are assigned structures, presented in Table S7.1, that fit the observed mass while remaining exactly consistent with the master m/z 1567 structures given in line 1 of Table S.1 There are in fact two m/z 1567 isomers under consideration, as shown in the first entry of Table S7.1. In this analysis it was not possible to rule out either one of these. They may be able to interchange during high-energy excitation.

There is support in this MSMS analysis for many of the fundamental features of the 1494Da root entity that is the basis of m/z 1567 (Figures 12 and 13 in the main text).
a) polymers of glycine and hydroxy-glycine appear throughout, often attached to iron atoms, but also on their own in the mass range from 515 to 693.
b) many oxygen atoms are contained in three "reservoirs":
R1 as part of each peptide residue at the carbonyl group;
R2 as part of hydroxy-glycine;
R3 as part of the one or two-atom iron complex.
c) A residue count of 22 exists in 1567, exactly the same as for 1331, 1417, 1449 and 1639. Additionally, the m/z 1080 fragment has two Fe atoms according to its isotope analysis, as does m/z 405.

In the starting m/z 1567 structure there is an excess proton, which could be attached in many possible places. The fragment structures mostly have an excess proton, which could be the same one, again written separately. Cases in which there are zero additional mass units are possibly the result of the original excess proton having combined with a hydrogen atom to form $H_2$, which leaves the fragment and is not separately observed.

This analysis gives strong support to the 1567 and 1494Da core structures described in the main text that were originally reached via synthesis from the most dominant peaks in the main spectrum. As discussed, the standard automated peptide analysis was inadequate when faced with anti-parallel beta chains of glycine terminated by iron atoms such as we find here. Because of the closeness of the glycine residue mass (57Amu) to the dominant iron isotope (56Amu) it had been difficult to accurately separate these prior to the observation of $^{54}Fe$ isotopes in the (-1) and (-2) positions relative to a mass spectrum peak in the "mono-isotopic" (0) position.

The MS/MS spectrum from m/z 1639 was helpful in that it showed m/z 1567 and the core of m/z 1417 to be sub-components. However, unlike the peak at m/z 1567, the peak at m/z 1639.619 was accompanied by a known CHCA matrix cluster (at m/z 1639.232) that gave interfering fragments and hence was more difficult to completely interpret.

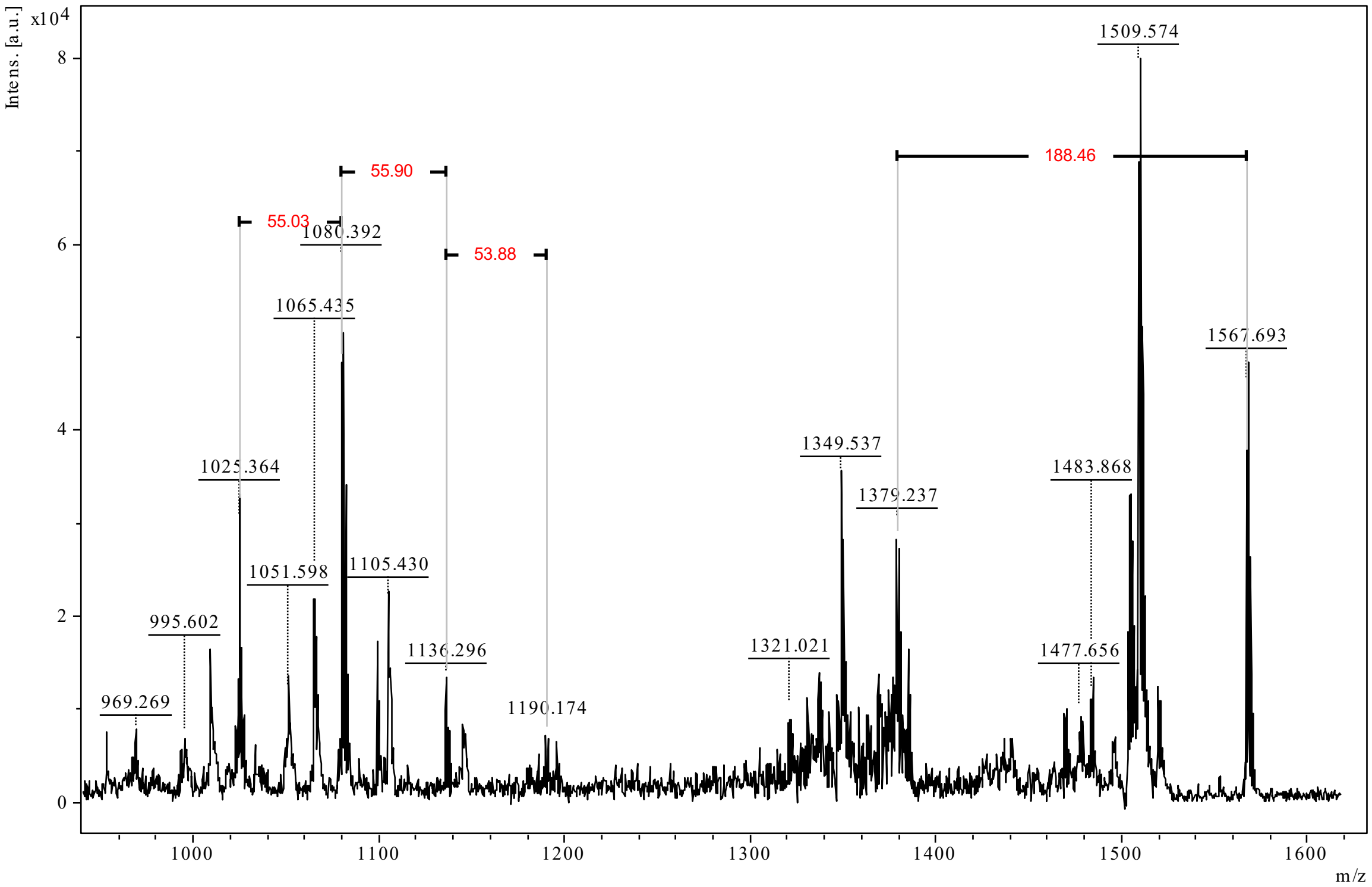

**Figure S7.1 The 1567 MSMS spectrum in the m/z range 1000 to 1600.**

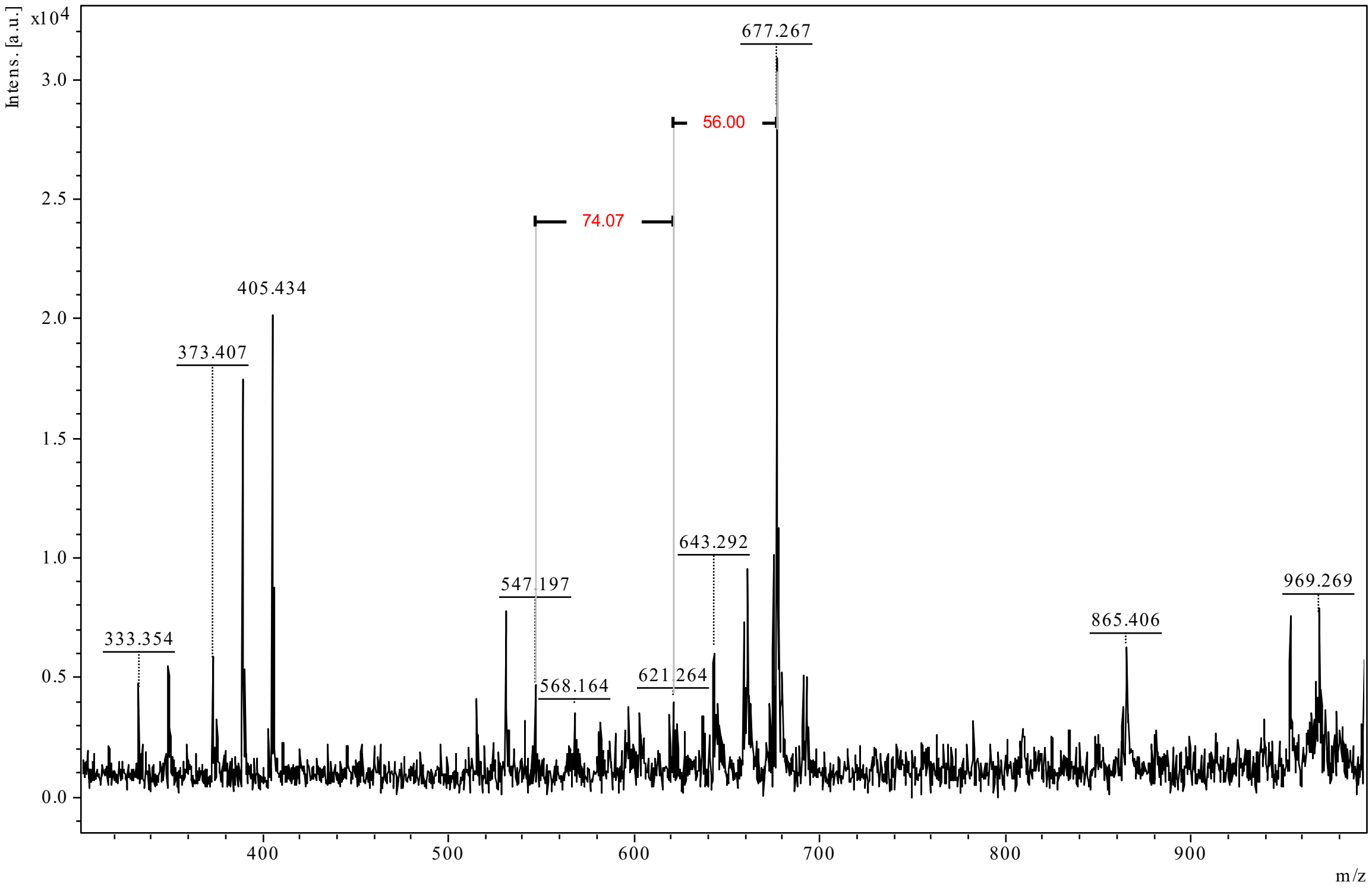

**Figure S7.2 The 1567 MSMS spectrum in the m/z range 300 to 1000**

**Table S7.1 Listing of peaks in the 1567 MS/MS spectrum.**

| Observed m/z | Intensity | Structure | S/N |
|---|---|---|---|
| 1567.691 | 241,061 | 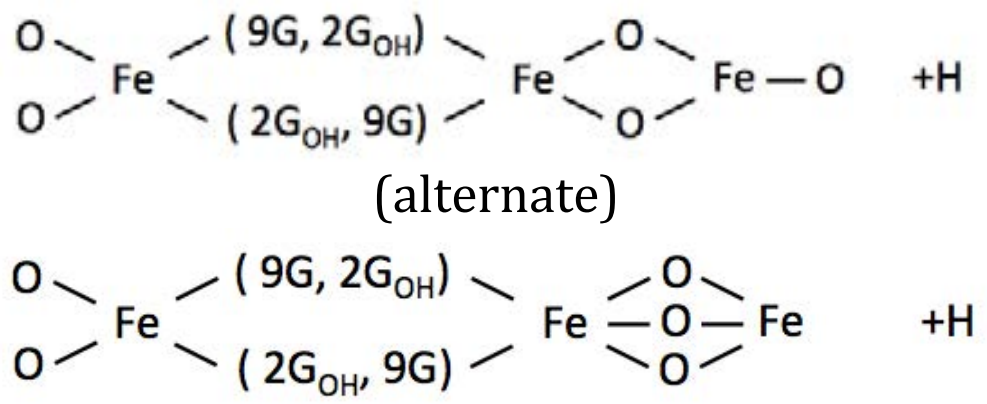 | 20 |
| 1519.921 | 12,433 | 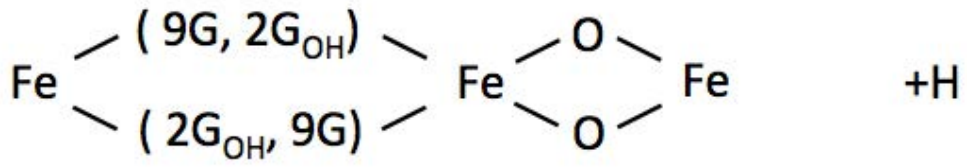 | 6 |
| 1509.574 | 68,734 | 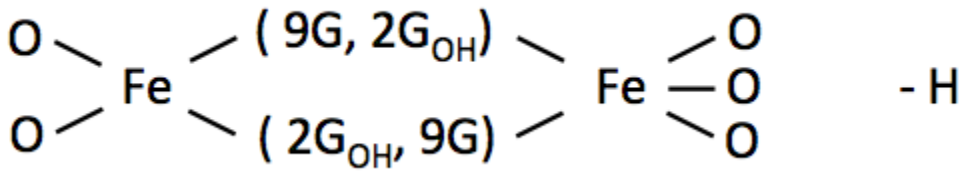 | 35 |
| 1503.819 | 18,359 | 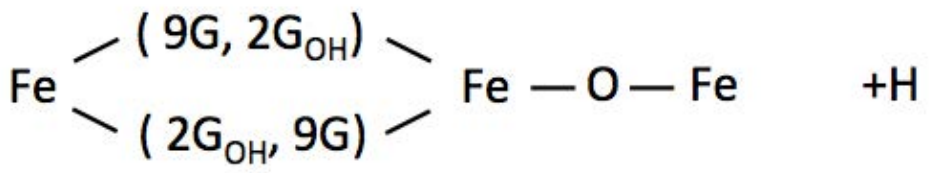 | 15 |
| 1478.658 | 9,161 | 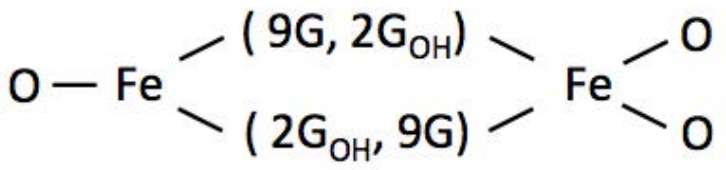 | 4 |
| 1379.237 | 28,185 | 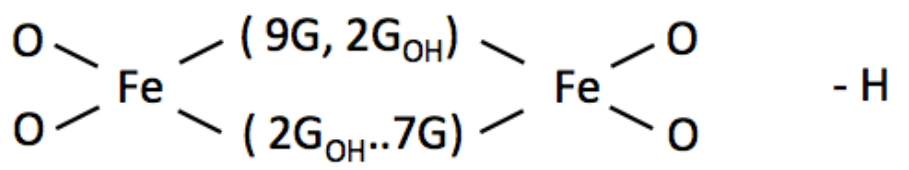 | 13 |
| 1349.537 | 35,711 | 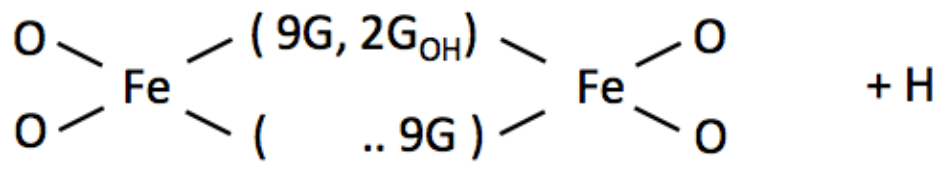 | 15 |
| 1136.296 | 12,378 | 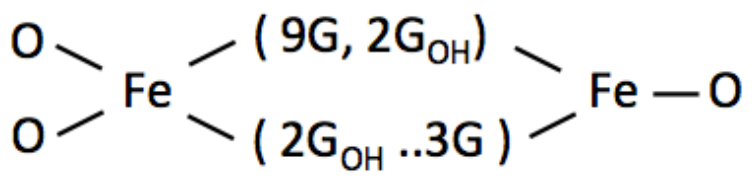 | 6 |
| 1105.326 | 21,809 | 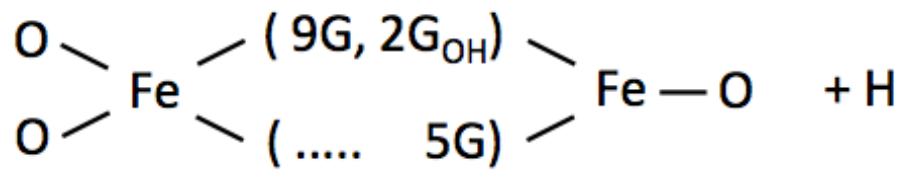 | 11 |
| 1080.392 | 47,283 | 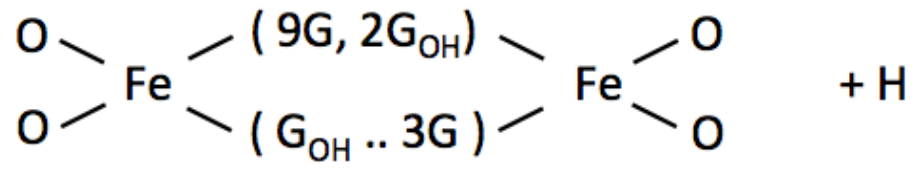 | 30 |
| 1065.435 | 21,632 | 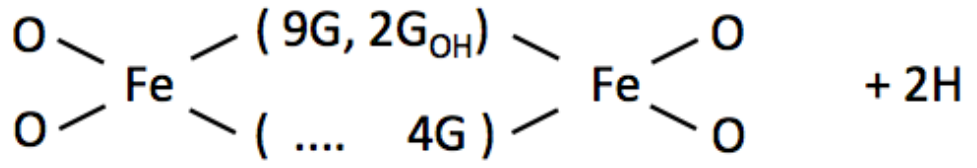 | 12 |
| 1049.534 | 19,080 | 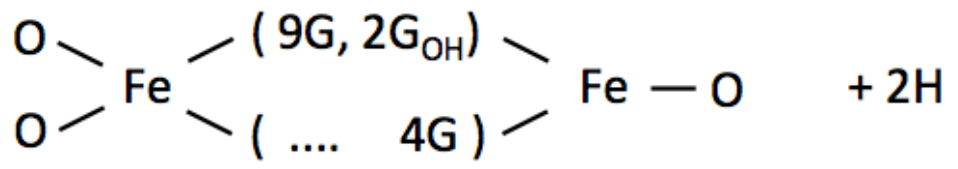 | 10 |
| 1025.364 | 33,140 | 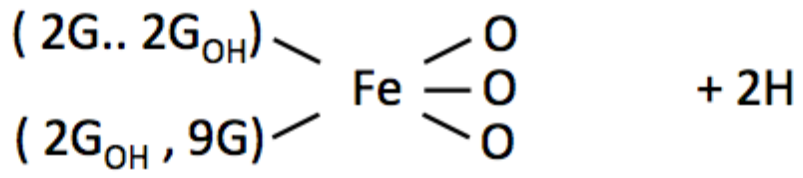 | 16 |
| 1009.431 | 15,864 | 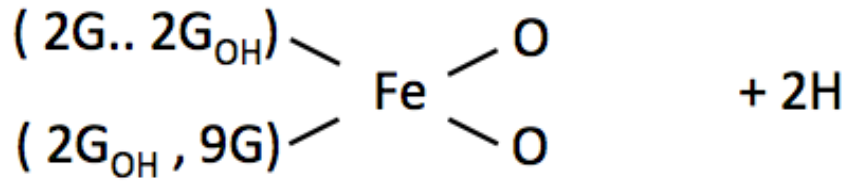 | 9 |
| 693.290 | 5,022 | ( 7G, 4G$_{OH}$) + 2H | 3 |
| 691.312 | 5,063 | ( 7G, 4G$_{OH}$) | 3 |
| 677.267 | 30,802 | ( 8G, 3G$_{OH}$) + 2H | 23 |
| 675.283 | 10,140 | ( 8G, 3G$_{OH}$) | 8 |
| 661.289 | 9,488 | ( 9G, 2G$_{OH}$) + 2H | 8 |
| 659.259 | 7,276 | ( 9G, 2G$_{OH}$) | 6 |
| 645.221 | 3,775 | ( 10G, G$_{OH}$) + 2H | 3 |
| 643.292 | 5,982 | ( 10G, $G_{OH}$) | 5 |
| 619.105 | 3,356 | ( 7G, $3G_{OH}$) + H | 3 |
| 547.197 | 4,694 | ( 7G, $2G_{OH}$) + 2H | 4 |
| 531.291 | 7,574 | ( 8G, $G_{OH}$) + 2H | 7 |
| 515.476 | 4,117 | (9G, $0G_{OH}$ ) + 2H | 4 |
| 405.437 | 20,145 | 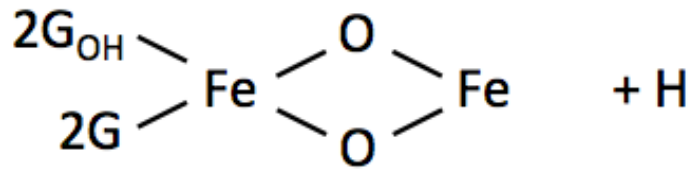 | 22 |
| 389.426 | 17,330 | 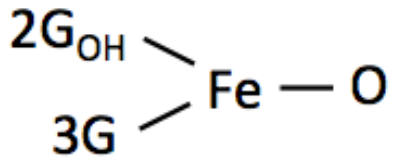 | 18 |
| 373.407 | 5,872 | 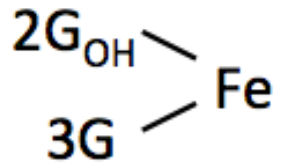 | 6 |
| 349.357 | 5,408 | 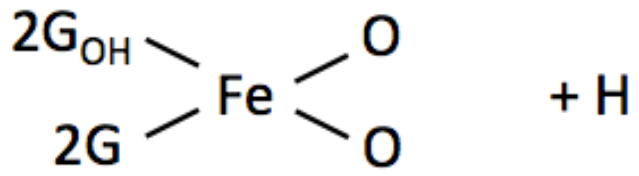 | 5 |
| 333.397 | 4,731 | 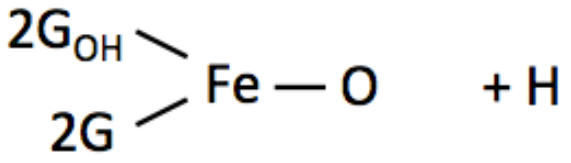 | 5 |

# References for Supplementary Information

(References are listed here with the same numbers as in the main text)

1. McGeoch JEM and McGeoch MW. Polymer amide in the Allende and Murchison meteorites. *Meteorit and Planet Sci.* 2015; **50**: 1971-1983.

2. McGeoch JEM and McGeoch MW. A 4641Da. Polymer of Amino Acids in Allende and Acfer 086 Meteorites. arXiv:1707.09080 (28th July 2017).

7. Harris WA, Janecki DJ and Reilly JP. The use of matrix clusters and trypsin autolysis fragments as mass calibrants in matrix-assisted laser desorption/ ionization time-of-flight mass spectrometry. *Rapid Commun. Mass Spectrom.* 2002; **16**: 1714-1722. doi: 10.1002/rcm.775.

8. Atomic Weights and Isotopic Compositions for All Elements. https://physics.nist.gov/cgi-bin/Compositions/stand_alone.pl

9. McGeoch MW, Samoril T, Zapotok D and McGeoch JEM. Polymer amide as a carrier of $^{15}N$ in Allende and Acfer 086 meteorites. www.ArXiv.org/abs/1811.06578

10. Lu I-C, Chu KY, Lin CY et al. Ion-to-Neutral Ratios and Thermal Proton Transfer in Matrix-Assisted Laser Desorption/Ionization. *J. Am. Soc. Mass Spectrom.,* (2015) **26**:1242-1251. DOI: 10.1007/s13361-015-1112-3

18. Duprat J et al. Extreme Deuterium Excesses in Ultracarbonaceous Micrometeorites from Central Antarctic Snow. *Science* 2010; **328**: 742-745.

19. Hashizume K, Takahata N, Naraoka H and Sano Y. Extreme oxygen isotope anomaly with a solar origin detected in meteoritic organics. *Nature Geosci.* 2011; **4**: 165-168.

23. Reference and inter-comparison materials for stable isotopes of light elements. IAEA, Vienna, 1995. IAEA-TECDOC-825, ISSN 1011-4289.

24. Sephton MA, James RH, Zolensky ME, The origin of dark inclusions in Allende: New evidence from lithium isotopes. *Meteorit and Planet Sci* 2006; **41**: 1039-1043.

25. Seitz H-M, Brey GP, Zipfel J, Ott U, Weyer S, Durali S, et al. Lithium isotope composition of ordinary and carbonaceous chondrites, and differentiated planetary bodies: Bulk solar system and solar reservoirs. *Earth and Planet Sci Lett.* 2007; **260**: 582-596.

26. Dauphas N, Cook DL, Sacarabany A, Frohlich C et al. Iron 60 evidence for early injection and efficient mixing of stellar debris in the protosolar nebula. *Astrophys J* 2008; **686**: 560-569.

27. Dauphas N, Janney PE, Mendybaev RA, Wadhwa M et al. Chromatographic separation and multicollection ICPMS analysis of iron. Investigating mass-dependent and -independent isotope effects. *Anal Chem* 2004; **76**: 5855-5863.

28. Clayton RN, Mayeda TK. Oxygen isotope studies of carbonaceous chondrites. *Geochim et Cosmochim Acta* 1999; **63**: 2089-2104.

29. Pearson VK, Sephton MA, Franchi IA, Gibson JM and Gilmour I. 2006. Carbon and nitrogen in carbonaceous chondrites: Elemental abundances and stable isotope compositions. *Meteorit and Planet Sci.* 2006; **41**:1899-1918.

30. Pizzarello S, Huang Y and Fuller M. The carbon isotopic distribution of Murchison amino acids. *Geochim et Cosmochim Acta*. 2004; **68**: 4963 – 4969.

31. Prombo CA and Clayton RN. A Striking Nitrogen Isotope Anomaly in the Bencubbin and Weatherford Meteorites. *Science* 1985; **230**: 935-937.

32. Pizzarello S and Holmes W. Nitrogen-containing compounds in two CR2 meteorites: $^{15}N$ composition, molecular distribution and precursor molecules. *Geochim et Cosmochim Acta*. 2009; **73**: 2150-2162.

33. Arpigny C, Jehin E, Manfroid J, Hutsemekers D, Schulz R, Stuwe JA, Zucconi JM and Ilyin I. Anomalous Nitrogen Isotope Ratio in Comets. *Science* 2003; **301**:1522-1524.

34. Wirstrom ES, Charnley SB, Cordiner MA, Milam SN. Isotopic Anomalies in Primitive Solar System Matter: Spin-state Dependent Fractionation of Nitrogen and Deuterium in Interstellar Clouds. *Astrophys J.* 2012; **757**: 1208.0192

35. Pizzarello S and Huang Y. The deuterium enrichment of individual amino acids in carbonaceous meteorites: A case for the presolar distribution of biomolecule precursors. *Geochim et Cosmochim Acta*. 2005; **69**: 599 – 605.

36. Hily-Blant P, Bonal L, Faure A, Quirico E, The $^{15}N$-enrichment in dark clouds and Solar System objects, *Icarus* 2013; **223**: 582-590.